\documentclass[12pt,a4paper,french,english]{article}
\usepackage[T1]{fontenc}
\usepackage[latin9]{inputenc}
\usepackage{color}
\definecolor{shadecolor}{rgb}{1, 0, 0}
\usepackage{babel}
\makeatletter
\addto\extrasfrench{%
   \providecommand{\fg}{\ifdim\lastskip>\z@\unskip\fi~\frqq}%
}

\makeatother
\usepackage{framed}
\usepackage{textcomp}
\usepackage{amsthm}
\usepackage{amsmath}
\usepackage{amssymb}
\usepackage{graphicx}
\usepackage[unicode=true,pdfusetitle,
 bookmarks=true,bookmarksnumbered=false,bookmarksopen=false,
 breaklinks=false,pdfborder={0 0 1},backref=false,colorlinks=false]
 {hyperref}
\usepackage{breakurl}

\makeatletter

\numberwithin{figure}{section}
\numberwithin{equation}{section}
\theoremstyle{plain}
\newtheorem{thm}{\protect\theoremname}[section]
  \theoremstyle{definition}
  \newtheorem{defn}[thm]{\protect\definitionname}
  \theoremstyle{remark}
  \newtheorem{rem}[thm]{\protect\remarkname}
  \theoremstyle{plain}
  \newtheorem{lem}[thm]{\protect\lemmaname}
  \theoremstyle{plain}
  \newtheorem{prop}[thm]{\protect\propositionname}
  \theoremstyle{plain}
  \newtheorem{assumption}[thm]{\protect\assumptionname}
  \theoremstyle{plain}
  \newtheorem{cor}[thm]{\protect\corollaryname}

\usepackage{a4wide}
\usepackage{color}
\usepackage{graphics}

\usepackage{amsmath, amsthm, amssymb}

\makeatother

  \addto\captionsenglish{\renewcommand{\assumptionname}{Assumption}}
  \addto\captionsenglish{\renewcommand{\corollaryname}{Corollary}}
  \addto\captionsenglish{\renewcommand{\definitionname}{Definition}}
  \addto\captionsenglish{\renewcommand{\lemmaname}{Lemma}}
  \addto\captionsenglish{\renewcommand{\propositionname}{Proposition}}
  \addto\captionsenglish{\renewcommand{\remarkname}{Remark}}
  \addto\captionsenglish{\renewcommand{\theoremname}{Theorem}}
  \addto\captionsfrench{\renewcommand{\assumptionname}{Hypothèse}}
  \addto\captionsfrench{\renewcommand{\corollaryname}{Corollaire}}
  \addto\captionsfrench{\renewcommand{\definitionname}{Définition}}
  \addto\captionsfrench{\renewcommand{\lemmaname}{Lemme}}
  \addto\captionsfrench{\renewcommand{\propositionname}{Proposition}}
  \addto\captionsfrench{\renewcommand{\remarkname}{Remarque}}
  \addto\captionsfrench{\renewcommand{\theoremname}{Théorème}}
  \providecommand{\assumptionname}{Assumption}
  \providecommand{\corollaryname}{Corollary}
  \providecommand{\definitionname}{Definition}
  \providecommand{\lemmaname}{Lemma}
  \providecommand{\propositionname}{Proposition}
  \providecommand{\remarkname}{Remark}
\providecommand{\theoremname}{Theorem}

\begin{document}

\title{Asymptotic spectral gap and Weyl law for Ruelle resonances of open
partially expanding maps}

\date{01-06-2013}

\author{Jean Francois Arnoldi\textit{}%
\thanks{Institut Fourier, UMR 5582, 100 rue des Maths, BP74 38402 St Martin
d'Hères.%
}, Frédéric Faure\textit{}%
\thanks{Institut Fourier, UMR 5582, 100 rue des Maths, BP74 38402 St Martin
d'Hères. frederic.faure@ujf-grenoble.fr http://www-fourier.ujf-grenoble.fr/\textasciitilde{}faure%
}\textit{,} Tobias Weich\textit{}%
\thanks{Fachbereich Mathematik, Philipps-Universit\"{ä}t Marburg, a Hans-Meerwein-Straße,
35032 Marburg, Germany. weich@mathematik.uni-marburg.de http://www.mathematik.uni-marburg.de/\textasciitilde{}weich/%
}}
\maketitle
\begin{abstract}
We consider a simple model of an open partially expanding map. Its
trapped set $\mathcal{K}$ in phase space is a fractal set. We first
show that there is a well defined discrete spectrum of Ruelle resonances
which describes the asymptotisc of correlation functions for large
time and which is parametrized by the Fourier component $\nu$ on
the neutral direction of the dynamics. We introduce a specific hypothesis
on the dynamics that we call ``minimal captivity''. This hypothesis
is stable under perturbations and means that the dynamics is univalued
on a neighborhood of $\mathcal{K}$. Under this hypothesis we show
the existence of an asymptotic spectral gap and a fractal Weyl law
for the upper bound of density of Ruelle resonances in the semiclassical
limit $\nu\rightarrow\infty$. Some numerical computations with the
truncated Gauss map illustrate these results.
\end{abstract}
\footnote{
operators, and other functional analytic techniques in dynamical systems
81Q20 Semi-classical techniques

}

\newpage

\tableofcontents{}

\section{Introduction}

A ``partially expanding map'' is a map which is expanding except
in some directions which are ``neutral''. An ``open map'' is a
map for which the non wandering set (or trapped set) is not the full
manifold but a relatively compact subset. The aim of this paper is
to study the dynamics of a class of open partially expanding maps
from the spectral approach initiated by D. Ruelle and R. Bowen. In
this approach, the pull back operator by the map, called transfer
operator, is shown to have some discrete spectrum in some specific
functional spaces. These eigenvalues called ``Ruelle resonances''
are very useful to describe the effective long time behavior of the
dynamics: to express dynamical correlation functions and deduce statistical
properties of the dynamics such as mixing and central limit theorems. 

In Section 2 we define the model of expanding maps on some union of
intervals $I\subset\mathbb{R}$ precisely called an iterated function
scheme (I.F.S.). This is a well studied class of dynamical systems
for which the trapped set $K\subset I$ is a Cantor set and has some
Hausdorff dimension $\mathrm{dim}_{H}K\in[0,1[$. In Section 2.4 we
extend this model by adding a neutral direction and obtain a ``partially
expanding map''. The transfer operator is defined in Section 2.5.
We can decompose the transfer operator into its Fourier components
$\nu\in\mathbb{Z}$ with respect to the neutral direction and obtain
a family of operators $\hat{F}_{\nu}$ also written $\hat{F}_{\hbar}$
with $\hbar:=1/\left(2\pi\nu\right)$ (if $\nu\neq0$). 

In Section 3 we present the main new results of this paper. Theorem
\ref{thm:discrete-spectrum} shows that each transfer operator $\hat{F}_{\nu}$
has some discrete spectrum of Ruelle resonances in specific Sobolev
spaces. Then Theorem \ref{th:gap_spectral} shows that in the limit
of large frequencies $\left|\nu\right|\rightarrow\infty$ the spectral
radius of $\hat{F}_{\nu}$ is bounded by some expression, under some
condition that we call ``minimal captivity''. In order to derive
this result we use a semiclassical approach which relies on microlocal
properties of the family of operators $\left(\hat{F}_{\nu}\right)_{\nu}$
in phase space $T^{*}I$ (precisely it is a Fourier integral operator).
This allows to consider the associated canonical map $F$ on $T^{*}I$.
This canonical map $F$ is multivalued and has a trapped set $\mathcal{K}$
which is also a Cantor set (which projects on $K$). We also obtain
an upper bound on the number of Ruelle resonances in the limit $\left|\nu\right|\rightarrow\infty$
in Theorem \ref{thm:fractal-weyl-upper}. This upper bound involves
the Hausdorff dimension $\mathrm{dim}_{H}K$ and is usually called
``fractal Weyl law'' after the work of J. Sjöstrand in \cite{sjostrand_90}.
The ``minimal captivity'' condition means that the dynamics of the
canonical map $F$ restricted to its trapped set $\mathcal{K}$ is
one-to-one (whereas the map $F$ on $T^{*}I$ is multivalued).

In Section 4 we illustrate our results by numerical computations with
two particular models: the truncated Gauss map and Bowen-Series map
for Schottky groups. We show that these models fullfill the minimal
captivity property. Sections 5 to 8 are devoted to the proofs of the
results.

The same semiclassical approach has been used before for ``closed
dynamical systems'' in \cite{fred_expanding_09} and \cite{arnoldi_skew_extension_12},
i.e. for systems in which the trapped set was the full manifold. In
these latter papers as well as in \cite{tsujii_05} a similar result
for the asymptotic spectral radius has been obtained. Technically
the open aspect here is overcame by using a truncation function $\chi$
as explained in Section 2.6. In \cite{arnoldi_skew_extension_12}
the author considers models for which the neutral direction is a non
commutative compact Lie group and shows discrete spectrum of Ruelle
resonances, asymptotic spectral radius and Weyl law. Let us remark
that we could extend the present results similarly by considering
extensions with compact groups.

As explained in Section 4.1 our results can be applied to ``Bowen
Series maps'' and ``Bowen Series transfer operators'' associated
to the geodesic flow of ``convex co-compact hyperbolic surfaces''
also called ``Schottky surfaces''. So our results give some results
for the zeroes of the Selberg zeta function and resonances of the
Laplacian of these surfaces. In that case the Weyl law of Theorem
\ref{thm:fractal-weyl-upper} is in close relation with the results
obtained by Lin, Guilloppe and Zworski in \cite{zworski_lin_guillope_02}
where they give an upper bound on the density of resonances for the
Laplace-Beltrami operator on open hyperbolic surfaces. We can also
apply our results to the quadratic maps and recover results already
obtained in \cite{zworski_04,naud_08}.

Also let us remark that with the condition of ``minimal captivity'',
the dynamics of the canonical map $F$ in the vicinity of the trapped
set $\mathcal{K}$ is univalued and can be identified with the classical
dynamics of a ``open quantum map''. Within this identification,
the results of S. Nonnenmacher et M. Zworski \cite{nonnenmacher_07}
about asymptotic spectral radius and Weyl law of these open quantum
maps are  very similar to the results presented in this paper.

\section{The transfer operator}

\subsection{Iterated function scheme}

The transfer operator studied in this paper is constructed from a
simple model of chaotic dynamics called ``an iterated function scheme,
I.F.S.''\cite[chap.9]{falconer_03_book}. We give the definition
below and refer to Section \ref{sec:Examples} where many standard
examples are presented.

\begin{center}{\color{red}\fbox{\color{black}\parbox{16cm}{
\begin{defn}
\label{def:IFSAn-iterated-function}``\textbf{An iterated function
scheme (I.F.S.)''. }Let $N\in\mathbb{N}$, $N\geq1$. Let $I_{1},\ldots I_{N}\subset\mathbb{R}$
be a finite collection of \textbf{disjoint and closed} intervals.
Let $A$ be a $N\times N$ matrix, called adjacency matrix, with $A_{i,j}\in\left\{ 0,1\right\} $.
We will note $i\rightsquigarrow j$ if $A_{i,j}=1.$ Assume that for
each pair $i,j\in\left\{ 1,\ldots,N\right\} $ such that $i\rightsquigarrow j$,
we have a smooth invertible map $\phi_{i,j}:I_{i}\rightarrow\phi_{i,j}\left(I_{i}\right)$
with $\phi_{i,j}\left(I_{i}\right)\subset\mbox{Int}\left(I_{j}\right)$.
Assume that the map $\phi_{i,j}$ is a \textbf{strict contraction},
i.e. there exists $0<\theta<1$ such that for every $x\in I_{i}$,
\begin{equation}
\left|\phi_{i,j}'\left(x\right)\right|\leq\theta\label{eq:theta_contraction}
\end{equation}
We suppose that different images of the maps $\phi_{i,j}$ do not
intersect (this is the ``strong separation condition'' in \cite[p.35]{Falconer_97}):
\begin{equation}
\phi_{i,j}\left(I_{i}\right)\cap\phi_{k,l}\left(I_{k}\right)\neq\emptyset\quad\Rightarrow\quad i=k\mbox{ and }j=l.\label{eq:hyp_non_intersect}
\end{equation}

\end{defn}
}}}\end{center}

\begin{figure}
\begin{centering}
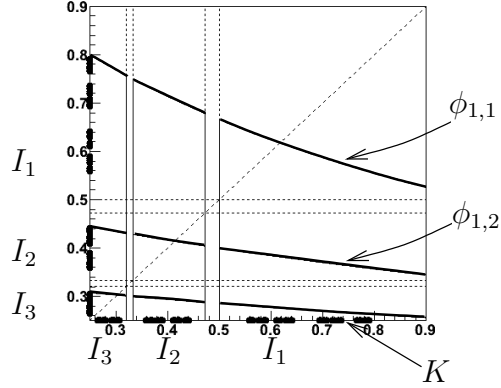
\par\end{centering}

\caption{\label{fig:Gauss_Cutter}The iterated functions scheme (IFS) defined
from the truncated Gauss map (\ref{eq:Gauss_map}). Here we have $N=3$
branches. The maps $\phi$: $\phi_{i,j}:I_{i}\rightarrow I_{j}$,
$i,j=1\ldots N$ are contracting and given by $\phi_{i,j}\left(x\right)=\frac{1}{x+j}$.
The trapped set $K$ defined in (\ref{eq:trapped_set_K_def1}) is
a $N$-adic Cantor set. It is obtained as the limit of the sets $K_{0}=\left(I_{1}\cup I_{2}\ldots\cup I_{N}\right)\supset K_{1}=\phi\left(K_{0}\right)\supset K_{2}=\phi\left(K_{1}\right)\supset\ldots\supset K$.}
\end{figure}

Remark that the derivative $\phi_{i,j}'\left(x\right)$ may be negative.
Figure \ref{fig:Gauss_Cutter} illustrates Definition \ref{def:IFSAn-iterated-function}
on a specific example.

\subsection{\label{sub:Example-with-the}Example with the truncated Gauss map}

The Gauss map is
\begin{equation}
G:\begin{cases}
\left]0,1\right] & \rightarrow\left]0,1\right[\\
y & \rightarrow\left\{ \frac{1}{y}\right\} 
\end{cases}\label{eq:Gauss_map}
\end{equation}
where $\left\{ a\right\} :=a-\left[a\right]\in[0,1[$ denotes the
fractional part of $a\in\mathbb{R}$. For $j\in\mathbb{N}\backslash\left\{ 0\right\} $,
and $y\in\mathbb{R}$ such that $\frac{1}{j+1}<y\leq\frac{1}{j}$
then $G\left(y\right)=G_{j}\left(y\right):=\frac{1}{y}-j$. Notice
that $dG/dy<0$. The inverse map is $y=G_{j}^{-1}\left(x\right)=\frac{1}{x+j}$.

Let $N\geq1$. We will consider only the first $N$ ``branches''
$\left(G_{j}\right)_{j=1,\ldots N}$. In order to have a well defined
I.F.S according to definition \ref{def:IFSAn-iterated-function},
for $1\leq i\leq N$, let $\alpha_{i}:=G_{i}^{-1}\left(\frac{1}{N+1}\right)$
, $a_{i}=\frac{1}{1+i}$, $b_{i}$ such that $\alpha_{i}<b_{i}<\frac{1}{i}$,
and intervals $I_{i}:=\left[a_{i},b_{i}\right]$. On these intervals
$\left(I_{i}\right)_{i=1\ldots N}$, we define the maps
\begin{equation}
\phi_{i,j}\left(x\right)=G_{j}^{-1}\left(x\right)=\frac{1}{x+j},\quad j=1,\ldots N.\label{eq:Gauss_map-1}
\end{equation}
The adjacency matrix is $A=\left(A_{i,j}\right)_{i,j}$, the full
$N\times N$ matrix with all entries $A_{i,j}=1$. 

\subsection{The trapped set $K$}

We define
\begin{equation}
I:=\bigcup_{i=1}^{N}I_{i}\label{eq:def_I}
\end{equation}
and the multivalued map: 
\[
\phi:I\rightarrow I,\qquad\phi=\left(\phi_{i,j}\right)_{i,j}.
\]
$\phi$ can be iterated and generates a multivalued map $\phi^{n}:I\rightarrow I$
for $n\geq1$. From hypothesis (\ref{eq:hyp_non_intersect}) the inverse
map
\[
\phi^{-1}:\phi\left(I\right)\rightarrow I
\]
is uni-valued. Throughout the paper we will use the \textbf{``unstable
Jacobian function''} 
\begin{equation}
J\left(x\right):=\log\left|\frac{d\phi^{-1}}{dx}\left(x\right)\right|\label{eq:def_J}
\end{equation}
defined on $\phi\left(I\right)$. From (\ref{eq:theta_contraction}),
one has
\begin{equation}
\forall x,\quad J\left(x\right)>\log\frac{1}{\theta}>0.\label{eq:J_positive}
\end{equation}
Let 
\begin{equation}
K_{n}:=\phi^{n}\left(I\right)\label{eq:def_Kn}
\end{equation}
and $K_{0}=I$. Since $K_{n+1}\subset K_{n}$ we can defined the limit
set
\begin{equation}
K:=\bigcap_{n\in\mathbb{N}}K_{n}\label{eq:trapped_set_K_def1}
\end{equation}
called the \textbf{trapped set}. Then the map 
\begin{equation}
\phi^{-1}:K\rightarrow K\label{eq:phi_-1_on_K}
\end{equation}
 is well defined and uni-valued.

\subsubsection{The fractal dimension of the trapped set $K$ and the topological
pressure}

In this paper we will use the following definition of fractal dimension.

\begin{center}{\color{red}\fbox{\color{black}\parbox{16cm}{
\begin{defn}
\label{Minkowski}\cite[p.76]{mattila_1999},\cite[p.20]{Falconer_97}
If $B\subset\mathbb{R}^{d}$ is a non empty bounded set, its \textbf{\small{upper
Minkowski dimension}}{\small{ (or box dimension)}} is 
\begin{equation}
\mathrm{dim}_{M}B:=d-\mathrm{codim}_{M}B\label{eq:def_dim_M}
\end{equation}
 with 
\begin{equation}
\mathrm{codim}_{M}B:=\sup\left\{ s\in\mathbb{R}\,|\,\limsup_{\delta\downarrow0}\delta^{-s}\cdot\mathrm{Leb}\left(B_{\delta}\right)<+\infty\right\} .\label{eq:defdim-1-1}
\end{equation}
where $B_{\delta}:=\left\{ x\in\mathbb{R}^{d},\mathrm{dist}\left(x,B\right)\leq\delta\right\} $
and $\mathrm{Leb}\left(.\right)$ is the Lebesgue measure.
\end{defn}
}}}\end{center}
\begin{rem}
\label{rem3}In general 
\begin{equation}
\limsup_{\delta\downarrow0}\delta^{-\mathrm{codim}_{M}B}\cdot\mbox{Leb}\left(B_{\delta}\right)<+\infty\label{eq:pure_dim}
\end{equation}
 does not hold, but if it does, $B$ is said to be of \textbf{pure
dimension}%
\footnote{see \cite{sjostrand_90} for comments and further references.%
}. It is known that the trapped set $K$ defined in (\ref{eq:trapped_set_K_def1})
has pure dimension and that the above definition of Minkowski dimension
coincides with the more usual \textbf{Hausdorff dimension} of $K$
\cite[p.68]{Falconer_97}:
\begin{equation}
\mathrm{dim}_{M}K=\mathrm{dim}_{H}K\quad\in[0,1[\label{eq:dim_H_dim_M}
\end{equation}

An efficient way to calculate the fractal dimension $\mathrm{dim}_{H}K$
is given by the topological pressure. The topological pressure can
be defined from the periodic points as follows. A \textbf{periodic
point} of period $n\geq1$ is $x\in K$ such that $x=\phi^{-n}\left(x\right)$.
The topological pressure can be defined in terms of periodic points.
\end{rem}
\begin{center}{\color{red}\fbox{\color{black}\parbox{16cm}{
\begin{defn}
\cite[p.72]{Falconer_97} The \textbf{topological pressure} of a continuous
function $\varphi\in C\left(I\right)$ is
\[
\mathrm{Pr}\left(\varphi\right):=\lim_{n\rightarrow\infty}\frac{1}{n}\log\left(\sum_{x=\phi^{-n}\left(x\right)}e^{\varphi_{n}\left(x\right)}\right)
\]
where $\varphi_{n}\left(x\right)$ is the Birkhoff sum of $\varphi$
along the periodic orbit:
\[
\varphi_{n}\left(x\right):=\sum_{k=0}^{n-1}\varphi\left(\phi^{-k}\left(x\right)\right)
\]

\end{defn}
}}}\end{center}

It is interesting to consider the special case of the function $\varphi\left(x\right)=-\beta J\left(x\right)$
with some $\beta\in\mathbb{R}$ and $J\left(x\right)$ defined in
(\ref{eq:def_J}).  This gives the function $\mathrm{P}:\mathbb{R}\rightarrow\mathbb{R}$
defined by
\begin{equation}
\mathrm{P}\left(\beta\right):=\mathrm{Pr}\left(-\beta J\right)=\lim_{n\rightarrow\infty}\frac{1}{n}\log\left(\sum_{x=\phi^{-n}\left(x\right)}\prod_{k=0}^{n-1}\left|\phi'\left(\phi^{-k}\left(x\right)\right)\right|^{\beta}\right)\label{eq:def_P_beta-1}
\end{equation}

\begin{figure}
\begin{centering}
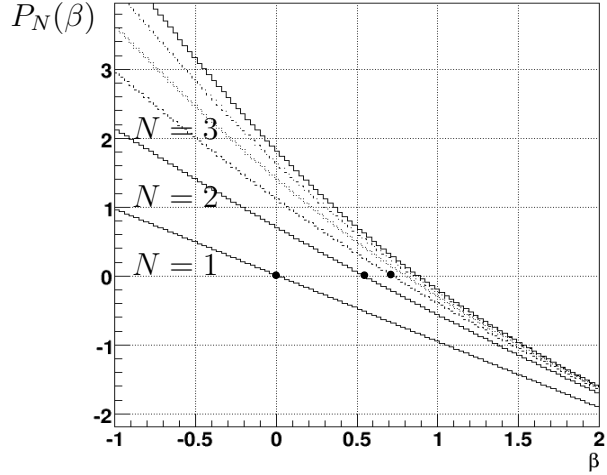
\par\end{centering}

\caption{\label{fig:Pressure_Gauss}Topological Pressure $\mathrm{P}_{N}\left(\beta\right)$
defined by (\ref{eq:def_P_beta-1}) for the truncated Gauss map example
of Section \ref{sub:Example-with-the} for each value of $N=1,2,3\ldots$
being the number of branches. The black points mark the zero of $\mathrm{P}_{N}\left(\beta\right)=0$
giving the fractal dimension of the trapped set $K_{N}$ for each
value of $N$: $\dim_{H}K_{1}=0$, $\dim_{H}K_{2}=0.531\ldots$, $\dim_{H}K_{3}=0.705\ldots$,
and $\dim_{H}K_{N}\protect\underset{N\rightarrow\infty}{\rightarrow1}$.}
\end{figure}

The following lemma provides an easy way to compute (numerically)
the dimension of $K$. See figure \ref{fig:Pressure_Gauss}.

\begin{center}{\color{blue}\fbox{\color{black}\parbox{16cm}{
\begin{lem}
\cite[p.77]{Falconer_97} $\mathrm{P}\left(\beta\right)$ is continuous
and strictly decreasing in $\beta$ and its unique zero is given by
$\beta=\dim_{H}K$.
\end{lem}
}}}\end{center}

\subsection{An extended partially expanding map and reduced transfer operators}

The map $\phi^{-1}:\phi\left(I\right)\rightarrow I$ is univalued
and expanding. Let $\tau\in C^{\infty}\left(I;\mathbb{R}\right)$
be a smooth real valued function called \textbf{roof function}. We
define the map
\begin{equation}
f:\begin{cases}
\phi\left(I\right)\times S^{1} & \rightarrow I\times S^{1}\\
\left(x,y\right) & \rightarrow\left(\phi^{-1}\left(x\right),y+\tau\left(x\right)\right)
\end{cases}\label{eq:def_extended_map_f}
\end{equation}
with $S^{1}:=\mathbb{R}/\mathbb{Z}$. Notice that the map $f$ is
expanding in the $x$ variable whereas it is neutral in the $y$ variable
in the sense that $\frac{\partial f}{\partial y}=1$. This is called
a partially expanding map and is a very simple model of more general
partially hyperbolic dynamics \cite{pesin_04}.

Let $V\in C^{\infty}\left(I;\mathbb{C}\right)$ be a smooth complex
valued function called a \textbf{potential} function.
\begin{defn}
The transfer operator of the map $f$ with potential $V$ is
\begin{equation}
\mathcal{\hat{F}}:\begin{cases}
C^{\infty}\left(I\times S^{1}\right) & \rightarrow C^{\infty}\left(\phi\left(I\right)\times S^{1}\right)\\
\psi\left(x,y\right) & \rightarrow e^{V\left(x\right)}\psi\left(f\left(x,y\right)\right)
\end{cases}.\label{eq:def_extended_transfer_op}
\end{equation}

\end{defn}
Notice that $\psi\left(x,y\right)$ can be decomposed into Fourier
modes in the $y$ direction. For $\nu\in\mathbb{Z}$, a Fourier mode
is
\[
\psi_{\nu}\left(x,y\right)=\varphi\left(x\right)e^{i2\pi\nu y}
\]
and we have
\begin{eqnarray*}
\left(\mathcal{\hat{F}}\psi_{\nu}\right)\left(x,y\right) & = & e^{V\left(x\right)}\psi_{\nu}\left(f\left(x,y\right)\right)=e^{V\left(x\right)}\varphi\left(\phi^{-1}\left(x\right)\right)e^{i2\pi\nu\left(y+\tau\left(x\right)\right)}\\
 & = & \left(\hat{F}_{\nu}\varphi\right)\left(x\right)e^{i2\pi\nu y}
\end{eqnarray*}
where the Standard reduced transfer operator $\hat{F}_{\nu}:C^{\infty}\left(I\right)\rightarrow C^{\infty}\left(\phi\left(I\right)\right)$
defined by
\begin{equation}
\left(\hat{F}_{\nu}\varphi\right)\left(x\right):=e^{V\left(x\right)}e^{i2\pi\nu\tau\left(x\right)}\varphi\left(\phi^{-1}\left(x\right)\right).\label{eq:def_F_nu}
\end{equation}
So the operator $\hat{\mathcal{F}}$ is the direct sum of operators
$\bigoplus_{\nu\in\mathbb{Z}}\hat{F}_{\nu}$. From the next section
we will study the individual operator $\hat{F}_{\nu}$ in (\ref{eq:def_transfert_op_F}).
Since our main interest is the limit $\nu\rightarrow\infty$ of large
frequencies in the neutral direction, we will suppose $\nu\neq0$
and write $\hbar:=\frac{1}{2\pi\nu}$. In Section \ref{sub:Decay-of-correlations}
we will deduce from our principal results, some asymptotic expansions
for time correlation functions of the map (\ref{eq:def_extended_map_f}).

\subsection{The transfer operator}

\paragraph{Notations:}

We denote $C_{0}^{\infty}\left(\mathbb{R}\right)$ the space of smooth
function on $\mathbb{R}$ with compact support. If $B\subset\mathbb{R}$
is a compact set, we denote $C_{B}^{\infty}\left(\mathbb{R}\right)\subset C_{0}^{\infty}\left(\mathbb{R}\right)$
the space of smooth functions on $\mathbb{R}$ with support included
in $B$. Recall that the inverse map $\phi^{-1}:\phi\left(I\right)\rightarrow I$
is uni-valued.

\begin{center}{\color{red}\fbox{\color{black}\parbox{16cm}{
\begin{defn}
\label{def_Transfer_op}Let $\tau\in C^{\infty}\left(I;\mathbb{R}\right)$
and $V\in C^{\infty}\left(I;\mathbb{C}\right)$ be smooth functions
called respectively \textbf{roof function} and \textbf{potential}
function. Let $\hbar>0$. We define the \textbf{transfer operator}:
\begin{equation}
\hat{F}:\begin{cases}
C_{I}^{\infty}\left(\mathbb{R}\right) & \rightarrow C_{I}^{\infty}\left(\mathbb{R}\right)\\
\varphi=\left(\varphi_{i}\right)_{i} & \rightarrow\left(\sum_{i=1}^{N}\hat{F}_{i,j}\varphi_{i}\right)_{j}
\end{cases}\label{eq:def_transfert_op_F}
\end{equation}
 with 
\begin{equation}
\hat{F}_{i,j}:\begin{cases}
C_{I_{i}}^{\infty}\left(\mathbb{R}\right) & \rightarrow C_{I_{j}}^{\infty}\left(\mathbb{R}\right)\\
\varphi_{i} & \rightarrow\left(\hat{F}_{i,j}\varphi_{i}\right)\left(x\right)=\begin{cases}
e^{V\left(x\right)}e^{i\frac{1}{\hbar}\tau\left(x\right)}\varphi_{i}\left(\phi_{i,j}^{-1}\left(x\right)\right) & \quad\mbox{if }i\rightsquigarrow j\mbox{ and }x\in\phi_{i,j}\left(I_{i}\right)\\
0 & \quad\mbox{otherwise}
\end{cases}
\end{cases}\label{e:def_F_op_ij}
\end{equation}

\end{defn}
}}}\end{center}

\begin{figure}
\begin{centering}
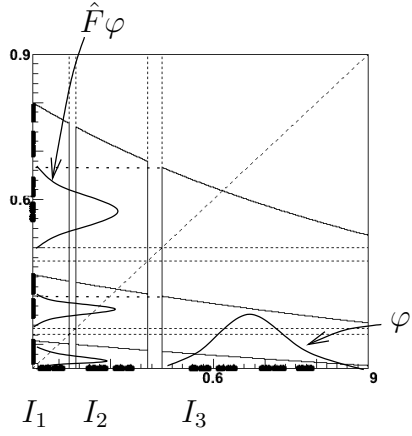
\par\end{centering}

\caption{\label{fig:op_F_hat}Action of the transfer operator $\hat{F}$ on
a function $\varphi$ as defined in (\ref{eq:def_transfert_op_F}).
In this schematic figure we have $V=0$ and $\tau=0$. In general
the factor $e^{V\left(x\right)}$ changes the amplitude and $e^{i\frac{1}{\hbar}\tau\left(x\right)}$
creates some fast oscillations if $\hbar\ll1$.}
\end{figure}

\paragraph{Remark:}
\begin{itemize}
\item From assumption (\ref{eq:hyp_non_intersect}), for any $x\in I$,
the sum $\sum_{i=1}^{N}\left(\hat{F}_{i,j}\varphi_{i}\right)\left(x\right)$
which appears on the right hand side of (\ref{eq:def_transfert_op_F})
contains at most one non vanishing term. See figure \ref{fig:op_F_hat}.
\item For short we can write that
\begin{equation}
\hat{F}:\begin{cases}
C_{I}^{\infty}\left(\mathbb{R}\right) & \rightarrow C_{I}^{\infty}\left(\mathbb{R}\right)\\
\varphi & \rightarrow\begin{cases}
e^{V\left(x\right)}e^{i\frac{1}{\hbar}\tau\left(x\right)}\varphi\left(\phi^{-1}\left(x\right)\right) & \quad\mbox{if }x\in\phi\left(I\right)\\
0 & \quad\mbox{otherwise}
\end{cases}
\end{cases}\label{eq:def_transfert_op_F-1}
\end{equation}

\end{itemize}

\begin{rem}
~\end{rem}
\begin{itemize}
\item For any $\varphi\in C_{I}^{\infty}\left(\mathbb{R}\right)$, $n\geq0$
we have 
\begin{equation}
\mbox{supp}\left(\hat{F}^{n}\varphi\right)\subset K_{n}\label{eq:supp_F_phi}
\end{equation}
 with $K_{n}$ defined in (\ref{eq:def_Kn}).
\item In the definition (\pageref{eq:def_transfert_op_F}) we can write
$e^{V\left(x\right)}e^{i\frac{1}{\hbar}\tau\left(x\right)}=\exp\left(i\frac{1}{\hbar}\mathcal{V}\left(x\right)\right)$
with $\mathcal{V}\left(x\right):=\tau\left(x\right)+\hbar\left(-iV\left(x\right)\right)$.
More generally we may consider a finite series $\mathcal{V}\left(x\right)=\sum_{j=0}^{n}\hbar^{j}\mathcal{V}_{j}\left(x\right)$
with leading term $\mathcal{V}_{0}\left(x\right)=\tau\left(x\right)$
and complex valued sub-leading terms $\mathcal{V}_{j}:I\rightarrow\mathbb{C}$,
$j\geq1$.
\end{itemize}

\subsection{Extension of the transfer operator to distributions}

The transfer operator $\hat{F}$ has been defined on smooth functions
$C_{I}^{\infty}\left(\mathbb{R}\right)$ in (\ref{eq:def_transfert_op_F}).
We will need to extend it to the space of distributions. For that
purpose we first introduce a cut-off function $\chi\in C_{I}^{\infty}\left(\mathbb{R}\right)$
such that $\chi\left(x\right)=1$ for every $x\in K_{1}=\phi\left(I\right)$,
i.e. $\chi\left(\phi_{i,j}\left(x\right)\right)=1$ for every $x\in I_{i}$
and $j$ such that $i\rightsquigarrow j$. We denote $\hat{\chi}$
the multiplication operator by the function $\chi$. Let us define:
\begin{equation}
\hat{F}_{\chi}:=\hat{F}\hat{\chi},\qquad\hat{F}_{i,j,\chi}:=\hat{F}_{i,j}\hat{\chi}.\label{eq:def_F_Chi}
\end{equation}

Note that for any $\varphi\in C_{K_{1}}^{\infty}\left(\mathbb{R}\right)$
we have $\hat{\chi}\varphi=\varphi$ hence $\left(\hat{F}\hat{\chi}\right)\varphi=\hat{F}\varphi$.)
has $\hat{\chi}:C_{0}^{\infty}\left(\mathbb{R}\right)\rightarrow C_{I}^{\infty}\left(\mathbb{R}\right)$
hence $\hat{F}_{\chi}$ is defined on .

The formal adjoint operator $\hat{F}_{i,j,\chi}^{*}:C_{0}^{\infty}\left(\mathbb{R}\right)\rightarrow C_{I_{i}}^{\infty}\left(\mathbb{R}\right)$
is defined by 
\begin{equation}
\langle\varphi_{i}|\hat{F}_{i,j,\chi}^{*}\psi_{j}\rangle=\langle\hat{F}_{i,j,\chi}\varphi_{i}|\psi_{j}\rangle,\qquad\forall\varphi_{i}\in C_{0}^{\infty}\left(\mathbb{R}\right),\psi_{j}\in C_{0}^{\infty}\left(\mathbb{R}\right),\label{eq:def_F*}
\end{equation}
with the $L^{2}$-scalar product $\langle u|v\rangle:=\int\overline{u}\left(x\right)v\left(x\right)dx$.

\begin{center}{\color{blue}\fbox{\color{black}\parbox{16cm}{
\begin{lem}
For $i\rightsquigarrow j$, the adjoint operator $\hat{F}_{i,j,\chi}^{*}:C_{0}^{\infty}\left(\mathbb{R}\right)\rightarrow C_{I_{i}}^{\infty}\left(\mathbb{R}\right)$
is given, for $y\in I_{i}$ by
\begin{equation}
\left(\hat{F}_{i,j,\chi}^{*}\psi_{j}\right)\left(y\right)=\chi\left(y\right)\left|\phi'_{i,j}\left(y\right)\right|e^{\overline{V\left(\phi_{i,j}\left(y\right)\right)}}e^{-\frac{i}{\hbar}\tau\left(\phi_{i,j}\left(y\right)\right)}\psi_{j}\left(\phi_{i,j}\left(y\right)\right)\label{eq:express_dual_1}
\end{equation}
The adjoint operator $\hat{F}_{\chi}^{*}:C_{0}^{\infty}\left(\mathbb{R}\right)\rightarrow C_{I}^{\infty}\left(\mathbb{R}\right)$
is given by
\[
\psi=\left(\psi_{j}\right)_{j}\rightarrow\left(\hat{F}_{\chi}^{*}\psi\right)_{i}\left(y\right)=\sum_{j\mbox{ s.t. }i\rightsquigarrow j}\left(\hat{F}_{i,j,\chi}^{*}\psi_{j}\right)\left(y\right)
\]

\end{lem}
}}}\end{center}
\begin{proof}
Using the change of variables $x=\phi_{i,j}\left(y\right)$ and definition
(\ref{eq:def_F*}), we write 
\begin{eqnarray*}
\langle\varphi_{i}|\hat{F}_{i,j,\chi}^{*}\psi_{j}\rangle & = & \int\overline{\varphi}_{i}\left(y\right)\left(\hat{F}_{i,j,\chi}^{*}\psi_{j}\right)\left(y\right)dy\\
 & = & \langle\hat{F}_{i,j,\chi}\varphi_{i}|\psi_{j}\rangle=\int_{\phi_{i,j}\left(I_{i}\right)}\overline{e^{V\left(x\right)}e^{i\frac{1}{h}\tau\left(x\right)}\varphi_{i}\left(\phi_{i,j}^{-1}\left(x\right)\right)\chi\left(\phi_{i,j}^{-1}\left(x\right)\right)}\psi_{j}\left(x\right)dx\\
 & = & \int_{I_{i}}\overline{\varphi_{i}}\left(y\right)\chi\left(y\right)\left|\phi'_{i,j}\left(y\right)\right|e^{\overline{V\left(x\right)}}e^{-i\frac{1}{h}\tau_{j}\left(x\right)}\psi_{j}\left(x\right)dy
\end{eqnarray*}
and deduce (\ref{eq:express_dual_1}).\end{proof}
\begin{rem}
~\end{rem}
\begin{itemize}
\item Without the cut-off function $\chi$ the image of $\hat{F}_{i,j}^{*}$
may not be continuous on the boundary of $I_{i}$.
\item An other more general possibility would have been to consider $\chi\in C_{I}^{\infty}\left(\mathbb{R}\right)$
such that $0<\chi\left(x\right)$ for $x\in\mbox{Int}\left(I\right)$
(without assumption that $\chi\equiv1$ on $K_{1}$) and define
\begin{equation}
\hat{F}_{i,j,\chi}:=\hat{\chi}^{-1}\hat{F}_{i,j}\hat{\chi}:C_{I_{i}}^{\infty}\left(\mathbb{R}\right)\rightarrow C_{I_{j}}^{\infty}\left(\mathbb{R}\right)\label{eq:def_F_Chi_Chim}
\end{equation}
which is well defined since $\mbox{supp}\left(\hat{F}_{i,j}\hat{\chi}\varphi\right)\subset\mbox{Int}\left(I_{j}\right)$
where $\chi$ does not vanish. This more general definition (\ref{eq:def_F_Chi_Chim})
may be more useful in some cases, e.g. we use it in numerical computation.
We recover the previous definition (\ref{eq:def_F_Chi}) if we make
the additional assumption that $\chi\equiv1$ on $K_{1}$.
\end{itemize}
\begin{center}{\color{blue}\fbox{\color{black}\parbox{16cm}{
\begin{prop}
By duality the transfer operator (\ref{eq:def_F_Chi}) extends to
distributions:
\begin{equation}
\hat{F}_{\chi}:\mathcal{D}'\left(\mathbb{R}\right)\rightarrow\mathcal{D}'\left(\mathbb{R}\right)\label{eq:F_Chi_Distributions}
\end{equation}
\[
\hat{F}_{\chi}^{*}:\mathcal{D}'\left(\mathbb{R}\right)\rightarrow\mathcal{D}'\left(\mathbb{R}\right)
\]
Similarly to (\ref{eq:supp_F_phi}) we have that for any $n\geq1$,
any $\alpha\in\mathcal{D}'\left(\mathbb{R}\right)$,
\begin{equation}
\mathrm{supp}\left(\hat{F}_{\chi}^{n}\alpha\right)\subset K_{n}\label{eq:supp_F_phi-1}
\end{equation}
with $K_{n}$ defined in (\ref{eq:def_Kn}).
\end{prop}
}}}\end{center}
\begin{proof}
The extension is defined by%
\footnote{The complex conjugation appears in (\ref{eq:duality-1}) because duality
is related to scalar product on $L^{2}$ by $\alpha\left(\overline{\varphi}\right):=\int\overline{\varphi}\alpha=\langle\varphi,\alpha\rangle_{L^{2}}$.%
}
\begin{equation}
\hat{F}_{i,j,\chi}\left(\alpha_{i}\right)\left(\psi_{j}\right)=\alpha_{i}\left(\overline{\hat{F}_{i,j,\chi}^{*}\overline{\psi_{j}}}\right),\qquad\alpha_{i}\in\mathcal{D}'\left(\mathbb{R}\right),\psi_{j}\in C_{0}^{\infty}\left(\mathbb{R}\right),\label{eq:duality-1}
\end{equation}
Then the transfer operator extends to: $\hat{F}_{\chi}:\mathcal{D}'\left(\mathbb{R}\right)\rightarrow\mathcal{D}'\left(\mathbb{R}\right)$.

If $\psi_{j}\left(\phi_{i,j}\left(y\right)\right)=0,\forall y\in I_{i}$,
Eq.(\ref{eq:express_dual_1}) shows that $\hat{F}_{i,j,\chi}^{*}\psi_{j}\equiv0$.
More generally let $\psi\in C_{0}^{\infty}\left(\mathbb{R}\right)$
with $\mbox{supp}\left(\psi\right)\cap K_{n}=\emptyset$ with $n\geq1$
and $K_{n}$ defined in (\ref{eq:def_Kn}). Then
\begin{equation}
\left(\hat{F}_{\chi}^{*}\right)^{n}\psi\equiv0.\label{eq:F*^nPsi=00003D0}
\end{equation}
For any $\alpha\in\mathcal{D}'\left(\mathbb{R}\right)$, we deduce
that $\left(\hat{F}_{\chi}^{n}\alpha\right)\left(\overline{\psi}\right)=\alpha\left(\overline{\left(\hat{F}_{\chi}^{*}\right)^{n}\psi}\right)=0$.
By definition, this means that $\mbox{supp}\left(\hat{F}_{\chi}^{n}\alpha\right)\subset K_{n}$. \end{proof}

\section{Main results}

\subsection{Discrete spectrum of Ruelle resonances}

Theorem \ref{thm:discrete-spectrum} below shows that the transfer
operator $\hat{F}_{\chi}$ (for any $\hbar$) has discrete spectrum
called ``Ruelle resonances'' in ordinary Sobolev spaces with negative
order and that the spectrum does not depend on the choice of $\chi$.
Recall that for $m\in\mathbb{R}$, the \textbf{Sobolev space} $H^{-m}\left(\mathbb{R}\right)\subset\mathcal{D}'\left(\mathbb{R}\right)$
is defined by (\cite{taylor_tome1} p.271)
\begin{equation}
H^{-m}\left(\mathbb{R}\right):=\left\langle \hat{\xi}\right\rangle ^{m}\left(L^{2}\left(\mathbb{R}\right)\right)\label{eq:def_Hm}
\end{equation}
with the differential operator $\hat{\xi}:=-i\frac{d}{dx}$ and the
notation $\left\langle x\right\rangle :=\left(1+x^{2}\right)^{1/2}$.

\begin{center}{\color{blue}\fbox{\color{black}\parbox{16cm}{
\begin{thm}
\label{thm:discrete-spectrum}\textbf{''Discrete spectrum of resonances''.}
For any fixed $\hbar$, any $m\in\mathbb{R}$, the transfer operator
$\hat{F}_{\chi}$ in (\ref{eq:F_Chi_Distributions}) is bounded in
the Sobolev space $H^{-m}\left(\mathbb{R}\right)$ and can be written
\begin{equation}
\hat{F}_{\chi}=\hat{K}+\hat{R}\label{eq:F=00003DK+R}
\end{equation}
where $\hat{K}$ is a compact operator and $\hat{R}$ is such that:
\begin{equation}
\left\Vert \hat{R}\right\Vert _{H^{-m}\left(\mathbb{R}\right)}\leq r_{m},\qquad r_{m}:=C\theta^{m}\label{eq:bound_norm_R_def_rm}
\end{equation}
where $0<\theta<1$ is given in (\ref{eq:theta_contraction}) and
$C$ does not depend on $m$. Notice that $r_{m}\rightarrow0$ as
$m\rightarrow+\infty$ and that the operator $\hat{F}_{\chi}$ has
discrete spectrum on the domain $\left|z\right|>r_{m}$. These eigenvalues
of $\hat{F}_{\chi}$ and their eigenspace do not depend on $m$ nor
on $\chi$\textbf{.} The support of the eigendistributions is contained
in the trapped set $K$. These discrete eigenvalues are denoted
\begin{equation}
\mathrm{Res}\left(\hat{F}\right):=\left\{ \lambda_{i}\right\} _{i}\subset\mathbb{C}^{*}\label{eq:def_Res_F}
\end{equation}
and are called \textbf{Ruelle resonances.}
\end{thm}
}}}\end{center}

In Section \ref{sec:Examples} we show the discrete spectrum of Ruelle
resonances computed numerically for different examples.

\subsection{Asymptotic spectral gap and Fractal Weyl law}

We will give some partial description of the discrete spectrum of
Ruelle resonances of the operator $\hat{F}_{\chi,\hbar}$, Eq.(\ref{eq:F_Chi_Distributions}),
in the limit $\hbar\rightarrow0$. For brevity we will drop the index
$\chi$ and simply write $\hat{F}_{\hbar}$. In Theorem \ref{th:gap_spectral}
below we present a result giving an upper bound for the spectral radius
of $\hat{F}_{\hbar}$ in the semiclassical limit $\hbar\rightarrow0$.
In Theorem \ref{thm:fractal-weyl-upper} we provide an upper bound
for the number of resonances outside any radius $\varepsilon>0$ as
$\hbar\rightarrow0$. This is called ``fractal Weyl law''. These
results rely on the study of the dynamics of a symplectic map or canonical
map $F:T^{*}I\rightarrow T^{*}I$ associated to the family of operators
$\left(\hat{F}_{\hbar}\right)_{\hbar}$, that we describe first.

\begin{center}{\color{blue}\fbox{\color{black}\parbox{16cm}{
\begin{lem}
\label{prop:Fh_is_a_FIO}The family of operators $\left(\hat{F}_{\hbar}\right)_{\hbar}$
restricted to $C_{I}^{\infty}\left(\mathbb{R}\right)$ is a $\hbar$-Fourier
integral operator (FIO). Its \textbf{canonical transform} is a multi-valued
symplectic map $F:T^{*}I\rightarrow T^{*}I$ (with $T^{*}I\cong I\times\mathbb{R}$)
given by:
\begin{equation}
F:\begin{cases}
T^{*}I & \rightarrow T^{*}I\\
\left(x,\xi\right) & \rightarrow\left\{ F_{i,j}\left(x,\xi\right)\quad\quad\mbox{with }i,j\mbox{ s.t. }x\in I_{i},\, i\rightsquigarrow j\right\} 
\end{cases}\label{eq:canonical_map_Fij}
\end{equation}
with
\begin{equation}
F_{i,j}:\begin{cases}
x' & =\phi_{i,j}\left(x\right)\\
\xi' & =\frac{1}{\phi_{i,j}'\left(x\right)}\xi+\tau'\left(x'\right)
\end{cases}.\label{eq:def_symplectic_map_Fij-1}
\end{equation}

\end{lem}
}}}\end{center}

The proof of Lemma \ref{prop:Fh_is_a_FIO} will be given in the beginning
of Section \ref{sec:6}.
\begin{rem}
For short, we can write
\begin{equation}
F:\begin{cases}
T^{*}I & \rightarrow T^{*}I\\
\left(x,\xi\right) & \rightarrow\left(\phi\left(x\right),\frac{1}{\phi'\left(x\right)}\xi+\tau'\left(\phi\left(x\right)\right)\right)
\end{cases}.\label{eq:def_canonical_map_F}
\end{equation}

\end{rem}
We will study the dynamics of $F$ in detail in later Sections, but
we can already make some remarks. The term $\frac{d\tau_{j}}{dx}\left(x'\right)$
in the expression of $\xi'$, Eq.(\ref{eq:def_symplectic_map_Fij-1}),
complicates significantly the dynamics near the zero section $\xi=0$.
However the next Lemma shows that a trajectory from an initial point
$\left(x,\xi\right)$ with $\left|\xi\right|$ large enough, escape
towards infinity:

\vspace{0.cm}\begin{center}{\color{blue}\fbox{\color{black}\parbox{16cm}{
\begin{lem}
\label{lem:escape_F}For any $1<\kappa<1/\theta$, there exists $R\geq0$
such that for any $\left|\xi\right|>R$ and any $i\rightsquigarrow j$,
\begin{equation}
\left|\xi'\right|>\kappa\left|\xi\right|\label{eq:expand_nu}
\end{equation}
where $\left(x',\xi'\right)=F_{i,j}\left(x,\xi\right)$.
\end{lem}
}}}\end{center}\vspace{0.cm}
\begin{proof}
From (\ref{eq:def_symplectic_map_Fij-1}), one has $\xi'=\frac{1}{\phi_{i,j}'\left(x\right)}\xi+\tau'\left(x'\right)$.
Also $\left|\frac{1}{\phi_{i,j}'\left(x\right)}\right|\geq\theta$
hence 
\begin{eqnarray*}
\left|\xi'\right|-\kappa\left|\xi\right| & = & \left|\frac{1}{\phi_{i,j}'\left(x\right)}\xi+\tau'\left(x'\right)\right|-\kappa\left|\xi\right|\geq\left|\frac{1}{\phi_{i,j}'\left(x\right)}\xi\right|-\left|\mathcal{\tau}'\left(x'\right)\right|-\kappa\left|\xi\right|\\
 & \geq & \left(\frac{1}{\theta}-\kappa\right)\left|\xi\right|-\max_{x}\left|\tau'\left(x\right)\right|>0.
\end{eqnarray*}
The last inequality holds true if $\left|\xi\right|>R:=\left(\frac{1}{\theta}-\kappa\right)^{-1}\max_{x}\left|\tau'\right|$.
\end{proof}
\begin{center}{\color{red}\fbox{\color{black}\parbox{16cm}{
\begin{defn}
The \textbf{trapped set} in phase space $T^{*}I$ is defined as
\begin{equation}
\mathcal{K}=\left\{ \left(x,\xi\right)\in T^{*}I,\quad\exists C\Subset T^{*}I\mbox{ compact},\ \forall n\in\mathbb{Z},\ F^{n}\left(x,\xi\right)\cap C\neq\emptyset\right\} .\label{eq:def_K_phase_space}
\end{equation}

\end{defn}
}}}\end{center}
\begin{rem}
Since the map $F:T^{*}I\rightarrow T^{*}I$ is a lift of the map,
we have $\mathcal{K}\subset\left(K\times\mathbb{R}\right)$. For any
$R$ given from Lemma \ref{lem:escape_F} we can precise this and
obtain
\[
\mathcal{K}\subset\left(K\times\left[-R,R\right]\right).
\]

\end{rem}
For $\varepsilon>0$, let $\mathcal{K}_{\varepsilon}$ denote a $\varepsilon-$neighborhood
of the trapped set $\mathcal{K}$, namely
\[
\mathcal{K}_{\varepsilon}:=\left\{ \left(x,\xi\right)\in T^{*}I,\quad\exists\left(x_{0},\xi_{0}\right)\in\mathcal{K},\quad\max\left(\left|x-x_{0}\right|,\left|\xi-\xi_{0}\right|\right)\leq\varepsilon\right\} .
\]

From now on we will make the following hypothesis on the multi-valued
map $F$.

\vspace{0.cm}\begin{center}{\color{red}\fbox{\color{black}\parbox{16cm}{
\begin{assumption}
\label{hyp:minimal_capt}We assume the following property called ``\textbf{minimal
captivity'':}
\begin{equation}
\exists\varepsilon>0,\quad\forall\left(x,\xi\right)\in\mathcal{K}_{\varepsilon},\quad\sharp\left\{ F\left(x,\xi\right)\bigcap\mathcal{K}_{\varepsilon}\right\} \leq1.\label{eq:hyp_minimal_captivity_epsilon}
\end{equation}
This means that the dynamics of $F$ is univalued on the trapped set
$\mathcal{K}$.
\end{assumption}
}}}\end{center}\vspace{0.cm}
\begin{rem}
In the paper \cite{fred_expanding_09} we introduced the property
of \textbf{``partial captivity''} which is weaker than \textbf{``minimal
captivity''}: partial captivity roughly states that most of trajectories
escape from the trapped set $\mathcal{K}$ whereas minimal captivity
states that every trajectory except one, escapes from the trapped
set $\mathcal{K}$.
\end{rem}
In Section \ref{sec:6} we provide more details on the dynamics of
the map $F:T^{*}I\rightarrow T^{*}I$, namely we provide a more precise
description of the trapped set $\mathcal{K}$, a detailed symbolic
coding for this dynamics and some equivalent statements to the property
of minimal captivity.

For the next theorem, let us define the function $V_{0}\in C^{\infty}\left(I\right)$
\begin{equation}
V_{0}\left(x\right):=\frac{1}{2}J\left(x\right)\label{eq:def_V0}
\end{equation}
called ``\textbf{potential of reference}'' with $J\left(x\right)$
defined in (\ref{eq:def_J}). Recall from (\ref{eq:J_positive}) that
\[
V_{0}\left(x\right)\geq\frac{1}{2}\log\left(\frac{1}{\theta}\right)>0.
\]

Let us define the function $D\in C^{\infty}\left(I\right)$
\begin{equation}
D\left(x\right):=\mathrm{Re}\left(V\left(x\right)\right)-V_{0}\left(x\right)\label{eq:def_damping_function}
\end{equation}
called ``\textbf{effective damping function}''.

\vspace{0.cm}\begin{center}{\color{blue}\fbox{\color{black}\parbox{16cm}{
\begin{thm}
\label{th:gap_spectral}\textbf{Spectral gap in the semi-classical
limit.} With assumption \ref{hyp:minimal_capt} of ``minimal captivity''
(and $m$ sufficiently large so that $r_{m}\ll1$ in (\ref{eq:bound_norm_R_def_rm})),
the spectral radius of the operators $\hat{F}_{\hbar}:H^{-m}\left(\mathbb{R}\right)\rightarrow H^{-m}\left(\mathbb{R}\right)$
satisfies in the semi-classical limit $\hbar\rightarrow0$:
\begin{equation}
r_{s}\left(\hat{F}_{\hbar}\right)\leq e^{\gamma_{+}}+o\left(1\right)\label{eq:upper_bound}
\end{equation}
with
\begin{equation}
\gamma_{+}:=\limsup_{n\rightarrow\infty}\left(\sup_{x,w_{0,n}}\frac{1}{n}D_{w_{0,n}}\left(x\right)\right)\label{eq:gamma_+}
\end{equation}
where $D_{w_{0,n}}\left(x\right):=\sum_{k=1}^{n}D\left(\phi_{w_{k,n}}\left(x\right)\right)$
is the Birkhoff average of the damping function $D$ along a trajectory
of length $n$, starting from the point $x$. Moreover the norm of
the resolvent is controlled uniformly with respect to $\hbar$: for
any $\rho>e^{\gamma_{+}}$, there exist $C_{\rho}>0$, $\hbar_{\rho}>0$
such that $\forall\hbar<\hbar_{\rho}$, $\forall\left|z\right|>\rho$
then
\begin{equation}
\left\Vert \left(z-\hat{F}_{\hbar}\right)^{-1}\right\Vert _{H^{-m}\left(\mathbb{R}\right)}\leq C_{\rho}.\label{eq:bound_on_resolvent}
\end{equation}

\end{thm}
}}}\end{center}\vspace{0.cm}

\paragraph{Remark}
\begin{itemize}
\item Notice that Theorem \ref{th:gap_spectral} depends on the roof function
$\tau$ only implicitly through assumption \ref{hyp:minimal_capt}.
The value of the upper bound (\ref{eq:upper_bound}) does not depend
on $\tau$.
\item Eq.(\ref{eq:bound_on_resolvent}) implies (\ref{eq:upper_bound})
and is equivalent%
\footnote{Let us show the equivalence. In one sense, let $\rho_{2}>\rho_{1}>e^{\gamma_{+}}$,
suppose that $\left\Vert \hat{F}_{\hbar}^{n}\right\Vert _{H^{-m}\left(\mathbb{R}\right)}\leq c_{\rho_{1}}.\rho_{1}^{n}$.
Let $\left|z\right|>\rho_{2}$. The relation $\left(z-\hat{F}_{\hbar}\right)^{-1}=z^{-1}\sum_{n\geq0}\left(\frac{\hat{F}_{h}}{z}\right)^{n}$
gives that 
\[
\left\Vert \left(z-\hat{F}_{\hbar}\right)^{-1}\right\Vert \leq\left|z\right|^{-1}\sum_{n\geq0}\frac{\left\Vert \hat{F}_{h}^{n}\right\Vert }{\left|z\right|^{n}}\leq\left|z\right|^{-1}c_{\rho_{1}}\sum_{n\geq0}\frac{\rho_{1}^{n}}{\left|z\right|^{n}}=\frac{c_{\rho_{1}}}{\left|z\right|-\rho_{1}}\leq\frac{c_{\rho_{1}}}{\rho_{2}-\rho_{1}}=:C_{\rho_{2}}
\]
For the other sense, suppose that for $\left|z\right|>\rho$, $\left\Vert \left(z-\hat{F}_{\hbar}\right)^{-1}\right\Vert \leq C_{\rho}$.
From the Cauchy formula $\hat{F}_{\hbar}^{n}=\frac{1}{2\pi i}\oint_{\gamma}z^{n}\left(z-\hat{F}_{\hbar}\right)^{-1}dz$
where $\gamma$ is the circle of radius $\rho$ one deduces that$\left\Vert \hat{F}_{\hbar}^{n}\right\Vert \leq\rho C_{\rho}\rho^{n}$.%
} to the following property that the norm $\left\Vert \hat{F}_{\hbar}^{n}\right\Vert $
is controlled uniformly with respect to $\hbar$: For any $\rho>e^{\gamma_{+}}$,
there exist $c_{\rho}>0$, $\hbar_{\rho}>0$ such that $\forall\hbar<\hbar_{\rho}$,
for any $n\in\mathbb{N}$,
\begin{equation}
\left\Vert \hat{F}_{\hbar}^{n}\right\Vert _{H^{-m}\left(\mathbb{R}\right)}\leq c_{\rho}.\rho^{n}\label{eq:bound_for_F^n}
\end{equation}
We will use (\ref{eq:bound_for_F^n}) later to compute asymptotics
of correlation functions.
\end{itemize}
\begin{center}{\color{blue}\fbox{\color{black}\parbox{16cm}{
\begin{lem}
\textbf{\label{thm:dim_of_K_tilde}}If assumption \ref{hyp:minimal_capt}
holds true and if the adjacency matrix $A$ is symmetric then 
\begin{equation}
\mathrm{dim}_{M}\mathcal{K}=2\mathrm{dim}_{M}K=2\mathrm{dim}_{H}K\label{eq:dim_Ktilde}
\end{equation}
where $\mathrm{dim}_{M}B$ stands for the Minkowski dimension of a
set $B$ as defined in Eq.(\ref{eq:def_dim_M}).
\end{lem}
}}}\end{center}

Recall from (\ref{eq:dim_H_dim_M}) that $\mathrm{dim}_{H}K=\mathrm{dim}_{M}K$.

\begin{center}{\color{blue}\fbox{\color{black}\parbox{16cm}{
\begin{thm}
\label{thm:fractal-weyl-upper}\textbf{''Fractal Weyl upper bound}''.
Suppose that the assumption of minimal captivity \ref{hyp:minimal_capt}
holds and that the adjacency matrix $A$ is symmetric. For any $\varepsilon>0$,
any $\eta>0$, we have for $\hbar\rightarrow0$ 
\begin{equation}
\sharp\left\{ \lambda_{i}^{\hbar}\in\mathrm{Res}\left(\hat{F}_{\hbar}\right)\,|\,\left|\lambda_{i}^{\hbar}\right|\geq\varepsilon\right\} =\mathcal{O}\left(\hbar^{-\dim_{H}\left(K\right)-\eta}\right).\label{eq:fractal_weyl-1}
\end{equation}

\end{thm}
}}}\end{center}

\subsection{\label{sub:Decay-of-correlations}Decay of correlations}

In this subsection we present a quite immediate consequence of the
existence of an asymptotic spectral radius $e^{\gamma_{+}}$ obtained
in Theorem \ref{th:gap_spectral}: we obtain a finite expansion for
correlation functions $\langle v|\mathcal{\hat{F}}^{n}u\rangle$ of
the extended transfer operator $\mathcal{\hat{F}}$ defined in (\ref{eq:def_extended_transfer_op}),
with $u,v\in C^{\infty}\left(I\times S^{1}\right)$.

We first introduce a notation: for a given $\nu\in\mathbb{Z}$, we
have  seen in Theorem \ref{thm:discrete-spectrum} that the transfer
operator $\mathcal{\hat{F}}_{\nu}\equiv\hat{F}_{N}$, that the transfer
operator $\mathcal{\hat{F}}_{\nu}\equiv\hat{F}_{N}$ has a discrete
spectrum of resonances. For $\rho>0$ such that there is no eigenvalue
on the circle $\left|z\right|=\rho$ for any $\nu\in\mathbb{Z}$,
we denote by $\Pi_{\rho,\nu}$ the spectral projector of the operator
$\hat{F}_{\nu}$ on the domain $\left\{ z\in\mathbb{C},\left|z\right|>\rho\right\} $.
These projection operators have obviously finite rank and commute
with $\hat{F}_{\nu}$.

\begin{center}{\color{blue}\fbox{\color{black}\parbox{16cm}{
\begin{thm}
\label{thm:correl_funct}For any $\rho>e^{\gamma_{+}}$, there exists
$\nu_{0}\in\mathbb{N}$ such that for any $u,v\in C^{\infty}\left(I\times S^{1}\right)$,
in the limit $n\rightarrow\infty$,
\begin{equation}
\langle v|\mathcal{\hat{F}}^{n}u\rangle=\sum_{\left|\nu\right|\leq\nu_{0}}\langle v|\left(\hat{F}_{\nu}\Pi_{\rho,\nu}\right)^{n}u\rangle+O\left(\rho^{n}\right)\label{eq:correlations_external_band}
\end{equation}

\end{thm}
}}}\end{center}
\begin{rem}
In the right hand side of (\ref{eq:correlations_external_band}) there
is a finite sum and each operator $\hat{F}_{\nu}\Pi_{\rho,\nu}$ has
finite rank. Using the spectral decomposition of $\hat{F}_{\nu}$
 we get an expansion of the correlation function $\langle v|\mathcal{\hat{F}}^{n}u\rangle$
with a finite number of terms which involve the leading Ruelle resonances
(i.e. those with modulus greater than $\rho$) plus the error term
$O\left(\rho^{n}\right)$.\end{rem}
\begin{proof}
Let $\rho>e^{\gamma_{+}}$. Recall that $\hbar=\frac{1}{2\pi\nu}$
and that we note $\hat{F}_{\hbar}=\hat{F}_{\nu}$. In Theorem \ref{thm:discrete-spectrum}
we have for $\hbar\rightarrow0$ that $r_{s}\left(\hat{F}_{\hbar}\right)\leq e^{\gamma_{+}}+o\left(1\right)$.
Let the value of $\nu_{0}$ be such that $r_{s}\left(\hat{F}_{\hbar}\right)<\rho$
for every $\nu>\nu_{0}$. Then
\begin{equation}
\langle v|\mathcal{\hat{F}}^{n}u\rangle=\sum_{\left|\nu\right|\leq\nu_{0}}\langle v|\left(\hat{F}_{\nu}\Pi_{\rho,\nu}\right)^{n}u\rangle+O_{\nu_{0}}\left(\rho^{n}\right)+\sum_{\left|\nu\right|>\nu_{0}}\langle v|\hat{F}_{\nu}^{n}u\rangle\label{eq:sum}
\end{equation}
But $\left|\langle v|\hat{F}_{\nu}^{n}u\rangle\right|\leq\left\Vert u_{\nu}\right\Vert _{H^{m}}\left\Vert v_{\nu}\right\Vert _{H^{-m}}\left\Vert \hat{F}_{\nu}^{n}\right\Vert _{H^{-m}}$
where $u_{\nu},v_{\nu}\in C^{\infty}\left(S^{1}\right)$ stand for
the Fourier components of the smooth functions $u,v\in C^{\infty}\left(I\times S^{1}\right)$.
On one hand, for smooth functions one has fast decay $\left\Vert u_{\nu}\right\Vert ,\left\Vert v_{\nu}\right\Vert =O\left(\nu^{-\infty}\right)$.
On the other hand from (\ref{eq:bound_for_F^n}), $\left\Vert \hat{F}_{\nu}^{n}\right\Vert _{H^{-m}}=O\left(\rho^{n}\right)$.
So $\left|\sum_{\left|\nu\right|>\nu_{0}}\langle v|\hat{F}_{\nu}^{n}u\rangle\right|=O\left(\rho^{n}\right)$.
Then (\ref{eq:sum}) gives (\ref{eq:correlations_external_band}).
\end{proof}
We recall the following result called ``Perron-Frobenius Lemma'':

\begin{center}{\color{blue}\fbox{\color{black}\parbox{16cm}{
\begin{lem}
\label{lem:perron_Frob}For real potential $V$, if the map $\phi^{-1}$
is ergodic then the transfer operator $\hat{\mathcal{F}}$ has a leading
and simple eigenvalue $\lambda_{0}>0$ in the Fourier mode $\nu=0$,
i.e.
\[
\hat{\mathcal{F}}_{\nu=0}=\lambda_{0}\Pi_{\lambda_{0}}+\hat{\mathcal{F}}'
\]
with $\Pi_{\lambda_{0}}$ being the rank 1 spectral projector associated
to $\lambda_{0}$, the remainder operator has $r_{s}\left(\hat{\mathcal{F}}'\right)<\lambda_{0}$
and for any $\nu\neq0$, we also have $r_{s}\left(\hat{\mathcal{F}}_{\nu}\right)<\lambda_{0}$.
\end{lem}
}}}\end{center}

For example without potential, i.e. $V=0$, then $\lambda_{0}=\exp\left(\mathrm{Pr}\left(-J\right)\right)$
with $J$ given in (\ref{eq:def_J}). As a consequence of Lemma \ref{lem:perron_Frob}
and Theorem \ref{thm:correl_funct} we obtain (a result already obtained
by Dolgopyat \cite{dolgopyat_02}):

\begin{center}{\color{blue}\fbox{\color{black}\parbox{16cm}{
\begin{thm}
\label{Th1.8}Let $u,v\in C^{\infty}\left(I\times S^{1}\right)$,
then for $n\rightarrow\infty$,
\[
\langle v|\hat{\mathcal{F}}^{n}u\rangle=\lambda_{0}^{n}\langle v|\Pi_{\lambda_{0}}u\rangle+O\left(\left|\lambda_{1}\right|^{n}\right)
\]
where $\lambda_{1}$ is the second eigenvalue with $\left|\lambda_{1}\right|<\lambda_{0}$.
This gives that the extended map $f:I\times S^{1}\rightarrow I\times S^{1}$
is mixing with exponential decay of correlations.
\end{thm}
}}}\end{center}

\section{\label{sec:Examples}Numerical results for the truncated Gauss map
and Bowen-Series maps}

In this section we will present numerical results for two important
classes of I.F.S.: the truncated Gauss map and the Bowen-Series maps
for convex co-compact hyperbolic surfaces.  We will show that both
examples satisfy the partially captive property. We will then give
some numerical illustrations of the main theorems presented in this
paper and finally discuss the connection between the spectrum of these
transfer operators with the resonance spectrum of the Laplacian on
hyperbolic surfaces.

\subsection{\label{sub:4.1}The truncated Gauss map}

In this section we consider the example of the I.F.S. defined from
the truncated Gauss map with $N$ intervals presented in Section \ref{sub:Example-with-the}.
We choose the roof function $\tau$ and the potential function $V$
which enter in the definition of the transfer operator (\ref{e:def_F_op_ij})
to be:
\begin{equation}
\tau\left(x\right)=-J\left(x\right),\qquad V\left(x\right)=\left(1-a\right)J\left(x\right),\quad a\in\mathbb{R}.\label{eq:tau_J_for_Laplacian}
\end{equation}
where $J\left(x\right)=\log\left(\left|\left(\phi^{-1}\right)'\left(x\right)\right|\right)=\log\left(\left|G'\left(x\right)\right|\right)=-2\log\left(x\right)$
has been defined in \ref{eq:def_J}. Let us write
\[
s=a+ib\in\mathbb{C},\quad b=\frac{1}{\hbar}>0.
\]
Then for every $s\in\mathbb{C}$, the transfer operator $\hat{F}$
given in (\ref{eq:def_transfert_op_F}) will be written $\hat{L}_{s}=\hat{F}$
and is given by:
\begin{equation}
\hat{L}_{s}\varphi=\hat{F}\varphi=e^{V\left(x\right)}e^{i\frac{1}{\hbar}\tau\left(x\right)}\varphi\circ\phi^{-1}=e^{\left(1-s\right)J}\varphi\circ\phi^{-1}\label{eq:op_L_s}
\end{equation}
As explained in Section \ref{sub:Relation-with-the-Selberg} below,
this choice is interesting due its relation with the dynamics on the
modular surface. The (adjoint of the) transfer operator $\hat{F}$
constructed in this way is usually called the \textbf{Gauss-Kuzmin-Wirsing
transfer operator} or ``Dieter-Mayer transfer operator'' for the
truncated Gauss map.

\begin{figure}
\begin{centering}
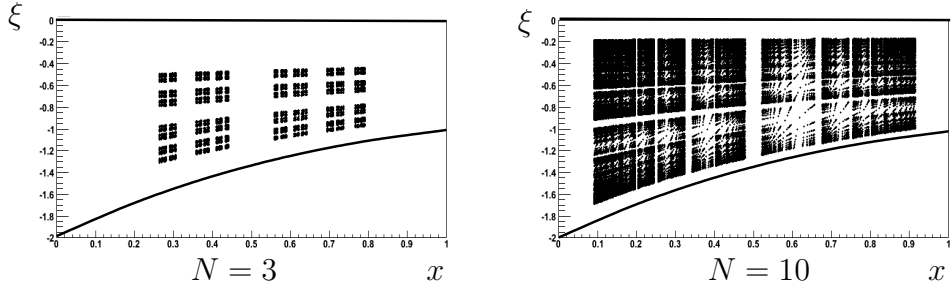
\par\end{centering}

\caption{\label{fig:trapped_set}The trapped set $\mathcal{K}_{N}:=\mathcal{K}$
for the truncated Gauss map with functions (\ref{eq:tau_J_for_Laplacian}),
for the cases of $N=3$ and $N=10$ branches. This corresponds to
the Gauss-Kuzmin-Wirsing transfer operator (\ref{eq:op_L_s}). We
have $\mathcal{K}_{N}\subset\mathcal{K}_{N+1}$ and for $N\rightarrow\infty$,
the limit trapped set $\mathcal{K}_{\infty}=\bigcup_{N\geq0}\mathcal{K}_{N}=\left\{ \left(x,\xi\right),x\in]0,1[,-\frac{2}{1+x}<\xi<0\right\} $
is the band between the marked black lines. (More precisely, we have
represented the periodic points with period $n=6$. That explains
the sparse aspect of the trapped set).}
\end{figure}

\begin{center}{\color{blue}\fbox{\color{black}\parbox{16cm}{
\begin{prop}
\label{prop:minimal_capt_for_model}For every $N\geq1$, the minimal
captivity assumption \ref{hyp:minimal_capt} holds true for the truncated
Gauss transfer operator defined by (\ref{eq:op_L_s}).
\end{prop}
}}}\end{center}

The proof is given in Section \ref{sub:Proof-of-minimal} below. In
this proof we explain the structure of the trapped set $\mathcal{K}$
with more details.

\begin{figure}
\begin{centering}
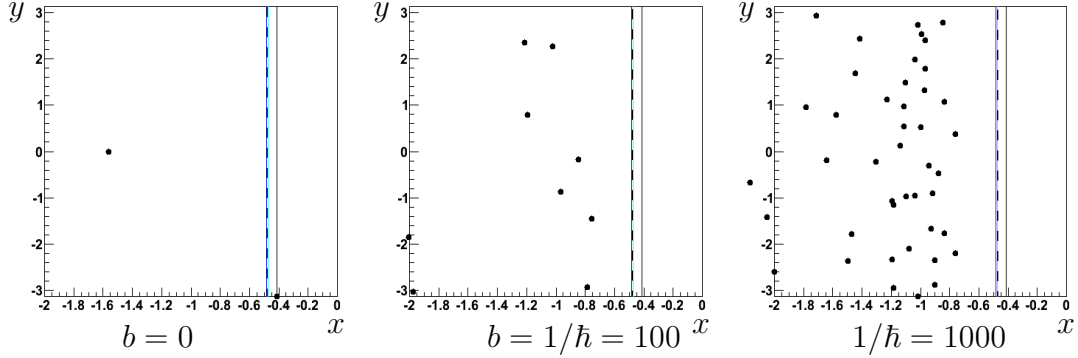
\par\end{centering}

\caption{\label{fig:Gauss_spectrum}The discrete spectrum of Ruelle resonances
$\lambda_{j}$ (in log scale writing: $\log\lambda=x+iy$) for the
truncated Gauss-Kuzmin-Wirsing transfer operator (\ref{eq:op_L_s})
associated to the Gauss map, for $N=3$ branches and parameters $a=1$,
$b=0,100,1000$. For $b=0$ there is the eigenvalue $\lambda=e^{\mathrm{Pr}\left(-J\right)}$
at $x=\mathrm{Pr}\left(-J\right)$, $y=-\pi$ corresponding to the
``equilibrium measure''. The full vertical line is at $x=\mathrm{Pr}\left(-J\right)$.
The dashed vertical line is at $x=\gamma_{+}$ which is shown in (\ref{eq:upper_bound})
to be an asymptotic upper bound for $b=1/\hbar\rightarrow\infty$.}
\end{figure}

\begin{figure}
\begin{centering}
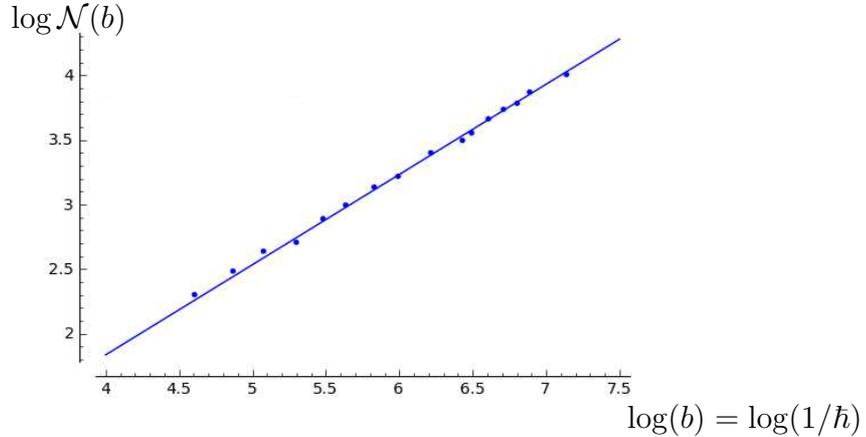
\par\end{centering}

\caption{\label{fig:Gauss_Weyl_law}This is the Weyl law for the model of Gauss
map with $N=3$ branches. The points represent the number of resonances
$\mathcal{N}\left(b\right)=\sharp\left\{ \lambda_{j}\in\mathrm{Res}\left(\hat{L}_{s}\right),\log\left|\lambda_{j}\right|>-3.5\right\} $
computed numerically, as a function of the semiclassical parameter
$b=1/\hbar$ in log scale. The linear fit gives $\log\mathcal{N}\left(b\right)=-0.70\cdot\log b-0.96$
which has to be compared to the fractal Weyl law (\ref{eq:fractal_weyl-1})
giving $\log\mathcal{N}\left(b\right)\leq-\dim_{H}\left(K\right)\cdot\log b+\mathrm{cste}$.
From (\ref{fig:Pressure_Gauss}) with have $\dim_{H}K_{3}=0.705$
giving an excellent agreement with the numerical results and suggesting
that the upper bound is in fact optimal.}
\end{figure}

Consequently, we can apply Theorem \ref{th:gap_spectral} and deduce
that there is an asymptotic spectral gap. See figure \ref{fig:Gauss_spectrum}
for numerical results and comments. We can also apply Theorem \ref{thm:fractal-weyl-upper}
and deduce an fractal Weyl upper bound for the density of resonances.
See figure \ref{fig:Gauss_Weyl_law} for numerical results and comments. 

\subsubsection{\label{sub:Relation-with-the-Selberg}Relation with the zeroes of
the Selberg zeta function}

For the geodesic flow on the modular surface $\mathrm{SL}_{2}\mathbb{Z}\backslash\mathrm{SL}_{2}\mathbb{R}$
it is possible to define the Selberg zeta function (see Section \ref{sub:4.2.1}
below for more comments and references):

\[
\zeta_{\mathrm{Selberg}}\left(s\right)=\prod_{\gamma}\prod_{m\geq0}\left(1-e^{-\left(s+m\right)\left|\gamma\right|}\right),\quad s\in\mathbb{C},
\]
where the product is over the primitives periodic orbits $\gamma$
of the geodesic flow and $\left|\gamma\right|$ denotes the length
of the orbit. This zeta function is absolutely convergent for $\mathrm{Re}(s)>1$.
Using the Gauss map and continued fractions, C. Series has shown that
a periodic orbit $\gamma$ is in one to one correspondence with a
periodic sequence $\left(w_{j}\right)_{j\in\mathbb{Z}}\in\left(\mathbb{N}\backslash\left\{ 0\right\} \right)^{\mathbb{Z}}$
where $w_{j}\in\mathbb{N}\backslash\left\{ 0\right\} $ is the index
of the branch of the Gauss map $G_{w_{j}}^{-1}$ in (\ref{eq:Gauss_map-1}).
Given $N\geq1$, we can restrict the product $\prod_{\gamma}$ over
periodic orbits above to orbits for which $w_{j}\leq N$, $\forall j\in\mathbb{Z}$,
and define a truncated Selberg zeta function as follows:
\[
\zeta_{\mathrm{Selberg},N}\left(s\right)=\prod_{\gamma,\ w_{j}\leq N.\forall j,}\quad\prod_{m\geq0}\left(1-e^{-\left(s+m\right)\left|\gamma\right|}\right),\quad s\in\mathbb{C},
\]

On the other hand, for fixed $s\in\mathbb{C}$, we have from Theorem
\ref{thm:discrete-spectrum} that the operator $\hat{L}_{s}$ has
discrete spectrum of Ruelle resonances. It is possible to define the
dynamical determinant of $\hat{L}_{s}$ by
\[
d\left(z,s\right):=\mathrm{Det}\left(1-z\hat{L}_{s}\right):=\exp\left(-\sum_{n\geq1}\frac{z^{n}}{n}\mathrm{Tr}^{\flat}\left(\hat{L}_{s}^{n}\right)\right),\quad z\in\mathbb{C}
\]
where $\mathrm{Tr}^{\flat}\left(\hat{L}_{s}^{n}\right)$ stands for
the flat trace of Atiyah-Bott. The sum is convergent for $\left|z\right|$
 small enough. It is known that for fixed $s$, the zeroes of $d\left(z,s\right)$
 (as a function of $z$) coincide with multiplies with the Ruelle
resonances of $\hat{L}_{s}$ \cite{baladi_06}. In the case $z=1$,
we also have that $d\left(1,s\right)$ coincides with the truncated
Selberg zeta function \cite{pollicott_rocha_1997}\cite{bortwick_book_07}:

\begin{equation}
\mathrm{Det}\left(1-\hat{L}_{s}\right)=\zeta_{\mathrm{Selberg},N}\left(s\right)\label{eq:det_dyn}
\end{equation}
which means that the zeroes of $\zeta_{\mathrm{Selberg},N}\left(s\right)$
are given (with multiplicity) by the event that $1$ is a Ruelle resonance
of the transfer operator $\hat{L}_{s}$. This also shows that $\zeta_{\mathrm{Selberg},N}\left(s\right)$
has a holomorphic extension to the complex plane $s\in\mathbb{C}$.
\begin{rem}
in \cite{pollicott_rocha_1997}\cite[p.306]{bortwick_book_07} they
consider the adjoint operator $\hat{L}_{s}^{*}$ called the Perron-Frobenius
operator.
\end{rem}

\subsection{Bowen Series maps for Schottky surfaces}

The second class of examples that we consider in this section are
Bowen-Series maps for Schottky surfaces \cite{bowen_series_1979}.
We will follow the notation of D. Borthwick's book \cite[chap.15]{bortwick_book_07}
and recall the definition of a Schottky group given there. Recall
that an element $S=\left(\begin{array}{cc}
a & b\\
c & d
\end{array}\right)\in\mathrm{SL}_{2}\mathbb{R}$ acts on $\mathbb{H}^{2}=\mathrm{SL}_{2}\mathbb{R}/\mathrm{SO}_{2}$
and $\overline{\mathbb{R}}=\partial\mathbb{H}^{2}$ by $S\left(x\right):=\frac{ax+b}{cx+d}$.

\begin{figure}
\centering{}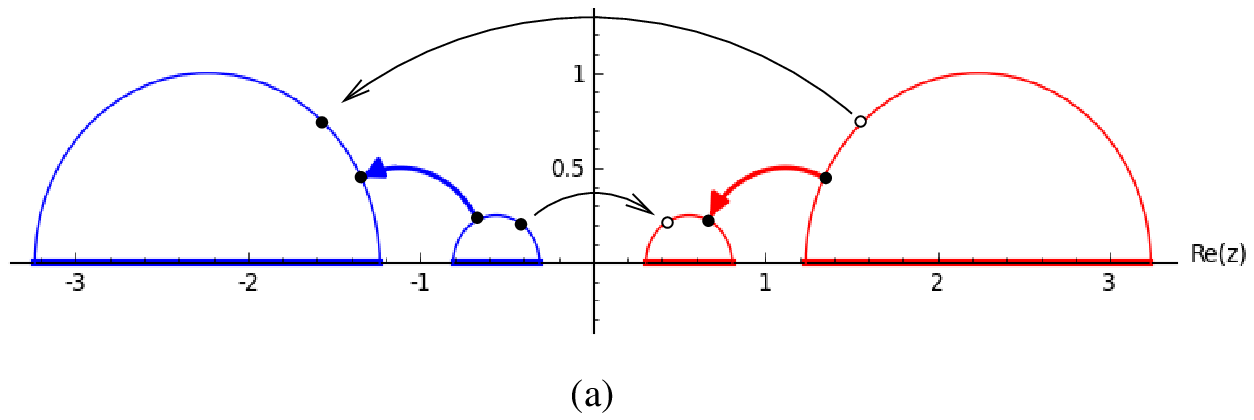\\
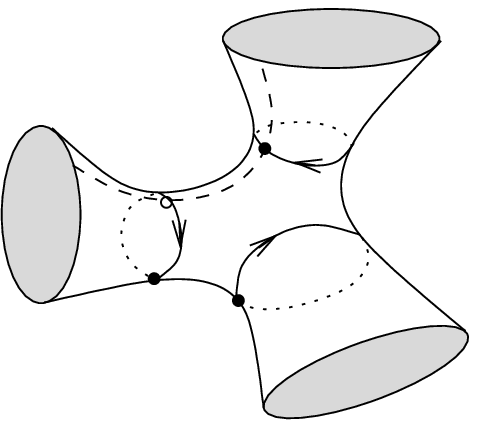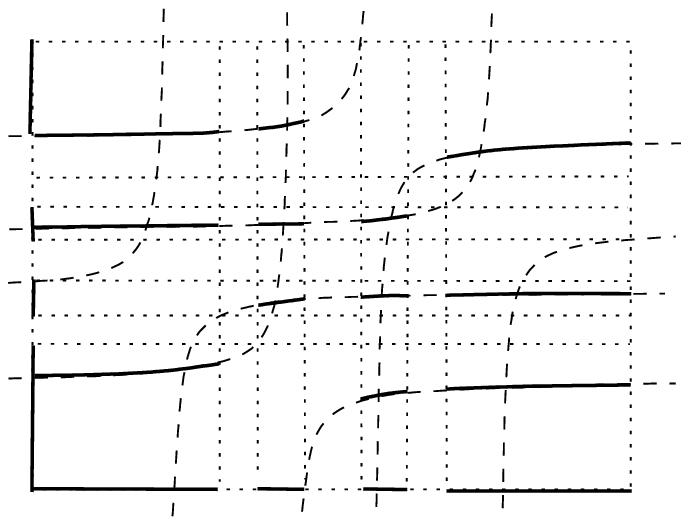 \caption{\label{fig:BowenSeries} In this (arbitrary) example, we have $r=2$
hyperbolic matrices of $\mathrm{SL}_{2}\mathbb{R}$: $S_{1}=\left(\protect\begin{array}{cc}
4 & \sqrt{5}\protect\\
-\sqrt{5} & -1
\protect\end{array}\right)$ and $S_{2}=\left(\protect\begin{array}{cc}
-1 & \sqrt{5}\protect\\
-\sqrt{5} & 4
\protect\end{array}\right)$ that generate a Schottky group $\Gamma=\left\langle S_{1},S_{2}\right\rangle $.
Figure (a) shows the Dirichlet fundamental domain $\mathbb{H}^{2}\backslash\left(D_{1}\cup D_{2}\right)$
with the intervals $I_{i}$,$i=1,2$, on which the I.F.S. is defined.
Figure (b) shows the resulting Schottky surface $\Gamma\backslash\mathbb{H}^{2}$.
It has three funnels. Figure (c) shows the graph of the generating
functions $\phi_{i,j}=g_{j}=S_{j+r}:I_{i}\rightarrow I_{j}$ of the
associated iterated function system. }
\end{figure}

\begin{center}{\color{red}\fbox{\color{black}\parbox{16cm}{
\begin{defn}
\label{def:SchottkyGroup} Let $D_{1},\dots D_{2r}$ be disjoint closed
half discs in the Poincaré half plane $\mathbb{H}^{2}=\mathrm{SL}_{2}\mathbb{R}/\mathrm{SO}_{2}$
with center in $\mathbb{R}=\partial\mathbb{H}^{2}\backslash\left\{ \infty\right\} $.
There exist elements $S_{i}\in\mathrm{SL_{2}\mathbb{R}}$, $i=1,\dots,r$
such that $S_{i}\left(\partial D_{i}\right)=\partial D_{i+r}$ and
$S_{i}\left(\mathrm{Int}\left(D_{i}\right)\right)=\mathbb{C}\setminus D_{i+r}$.
The group generated by the $S_{i}$ is called a \textbf{Schottky group}
$\Gamma=\langle S_{1},\dots,S_{r}\rangle$.
\end{defn}
}}}\end{center}
\begin{rem}
For convenience we will use a cyclic notation for the indices $i=1,\dots,2r$.
Then one can also define $S_{i}$ for $i=r+1,\dots,2r$ as in the
definition above and obtains $S_{i+r}=S_{i}^{-1}$.
\end{rem}
Let $I_{i}:=D_{i}\cap\partial\mathbb{H}$. Then $\left(I_{i}\right)_{i=1,\ldots2r}$
are $N=2r$ disjoint closed intervals. One has $S_{j}\left(\mathrm{Int}\left(I_{j}\right)\right)=\partial\mathbb{H}\backslash I_{j+r}$
and we assume that $S_{j}$ is expanding on $I_{j}$ (this can always
be obtained by taking iterations if necessary and localizing further
to the trapped set, see \cite[prop.15.4]{bortwick_book_07}). The
maps $S_{j}$ are usually called the Bowen series maps. Considering
the inverse maps one obtains an iterated function system according
to definition \ref{def:IFSAn-iterated-function} associated to this
Schottky group in the following way. For any $j=1,\ldots N$ and $i\neq j+r$
let: 
\[
\phi_{i,j}:=S_{j}^{-1}=S_{j+r}\quad:I_{i}\to S_{j}^{-1}\left(I_{i}\right)\subset\mathrm{Int}\left(I_{j}\right)
\]
The adjacency matrix $A_{i,j}$ has all entries are equal to one except
$A_{i,i+r}=0$. See Figure \ref{fig:BowenSeries}.

As in (\ref{eq:tau_J_for_Laplacian}) we make the following choice
for the potential and the roof function for $x\in I_{j}$

\begin{equation}
\tau\left(x\right)=-J\left(x\right),\qquad V\left(x\right)=\left(1-a\right)J\left(x\right),\quad a\in\mathbb{R}.\label{eq:RoofFunctionBowenSeries-1}
\end{equation}

where $J\left(x\right)=\log\left(\left|\left(\phi_{i,j}^{-1}\right)'\left(x\right)\right|\right)=\log\left(\left|g_{j}'\left(x\right)\right|\right)$
has been defined in \ref{eq:def_J}. Let us write
\[
s=a+ib\in\mathbb{C},\quad b=\frac{1}{\hbar}>0.
\]
Then for every $s\in\mathbb{C}$, the transfer operator $\hat{F}$
given in (\ref{eq:def_transfert_op_F}) will be written $\hat{L}_{s}=\hat{F}$
and is given by:
\begin{equation}
\hat{L}_{s}\varphi=\hat{F}\varphi=e^{V\left(x\right)}e^{i\frac{1}{\hbar}\tau\left(x\right)}\varphi\circ\phi^{-1}=e^{\left(1-s\right)J}\varphi\circ\phi^{-1}\label{eq:op_L_s-1}
\end{equation}

The adjoint of our transfer operator $L_{s}^{*}=\hat{F}^{*}$ is exactly
the Ruelle transfer operator defined in \cite[p.304]{bortwick_book_07}
and as we will discuss below, its spectrum is in a close connection
to the spectrum of the Laplace operator on the Schottky surface.

\begin{center}{\color{blue}\fbox{\color{black}\parbox{16cm}{
\begin{prop}
\label{prop:minimal_capt_for_model-1}The minimal captivity assumption
\ref{hyp:minimal_capt} holds true for the Bowen-Series transfer operator
defined by (\ref{eq:op_L_s-1}).
\end{prop}
}}}\end{center}

The proof is given in Section \ref{sub:Proof-of-minimal} below.

Consequently, we can apply Theorem \ref{th:gap_spectral} and deduce
that there is an asymptotic spectral gap. We can also apply Theorem
\ref{thm:fractal-weyl-upper} and deduce an fractal Weyl upper bound
for the density of resonances.

\subsubsection{\label{sub:4.2.1}Selberg zeta function and resonances of the Laplacian}

For the geodesic flow on a hyperbolic surface it is possible to define
the Selberg zeta function 
\[
\zeta_{\mathrm{Selberg}}\left(s\right)=\prod\limits _{\gamma}\prod\limits _{m\geq0}\left(1-e^{-(s+m)|\gamma|}\right)
\]
where the product is over primitive periodic orbits $\gamma$ of the
geodesic flow and $|\gamma|$ denotes the length of the orbit. This
zeta function is absolutely convergent for $\mathrm{Re}(s)>1$ and
has a meromorphic continuation to the whole complex plane. This continuation
is particularly interesting as its zeros are either ``topological
zeros'' located on the real axis or resonances of the Laplace operator
$\Delta$ on the corresponding hyperbolic surface $\Gamma\backslash\mathbb{H}^{2}$.
These resonances $s\in\mathrm{Res}\left(\Delta\right)$ are defined
as the poles of the meromorphic extension of the resolvent\cite{bortwick_book_07}:
\begin{equation}
R\left(s\right):=\left(\Delta-s\left(1-s\right)\right)^{-1},\quad s\in\mathbb{C}\label{eq:def_R_s}
\end{equation}
This correspondence follows from the Selberg trace formula for finite-area
surfaces, and has been shown by Patterson-Perry \cite{patterson_perry_2001}
for infinite volume surfaces without cusps and Borthwick, Judge and
Perry \cite{borthwick_05} for infinite volume surfaces with cusps
(see also \cite{bortwick_book_07} for an overview).

For the transfer operators as defined above, one can define a dynamical
zeta function by \cite[p.305]{bortwick_book_07} 
\[
d\left(z,s\right):=\mathrm{Det}\left(1-z\hat{L}_{s}\right)
\]
The dynamical and the Selberg zeta function are equal $\zeta_{\mathrm{Selberg}}\left(s\right)=d\left(1,s\right)$
(see \cite[th.15.8]{bortwick_book_07}). This implies immediately
that if $s\in\mathbb{C}$ is a resonance of the Laplacian on the Schottky
surface, then $1$ has to be an eigenvalue of $\hat{L}_{s}$:
\begin{equation}
s\in\mathrm{Res}\left(\Delta\right)\Leftrightarrow1\in\mathrm{Spec}\left(\hat{L}_{s}\right)\label{eq:correspondance}
\end{equation}

\begin{rem}
For the full Gauss map (i.e. with infinitely many branches) the same
correspondence between the resonances of the Laplacian on the modular
surface $\mathrm{SL}_{2}\mathbb{Z}\backslash\mathbb{H}^{2}$ and the
Dieter-Mayer transfer operator $\hat{L}_{s}$ is true and has been
developed by Dieter Mayer \cite{mayer_90}. For the truncated Gauss
map considered in Section \ref{sub:4.2.1}, to our knowledge, no such
corresponding surfaces are known.
\end{rem}
Using the relation (\ref{eq:correspondance}) between the Ruelle spectrum
of the transfer operator $\hat{L}_{s}$ and the resonances of the
Laplacian, it is possible to deduce from Theorem \ref{th:gap_spectral}
some estimate on the ``asymptotic spectral gap'' for the resonances
of the Laplacian as follows.

\begin{center}{\color{red}\fbox{\color{black}\parbox{16cm}{
\begin{defn}
The \textbf{asymptotic spectral gap} of resonances of the Laplacian
$\Delta$ is defined by
\[
a_{\mathrm{asymp}}:=\limsup_{b\rightarrow\infty}\left\{ \mathrm{Re}\left(s\right)\mbox{ s.t. }s\in\mathrm{Res}\left(\Delta\right),\left|\mathrm{Im}\left(s\right)\right|>b\right\} 
\]

\end{defn}
}}}\end{center}

The setting (\ref{eq:RoofFunctionBowenSeries-1}) gives $D\left(x\right)=V-\frac{1}{2}J=\left(\frac{1}{2}-a\right)J\left(x\right)$
hence our estimate (\ref{eq:gamma_+}) gives that $a_{\mathrm{asympt}}\leq\frac{1}{2}$.
However this result concerning the resonances of the hyperbolic Laplacian
is not new: from the self-adjoint properties of the Laplacian $\Delta$
in $L^{2}$ space we have that $\mathrm{Im}\left(s\left(1-s\right)\right)\leq0$
and this gives that 
\begin{equation}
a_{\mathrm{asympt}}\leq\frac{1}{2}.\label{eq:a_asympt_1/2}
\end{equation}

\begin{rem}
If $\delta$ denotes the dimension of the limit set (equal to the
dimension of the trapped set $K$)  a result from F. Naud gives \cite{naud_expanding_2005}:
$\exists\varepsilon>0$ s.t.
\[
a_{\mathrm{asymp}}\leq\left(1-\varepsilon\right)\delta
\]
which improves (\ref{eq:a_asympt_1/2}) if $\delta\leq1/2$.
\end{rem}

\subsection{\label{sub:Proof-of-minimal}Proof of minimal captivity for both
models}

We give now the proof of Propositions \ref{prop:minimal_capt_for_model}
and \ref{prop:minimal_capt_for_model-1}. Note first that in both
models the contracting maps are Möebius maps i.e. of the form $x'_{j}=\phi_{i,j}\left(x\right)=\frac{a_{j}x+b_{j}}{c_{j}x+d_{j}}=g_{j}\left(x\right)$
with $2\times2$ matrices $g_{j}=\left(\begin{array}{cc}
a_{j} & b_{j}\\
c_{j} & d_{j}
\end{array}\right)$ with $D_{j}:=\mathrm{det}g_{j}=\pm1$. For the truncated Gauss map
these matrices are 
\begin{equation}
g_{j}=\left(\begin{array}{cc}
0 & 1\\
1 & j
\end{array}\right)=G_{j}^{-1}\label{eq:gj_Gauss}
\end{equation}
with $j=1,\ldots N$ and $D:=D_{j}=-1$. For the Bowen-Series maps
we have 
\begin{equation}
g_{j}=S_{j}^{-1}\in\mathrm{SL}_{2}\mathbb{R}\label{eq:gj_Shottky}
\end{equation}
with $j=1,\ldots2r$ and $D:=D_{j}=+1$.

The following proposition shows that there exists coordinates $\left(x,\eta\right)$
on phase space such that the canonical map $F=\left(F_{j}\right)_{j=1\ldots N}$
is decoupled in a product of identical maps.

\selectlanguage{french}%
\vspace{0.cm}\begin{center}{\color{blue}\fbox{\color{black}\parbox{16cm}{
\selectlanguage{english}%
\begin{lem}
The canonical map $F$ defined in (\ref{eq:canonical_map_Fij}) is
the union of the following maps $F_{j}$, with $j=1\ldots N$: 
\begin{align}
\left(x'_{j},\xi'_{j}\right) & =F_{j}\left(x,\xi\right)=\left(g_{j}\left(x\right),\left(g_{j}^{-1}\right)'\left(x'_{j}\right)\xi+\tau'\left(x'_{j}\right)\right)\label{eq:expression_Fj}\\
 & =\left(g_{j}\left(x\right),D_{j}\cdot\left(c_{j}x+d_{j}\right)^{2}\xi-2c_{j}\left(c_{j}x+d_{j}\right)\right)\label{eq:4.12}
\end{align}
Using the change of variables $\left(x,\eta\right)=\Phi\left(x,\xi\right)\in\overline{\mathbb{R}}\times\overline{\mathbb{R}}$
with $\overline{\mathbb{R}}:=\mathbb{R}\cup\left\{ \infty\right\} $
and 
\begin{equation}
\eta:=x-\frac{2D}{\xi},\label{eq:eta_xi}
\end{equation}
the map $F_{j}$ gets the simpler ``decoupled expression''
\begin{equation}
\left(x'_{j},\eta'_{j}\right)=\left(\Phi\circ F_{j}\circ\Phi^{-1}\right)\left(x,\eta\right)=\left(g_{j}\left(x\right),g_{j}\left(\eta\right)\right)\label{eq:expression_F_eta}
\end{equation}

\end{lem}
\selectlanguage{french}%
}}}\end{center}\vspace{0.cm}
\selectlanguage{english}%
\begin{rem}
geometrically these new variables $\left(x,\eta\right)$ can be interpreted
as the limit points $\left(x,\eta\right)\in\partial\mathbb{H}$ of
a geodesic. The map $\left(x',\eta'\right)=\left(\Phi\circ F\circ\Phi^{-1}\right)\left(x,\eta\right)$
is simply the Poincaré map of the geodesic flow \cite{dalbo_horocycle_trajectoires_2012}. \end{rem}
\begin{proof}
One has $g_{j}^{-1}=D_{j}\cdot\left(\begin{array}{cc}
d_{j} & -b_{j}\\
-c_{j} & a_{j}
\end{array}\right)$, $\left(g_{j}^{-1}\right)\left(y\right)=\frac{d_{j}y-b_{j}}{-c_{j}y+a_{j}}$
and $\left(g_{j}^{-1}\right)'\left(y\right)=D_{j}\cdot\left(a_{j}-c_{j}y\right)^{-2}=D_{j}\cdot\left(c_{j}x+d_{j}\right)^{2}$
if $y=g_{j}\left(x\right)$. The roof function is given by (\ref{eq:tau_J_for_Laplacian}):
\begin{align*}
\tau\left(y\right) & =-J\left(y\right)=-\log\left(\left|\left(\phi_{i,j}^{-1}\right)'\left(y\right)\right|\right)=-\log\left(\left|\left(g_{j}^{-1}\right)'\left(y\right)\right|\right)\\
 & =2\log\left(a_{j}-c_{j}y\right)
\end{align*}
So $\tau'\left(y\right)=-2c_{j}\left(a_{j}-c_{j}y\right)^{-1}=-2c_{j}\left(c_{j}x+d_{j}\right)$
and
\begin{equation}
\left(x'_{j},\xi'_{j}\right)=F_{j}\left(x,\xi\right)\underset{(\ref{eq:expression_Fj})}{=}\left(g_{j}\left(x\right),D\cdot\left(c_{j}x+d_{j}\right)^{2}\xi-2c_{j}\left(c_{j}x+d_{j}\right)\right)\label{eq:expression_F_j}
\end{equation}
giving (\ref{eq:4.12}). Now we use the change of variable 
\begin{equation}
\xi=\frac{2D}{x-\eta}.\label{eq:eta_xi-1}
\end{equation}
So 
\begin{align*}
\xi'_{j} & =D\cdot\left(c_{j}x+d_{j}\right)^{2}\xi-2c_{j}\left(c_{j}x+d_{j}\right)\\
 & =D\cdot\left(c_{j}x+d_{j}\right)^{2}\frac{2D}{\left(x-\eta\right)}-2c_{j}\left(c_{j}x+d_{j}\right)\\
 & =\frac{2\left(c_{j}x+d_{j}\right)}{\left(x-\eta\right)}\left(c_{j}\eta+d_{j}\right)
\end{align*}
Then
\begin{align*}
\eta'_{j} & =x'_{j}-\frac{2D}{\xi'_{j}}=\frac{a_{j}x+b_{j}}{c_{j}x+d_{j}}-\frac{D\left(x-\eta\right)}{\left(c_{j}x+d_{j}\right)\left(c_{j}\eta+d_{j}\right)}\\
 & =\frac{\left(a_{j}x+b_{j}\right)\left(c_{j}\eta+d_{j}\right)-\left(a_{j}d_{j}-b_{j}c_{j}\right)\left(x-\eta\right)}{\left(c_{j}x+d_{j}\right)\left(c_{j}\eta+d_{j}\right)}=\frac{a_{j}\eta+b_{j}}{c_{j}\eta+d_{j}}=g_{j}\left(\eta\right)
\end{align*}

\end{proof}
Recall that the multivalued map $\phi=\left(\phi_{i,j}=g_{j}\right)_{j}$
has a trapped set $K$ defined in (\ref{eq:trapped_set_K_def1}) as
$K=\bigcap_{n\geq1}\phi^{n}\left(I\right)$. The basin of $K$ on
$\overline{\mathbf{\mathbb{R}}}$ is $\mathcal{B}\left(K\right):=\left\{ x\in\overline{\mathbb{R}},\exists n\geq0,\phi^{n}\left(x\right)\in I\right\} \subset\overline{\mathbb{R}}$. 

\selectlanguage{french}%
\vspace{0.cm}\begin{center}{\color{blue}\fbox{\color{black}\parbox{16cm}{
\selectlanguage{english}%
\begin{lem}
The trapped set in phase space $\mathcal{K}$ defined in (\ref{eq:def_K_phase_space})
in contained in the following set:
\begin{equation}
\mathcal{K}\subset\left\{ \left(x,\xi\right),x\in I,\eta\notin\mathcal{B}\left(K\right)\mbox{ with }\left(x,\eta\right)=\Phi\left(x,\xi\right)\right\} \label{eq:inclusion_K}
\end{equation}

\end{lem}
\selectlanguage{french}%
}}}\end{center}\vspace{0.cm}
\selectlanguage{english}%
\begin{proof}
Let $\left(x,\xi\right)\in I\times\mathbb{R}$ which does not belong
to the set defined on the right hand side of (\ref{eq:inclusion_K}).
Then $\eta\in\mathcal{B}\left(K\right)$. Hence for every admissible
word $w\in\mathcal{W}$, we have that $\left|\phi_{w_{0,n}}\left(x\right)-\phi_{w_{0,n}}\left(\eta\right)\right|\leq C.\theta^{n}\underset{n\rightarrow+\infty}{\rightarrow}0$.
From the change of variable (\ref{eq:eta_xi-1}) and the expression
(\ref{eq:expression_F_eta}) with the new variables, this gives that$\left(x_{n},\xi_{n}\right):=F_{w_{0,n}}\left(x,\xi\right)$
satisfies 
\[
\left|\xi_{n}\right|=\frac{2}{\left|\phi_{w_{0,n}}\left(x\right)-\phi_{w_{0,n}}\left(\eta\right)\right|}\geq C'.\theta^{-n}\rightarrow+\infty
\]
hence $\left(x,\xi\right)\notin\mathcal{K}$. We deduce (\ref{eq:inclusion_K}).
\end{proof}
Finally, we show minimal captivity of the canonical map $F$. According
to (\ref{eq:hyp_minimal_captivity_epsilon}), we have to show that
there exists a neighborhood $B$ of $\mathcal{K}$ such that $\forall\left(x,\xi\right)\in B$,
$\sharp\left\{ F\left(x,\xi\right)\cap B\right\} \leq1$. This is
true if $B_{j}:=F_{j}^{-1}\left(B\right),j=1\ldots N$ are disjoint
sets. Using the coordinates $\left(x,\eta\right)$ which decouple
the map $F_{j}$, in (\ref{eq:expression_F_eta}), it is equivalent
to show that there exists a neighborhood $\mathcal{B}$ of $K$ in
$\overline{\mathbb{R}}$ such that $\mathcal{B}_{j}:=g_{j}^{-1}\left(\mathcal{B}\right)\subset\overline{\mathbb{R}},j=1,\ldots N$
are disjoint sets. For this we consider both cases:

\paragraph{Minimal captivity of the truncated Gauss map.}

For this map, let $\mathcal{B}:=]-\infty,-1[$. Then the sets $g_{j}^{-1}\left(]-\infty,-1[\right)=]-j-1,j[$,
with $j=1\ldots N$, are mutually disjoint. From the argument above
this implies that the truncated Gauss map is minimal captive, i.e.
Proposition \ref{prop:minimal_capt_for_model}. Notice that, from
(\ref{eq:eta_xi-1}), in variables $\left(x,\xi\right)\in T^{*}[0,1]$
we have
\[
B=\left\{ x\in\left[0,1\right],\eta\in]-\infty,-1[\right\} =\left\{ \left(x,\xi\right),x\in\left[0,1\right],\frac{-2}{x+1}<\xi<0\right\} 
\]
This set $B$ contains the trapped set $\mathcal{K}_{N}$ and is depicted
in figure (\ref{fig:trapped_set}).

\paragraph{Minimal captivity of the Bowen Series map.}

For this case, let $\mathcal{B}:=I=\bigcup_{j=1}^{2r}I_{j}$. Then
$\mathcal{B}_{j}=g_{j}^{-1}\left(\mathcal{B}\right)=g_{j+r}\left(I\right)\subset I_{j+r}$.
Since the sets $I_{j+r}$ are mutually disjoint, the sets $\mathcal{B}_{j}$
are also disjoint. From the argument above this implies that the Bowen
Series map on phase space is minimal captive, i.e. Proposition \ref{prop:minimal_capt_for_model-1}.

\begin{figure}
\begin{centering}
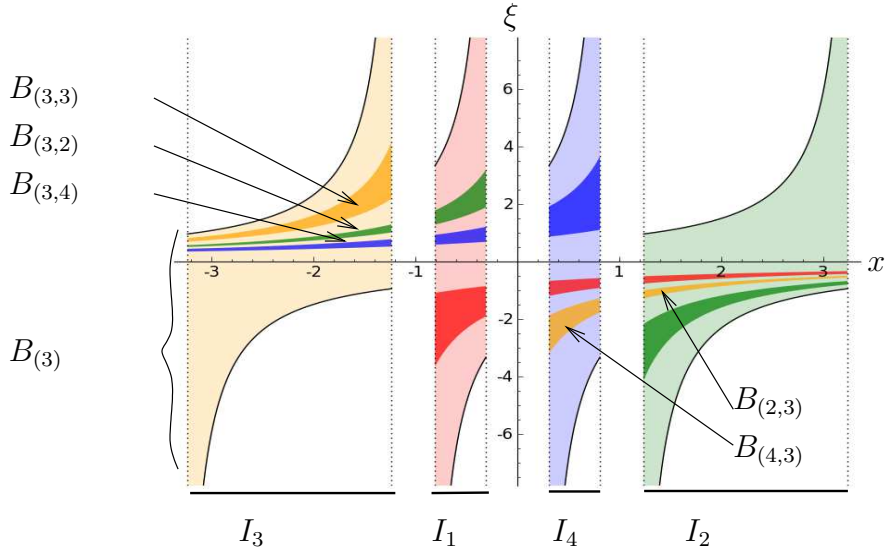
\par\end{centering}

\caption{\label{fig:MinCap-1}This figure illustrates the choice of the bounding
functions in the proof of the minimal captive property for the example
of a Schottky surface shown in figure \ref{fig:BowenSeries}. The
light shaded regions indicate the set $B_{\left(j\right)}:=B\cap\left(I_{j}\times\mathbb{R}\right)$
while the darker shaded regions indicate the different pre-images
$B_{\left(i,j\right)}:=F_{ij}^{-1}\left(B\cap\left(I_{j}\times\mathbb{R}\right)\right)\subset B_{\left(i\right)}$,
$i\neq j+2$ mod $4$. For example, that dark orange shaded regions
$B_{\left(3,3\right)},B_{\left(4,3\right)},B_{\left(2,3\right)}$
shows the three preimages of the light orange region $B_{\left(3\right)}$.
The trapped set $\mathcal{K}$ is contained in the union of these
$B_{\left(i,j\right)}$. }
\end{figure}

Figure (\ref{fig:MinCap-1}) shows the sets $B_{j}=F_{j}^{-1}\left(B\right)$
with $B:=\left\{ x\in I,\eta\in\mathcal{B}\right\} $and 
\[
B_{j}=\left\{ x\in I,\eta\in\mathcal{B}_{j}\right\} 
\]
that we have used in the proof of minimal captivity.

\section{\label{sec:Proof-of-Theorem_discrete_spectrum}Proof of Theorem \ref{thm:discrete-spectrum}
about the discrete spectrum}

For this proof we follow closely the proof%
\footnote{see also theorem 4 in \cite{fred-roy-sjostrand-07} although we are
dealing in this paper with expanding maps instead of hyperbolic maps
which simplifies the analysis, since we can work with ordinary Sobolev
spaces and not anisotropic Sobolev spaces.%
} of Theorem 2 in the paper \cite{fred_expanding_09} which uses semiclassical
analysis.

\subsection{Dynamics on the cotangent space $T^{*}I$}

In order to study the spectral properties of the transfer operator,
we have first to study the dynamics of the map $\phi:I\rightarrow I$
lifted on the cotangent space $T^{*}I$.

\begin{center}{\color{blue}\fbox{\color{black}\parbox{16cm}{
\begin{prop}
\label{prop:F_is_OIF}Considering $\hbar>0$ fixed, the transfer operator
$\hat{F}_{\chi}$ restricted to $C_{I}^{\infty}\left(\mathbb{R}\right)$
is a Fourier integral operator (FIO). Its \textbf{canonical transform}
is a multi-valued symplectic map $\mathfrak{F}:T^{*}I\rightarrow T^{*}I$
on the cotangent space $T^{*}I\equiv I\times\mathbb{R}$ given by:
\[
\mathfrak{F}:\begin{cases}
T^{*}I & \rightarrow T^{*}I\\
\left(x,\xi\right) & \rightarrow\left\{ \mathfrak{F}_{i,j}\left(x,\xi\right),\quad\mbox{with }i,j\mbox{ s.t. }x\in I_{i},\, i\rightsquigarrow j\right\} 
\end{cases}
\]
with
\begin{equation}
\mathfrak{F}_{i,j}:\begin{cases}
x' & =\phi_{i,j}\left(x\right)\\
\xi' & =\frac{1}{\phi_{i,j}'\left(x\right)}\xi
\end{cases}.\label{eq:def_symplectic_map_Fij}
\end{equation}

\end{prop}
}}}\end{center}

\paragraph{Remarks: }
\begin{itemize}
\item For short, we can write 
\begin{equation}
\mathfrak{F}:\begin{cases}
T^{*}I & \rightarrow T^{*}I\\
\left(x,\xi\right) & \rightarrow\left(\phi\left(x\right),\frac{1}{\phi'\left(x\right)}\xi\right)
\end{cases}.\label{eq:canonical_map_F_h1}
\end{equation}

\item Notice that the map $\mathfrak{F}$ differs from the canonical map
$F$ introduced in (\ref{eq:def_canonical_map_F}). The reason is
that this latter map is used in the asymptotic limit $\hbar\rightarrow0$,
whereas the study in this section is for fixed $\hbar$. \end{itemize}
\begin{proof}
of Proposition \ref{prop:F_is_OIF}. The operator $\varphi\rightarrow\varphi\circ\phi_{i,j}^{-1}$
is a pull back operator, one of the simplest example of Fourier integral
operator in the sense of semiclassical homogeneous theory \cite{taylor_tome2},
see also \cite{martinez-01} example 2 p.150. In that case the canonical
map $\mathfrak{F}_{i,j}$ is the map $\phi_{i,j}$ lifted on the cotangent
space $T^{*}I$ in the canonical way. In particular the action on
$\xi$ is linear. The term $e^{V\left(x\right)}e^{\frac{i}{\hbar}\tau\left(x\right)}\chi\left(\phi_{i,j}^{-1}\left(x\right)\right)$
in (\ref{e:def_F_op_ij}) does not contribute to the expression of
$\mathfrak{F}$ because it acts as a pseudodifferential operator,
equivalently as a FIO whose canonical map is the identity.
\end{proof}

\paragraph{Remarks}
\begin{itemize}
\item Observe that the dynamics of the map $\mathfrak{F}$ on $T^{*}I$
has a quite simple property: the zero section $\left\{ \left(x,\xi\right)\in I\times\mathbb{R},\,\xi=0\right\} $
is globally invariant and any other point $\left(x,\xi\right)$ with
$\xi\neq0$ escapes towards infinity $\left(\xi\rightarrow\pm\infty\right)$
in a controlled manner, because $\left|\phi_{i,j}'\left(x\right)\right|<\theta<1$,
with $\theta$ given in (\ref{eq:theta_contraction}), hence:
\begin{equation}
\left|\xi'\right|\geq\frac{1}{\theta}\left|\xi\right|\label{eq:expand}
\end{equation}

\item Due to hypothesis (\ref{eq:hyp_non_intersect}) the map $\phi_{i,j}^{-1}$
is uni-valued (when it is defined). Therefore the map $\mathfrak{F}^{-1}$
is also uni-valued and one has
\begin{equation}
\mathfrak{F}^{-1}\circ\mathfrak{F}=\mbox{Id}_{T^{*}I}\label{eq:Fm-F}
\end{equation}

\end{itemize}

\subsection{The escape function}

\begin{center}{\color{red}\fbox{\color{black}\parbox{16cm}{
\begin{defn}
\label{def:Class_S-m}\cite[p.2]{taylor_tome2} For $m\in\mathbb{R}$,
the \textbf{class of symbols} $S^{-m}\left(T^{*}\mathbb{R}\right)$,
with \textbf{order} $m$, is the set of functions on the cotangent
space $A\in C^{\infty}\left(T^{*}\mathbb{R}\right)$ such that for
any $\alpha,\beta\in\mathbb{N}$, there exists $C_{\alpha,\beta}>0$
such that 
\begin{equation}
\forall\left(x,\xi\right)\in T^{*}\mathbb{R},\quad\left|\partial_{x}^{\alpha}\partial_{\xi}^{\beta}A\left(x,\xi\right)\right|\leq C_{\alpha,\beta}\left\langle \xi\right\rangle ^{-m-\left|\beta\right|},\qquad\mbox{with }\left\langle \xi\right\rangle =\left(1+\xi^{2}\right)^{1/2}.\label{eq:def_symbol_class_S-m}
\end{equation}

\end{defn}
}}}\end{center}

\vspace{0.cm}\begin{center}{\color{blue}\fbox{\color{black}\parbox{16cm}{
\begin{lem}
Let $m>0$ and let
\[
A_{m}\left(x,\xi\right):=\left\langle \xi\right\rangle ^{-m}\quad\in S^{-m}\left(T^{*}\mathbb{R}\right).
\]
We have
\begin{equation}
\forall R>0,\forall\left|\xi\right|>R,\quad\forall i\rightsquigarrow j,\forall x\in I_{i},\qquad\frac{A_{m}\left(\mathfrak{F}_{i,j}\left(x,\xi\right)\right)}{A_{m}\left(x,\xi\right)}\leq C^{m}<1,\label{eq:A_decreases}
\end{equation}
with $C=\sqrt{\frac{R^{2}+1}{R^{2}/\theta^{2}+1}}<1$. We say that
$A_{m}$ is an \textbf{escape function}: (\ref{eq:A_decreases}) shows
that $A_{m}$ decreases strictly along the trajectories of $\mathfrak{F}$
outside the zero section.
\end{lem}
}}}\end{center}\vspace{0.cm}
\begin{proof}
From Eq. (\ref{eq:def_symplectic_map_Fij}) and (\ref{eq:expand})
we have
\[
\frac{A_{m}\left(\mathfrak{F}_{i,j}\left(x,\xi\right)\right)}{A_{m}\left(x,\xi\right)}=\frac{\left(1+\xi^{2}\right)^{m/2}}{\left(1+\left(\xi'\right)^{2}\right)^{m/2}}\leq\frac{\left(1+\xi^{2}\right)^{m/2}}{\left(1+\xi^{2}/\theta^{2}\right)^{m/2}}\leq\left(\frac{1+R^{2}}{1+R^{2}/\theta^{2}}\right)^{m/2}=C^{m}.
\]
The last inequality is because the function decreases with $\left|\xi\right|$.
\end{proof}
Using the standard quantization rule \cite[p.2]{taylor_tome2} the
symbol $A_{m}$ can be quantized into a pseudodifferential operator
$\hat{A}_{m}$ (PDO for short) which is self-adjoint and invertible
on $C_{0}^{\infty}\left(\mathbb{R}\right)$:
\begin{equation}
\left(\hat{A}_{m}\varphi\right)\left(x\right)=\frac{1}{2\pi}\int A_{m}\left(x,\xi\right)e^{i\left(x-y\right)\xi}\varphi\left(y\right)dyd\xi.\label{eq:quantiz_rule}
\end{equation}

Conversely $A_{m}$ is called the symbol of the PDO $\hat{A}_{m}$.
In our simple case, this is very explicit: in Fourier space, $\hat{A}_{m}$
is simply the multiplication by $\left\langle \xi\right\rangle ^{m}$.
Its inverse $\hat{A}_{m}^{-1}$ is the multiplication by $\left\langle \xi\right\rangle ^{-m}$.

\subsection{Use of the Egorov Theorem}

Let
\[
\hat{Q}_{m}:=\hat{A}_{m}\hat{F}_{\chi}\hat{A}_{m}^{-1}\quad:L^{2}\left(\mathbb{R}\right)\rightarrow L^{2}\left(\mathbb{R}\right),
\]
which is unitarily equivalent to $\hat{F}_{\chi}:H^{-m}\left(\mathbb{R}\right)\rightarrow H^{-m}\left(\mathbb{R}\right)$
(from the definition of $H^{-m}\left(\mathbb{R}\right)$, Eq.(\ref{eq:def_Hm})).
This is expressed by the following commutative diagram
\[
\begin{array}{ccc}
L^{2}\left(\mathbb{R}\right) & \overset{\hat{Q}_{m}}{\rightarrow} & L^{2}\left(\mathbb{R}\right)\\
\downarrow\hat{A}_{m}^{-1} &  & \downarrow\hat{A}_{m}^{-1}\\
H^{-m}\left(\mathbb{R}\right) & \overset{\hat{F}_{\chi}}{\rightarrow} & H^{-m}\left(\mathbb{R}\right)
\end{array}.
\]
We will therefore study the operator $\hat{Q}_{m}$ on $L^{2}\left(\mathbb{R}\right)$.
Notice that $\hat{Q}_{m}$ is defined a priori on a dense domain $C_{0}^{\infty}\left(\mathbb{R}\right)$.
Define
\begin{equation}
\hat{P}:=\hat{Q}_{m}^{*}\hat{Q}_{m}=\hat{A}_{m}^{-1}\left(\hat{F}_{\chi}^{*}\hat{A}_{m}^{2}\hat{F}_{\chi}\right)\hat{A}_{m}^{-1}=\hat{A}_{m}^{-1}\hat{B}\hat{A}_{m}^{-1}\label{eq:def_P}
\end{equation}
with 
\begin{equation}
\hat{B}:=\hat{F}_{\chi}^{*}\hat{A}_{m}^{2}\hat{F}_{\chi}=\hat{\chi}\hat{F}^{*}\hat{A}_{m}^{2}\hat{F}\hat{\chi}.\label{eq:def_B}
\end{equation}

Now, the crucial step in the proof is to use the Egorov Theorem.

\vspace{0.cm}\begin{center}{\color{blue}\fbox{\color{black}\parbox{16cm}{
\begin{lem}
\textbf{\label{lem:Egorov-theorem}(Egorov theorem}). $\hat{B}$ defined
in (\ref{eq:def_B}) is a pseudo-differential operator with symbol
in $S^{-2m}\left(T^{*}\mathbb{R}\right)$ given by:
\begin{equation}
B\left(x,\xi\right)=\left(\chi^{2}\left(x\right)\sum_{j\mbox{ s.t.}i\rightsquigarrow j}\left|\phi_{i,j}'\left(x\right)\right|e^{2\mathrm{Re}\left(V\left(\phi_{i,j}\left(x\right)\right)\right)}A_{m}^{2}\left(\mathfrak{F}_{i,j}\left(x,\xi\right)\right)\right)+R\label{eq:Egorov}
\end{equation}
where $R\in S^{-2m-1}\left(T^{*}\mathbb{R}\right)$ has a lower order,
$x\in I_{i}$, $\xi\in\mathbb{R}$.
\end{lem}
}}}\end{center}\vspace{0.cm}
\begin{proof}
$\hat{F}$ and $\hat{F}^{*}$ are Fourier integral operators (FIO)
whose canonical maps are respectively $\mathfrak{F}$ and $\mathfrak{F}^{-1}$.
The pseudodifferential operator (PDO) $\hat{A}_{m}$ can also be considered
as a FIO whose canonical map is the identity. By composition we deduce
that $\hat{B}=\hat{\chi}\hat{F}^{*}\hat{A}_{m}^{2}\hat{F}\hat{\chi}$
is a FIO whose canonical map is the identity since $\mathfrak{F}^{-1}\circ\mathfrak{F}=I$
from (\ref{eq:Fm-F}). Therefore $\hat{B}$ is a PDO. Using (\ref{e:def_F_op_ij}),
(\ref{eq:express_dual_1}) we obtain that the principal symbol of
$\hat{B}$ is the first term of (\ref{eq:Egorov}).
\end{proof}
Remark: contrary to (\ref{eq:def_B}), $\hat{F}\hat{A}_{m}\hat{F}^{*}$
is not a PDO, but a FIO whose canonical map $\mathfrak{F}\circ\mathfrak{F}^{-1}$
is multivalued.

Now by\textbf{ theorem of composition of PDO} \cite[p.11]{taylor_tome2},
Eq.(\ref{eq:def_P}) and Eq.(\ref{eq:Egorov}) imply that $\hat{P}$
is a PDO with symbol in $S^{0}\left(\mathbb{R}\right)$ and for $x\in I_{i},\xi\in\mathbb{R}$
the principal symbol is given by 
\begin{equation}
P\left(x,\xi\right)=\frac{B\left(x,\xi\right)}{A_{m}^{2}\left(x,\xi\right)}=\left(\chi^{2}\left(x\right)\sum_{j\mbox{ s.t. }i\rightsquigarrow j}\left|\phi_{i,j}'\left(x\right)\right|e^{2\mathrm{Re}\left(V\left(\phi_{i,j}\left(x\right)\right)\right)}\frac{A_{m}^{2}\left(\mathfrak{F}_{i,j}\left(x,\xi\right)\right)}{A_{m}^{2}\left(x,\xi\right)}\right).\label{eq:symb_P}
\end{equation}
The estimate (\ref{eq:A_decreases}) gives the following upper bound
for any $R>0$, $x\in I$ and $\left|\xi\right|>R$:
\[
\left|P\left(x,\xi\right)\right|\leq\chi^{2}\left(x\right)C^{2m}\sum_{j,i\rightsquigarrow j}\left|\phi_{i,j}'\left(x\right)\right|e^{2\mathrm{Re}\left(V\left(\phi_{i,j}\left(x\right)\right)\right)}\leq C^{2m}N\theta e^{2V_{\mathrm{max}}}
\]
with $V_{\mathrm{max}}=\mbox{max}_{x\in I}\mathrm{Re}\left(V\left(x\right)\right)$.

We apply%
\footnote{\label{fn:Properly_supported}Actually, we can not apply directly
the $L^{2}$-continuity theorem for PDO to $\hat{P}$ because $\hat{P}$
doesn't have a compactly supported Schwartz kernel. However $\hat{B}$
obviously has a compactly supported Schwartz kernel due to the presence
of $\hat{\chi}$ in Eq.(\ref{eq:def_B}). The trick is to approximate
$\hat{A}_{m}^{-1}$ by a properly supported operator $\Lambda_{m}$
as it is done in \cite[p.45]{grigis_sjostrand} and then apply the
$L^{2}$-continuity theorem to $\hat{\Lambda}_{m}\hat{B}\hat{\Lambda}_{m}$.%
} the\textbf{ $L^{2}$-continuity theorem for PDO} to $\hat{P}$ as
given in \cite[th 4.5 p.42]{grigis_sjostrand}. The result is that
for any $\varepsilon>0$,
\[
\hat{P}=\hat{k}_{\varepsilon}+\hat{p}_{\varepsilon}
\]
with $\hat{k}_{\varepsilon}$ a smoothing operator (hence compact)
and $\left\Vert \hat{p}_{\varepsilon}\right\Vert \leq C^{2m}N\theta e^{2V_{\mathrm{max}}}+\varepsilon$.

If $\hat{Q}_{m}=\hat{U}\left|\hat{Q}_{m}\right|$ is the polar decomposition
of $\hat{Q}_{m}$, with $\hat{U}$ unitary, then from (\ref{eq:def_P}),
$\hat{P}=\left|\hat{Q}_{m}\right|^{2}$, hence $\left|\hat{Q}_{m}\right|=\sqrt{\hat{P}}$
and the spectral theorem \cite[p.75]{taylor_tome2} gives that $\left|\hat{Q}_{m}\right|$
has a similar decomposition
\[
\left|\hat{Q}_{m}\right|=\hat{k}'_{\varepsilon}+\hat{q}_{\varepsilon}
\]
with $\hat{k}'_{\varepsilon}$ compact and $\left\Vert \hat{q}_{\varepsilon}\right\Vert \leq\sqrt{C^{2m}N\theta e^{2V_{\mathrm{max}}}}+\varepsilon$,
with any $\varepsilon>0$. Since $\left\Vert \hat{U}\right\Vert =1$
we deduce a similar decomposition for $\hat{Q}_{m}=\hat{U}\left|\hat{Q}_{m}\right|:L^{2}\left(I\right)\rightarrow L^{2}\left(I\right)$.
We also use the fact that $C\rightarrow\theta$ as $R\rightarrow\infty$
in (\ref{eq:A_decreases}) and we deduce (\ref{eq:F=00003DK+R}) and
(\ref{eq:bound_norm_R_def_rm}) for $\hat{F}_{\chi}:H^{-m}\left(\mathbb{R}\right)\rightarrow H^{-m}\left(\mathbb{R}\right)$.
Precisely we obtain that $r_{m}=\theta^{m}\sqrt{N\theta e^{2V_{\mathrm{max}}}}=C\theta^{m}$
with $C$ independent on $m$.

The fact that the eigenvalues $\lambda_{i}$ and their generalized
eigenspaces do not depend on the choice of space $H^{-m}\left(\mathbb{R}\right)$
is due to density of $C_{0}^{\infty}\left(\mathbb{R}\right)$ in Sobolev
spaces. We refer to the argument given in the proof of corollary 1
in \cite{fred-roy-sjostrand-07}.

Finally, if $\varphi$ is an eigendistribution of $\hat{F}_{\chi}$,
i.e. $\hat{F}_{\chi}\varphi=\lambda\varphi$ with $\lambda\neq0$,
we deduce that $\varphi=\frac{1}{\lambda^{n}}\hat{F}_{\chi}^{n}\varphi$
for any $n\geq1$, and (\ref{eq:supp_F_phi-1}) implies that $\mbox{supp}\left(\varphi\right)\subset K=\bigcap_{n\in\mathbb{N}}K_{n}$.
On the trapped set we have $\chi=1$ hence the eigendistribution and
eigenvalues of $\hat{F}_{\chi}$ do not depend on $\chi$. This finishes
the proof of Theorem \ref{thm:discrete-spectrum}.

\section{\label{sec:6}Dynamics of the canonical map $F:T^{*}I\rightarrow T^{*}I$}

In this Section we study the trapped set $\mathcal{K}$ and the symbolic
dynamics of the canonical map $F$ defined in Proposition \ref{prop:Fh_is_a_FIO}
as associated to the family transfer operator $\hat{F}_{\hbar}$.
Before we give:
\begin{proof}
of Proposition \ref{prop:Fh_is_a_FIO}. This is the same argument
as in the proof of Proposition \ref{prop:F_is_OIF} except that now
the family of operators $\left(e^{\frac{i}{\hbar}\tau\left(x\right)}\right)_{\hbar>0}$
which appears in (\ref{e:def_F_op_ij}) is a FIO. As explained in
\cite[ex.1,p.150]{martinez-01}, its canonical map is $\left(x,\xi\right)\rightarrow\left(x,\xi+\frac{d\tau}{dx}\right)$.
We compose this map with the previous one (\ref{eq:def_symplectic_map_Fij})
to get (\ref{eq:def_symplectic_map_Fij-1}).
\end{proof}

\subsection{The trapped set $\mathcal{K}$ in phase space}

We have provided a definition of the trapped set $\mathcal{K}$ in
(\ref{eq:def_K_phase_space}). We will give now a more precise description
of it. Recall that the inverse maps $\phi^{-1}$ and $F^{-1}$ are
uni-valued. For any integer $m\geq0$, let
\[
\tilde{K}_{m}:=F^{-m}\left(K_{m}\times\left[-R,R\right]\right)
\]
where $K_{m}=\phi^{m}\left(I\right)$ has been defined in (\ref{eq:def_Kn})
and $R$ is given by Lemma \ref{lem:escape_F}. In particular $\tilde{K}_{0}=I\times\left[-R,R\right]$.
Let $\pi:\left(x,\xi\right)\in T^{*}I\rightarrow x\in I$ be the projection
map. These sets have the following properties:
\[
\pi\left(\tilde{K}_{m}\right)=I,
\]
\begin{equation}
\tilde{K}_{m+1}\subset\tilde{K}_{m}\label{eq:Ktile_m_inclusion}
\end{equation}

\begin{proof}
of (\ref{eq:Ktile_m_inclusion}). From Lemma \ref{lem:escape_F} we
have
\[
\left(K_{m+1}\times\left[-R,R\right]\right)\subset F\left(K_{m}\times\left[-R,R\right]\right)
\]
hence
\[
\tilde{K}_{m+1}=F^{-m}\left(F^{-1}\left(K_{m+1}\times\left[-R,R\right]\right)\right)\subset F^{-m}\left(K_{m}\times\left[-R,R\right]\right)=\tilde{K}_{m}
\]

\end{proof}
Let us define
\begin{equation}
\tilde{K}:=\bigcap_{m}\tilde{K}_{m}\label{eq:def_K_tilde-2}
\end{equation}

Now we combine the sets $K_{n}$ defined in (\ref{eq:def_Kn}) with
the sets $\tilde{K}_{m}$ and define for any integers $a,b\geq0$
\begin{equation}
\mathcal{K}_{a,b}:=\pi^{-1}\left(K_{a}\right)\bigcap\tilde{K}_{b}\label{eq:def_K_a_b}
\end{equation}
We have
\begin{equation}
\mathcal{K}_{a+1,b}\subset\mathcal{K}_{a,b},\quad\mathcal{K}_{a,b+1}\subset\mathcal{K}_{a,b}\label{eq:inclusions_Kab}
\end{equation}
and
\begin{equation}
F^{-1}\left(\mathcal{K}_{a,b}\right)=\mathcal{K}_{a-1,b+1}.\label{eq:inclusions_of_K_ab}
\end{equation}

\begin{rem}
We can interpret the trapped set $K\subset I$ with respect to the
lifted map $F:T^{*}I\rightarrow T^{*}I$, as follows. The trapped
set $\pi^{-1}\left(K\right)\subset T^{*}I$ is characterized by
\[
\pi^{-1}\left(K\right)=\left\{ \left(x,\xi\right)\in T^{*}I,\quad\exists\mbox{compact }C\Subset T^{*}I,\forall n\geq0,F^{-n}\left(x,\xi\right)\in C\right\} 
\]
i.e. $\pi^{-1}\left(K\right)$ can be considered as the ``trapped
set of the map $F$ in the past''. Similarly $\tilde{K}\subset T^{*}I$
can be interpreted as the ``trapped set of the map $F$ in the future''
and $\mathcal{K}\subset T^{*}I$ as the full trapped set (past and
future) since they are characterized by
\[
\tilde{K}=\left\{ \left(x,\xi\right)\in T^{*}I,\quad\exists\mbox{compact }C\Subset T^{*}I,\forall n\geq0,F^{n}\left(x,\xi\right)\cap C\neq\emptyset\right\} 
\]
\begin{eqnarray}
\mathcal{K} & = & \left\{ \left(x,\xi\right)\in T^{*}I,\quad\exists\mbox{compact }C\Subset T^{*}I,\forall n\in\mathbb{Z},F^{n}\left(x,\xi\right)\cap C\neq\emptyset\right\} \label{eq:def2_trapped_K}\\
 & = & \pi^{-1}\left(K\right)\cap\tilde{K}\nonumber 
\end{eqnarray}

\end{rem}
From this previous remark, the next definition is equivalent to (\ref{eq:def_K_phase_space}).

\vspace{0.cm}\begin{center}{\color{red}\fbox{\color{black}\parbox{16cm}{
\begin{defn}
The \textbf{trapped set }$\mathcal{K}\subset T^{*}I$ of the map $F$
is 
\begin{equation}
\mathcal{K}:=\bigcap_{a=0}^{\infty}\mathcal{K}_{a,a}\label{eq:def_K_tilde}
\end{equation}

\end{defn}
}}}\end{center}\vspace{0.cm}

The hypothesis of minimal captivity has been defined in (\ref{hyp:minimal_capt}).
The following proposition gives equivalent, stronger and weaker definition
of minimal captivity. They  are convenient for practical purposes.

\begin{center}{\color{blue}\fbox{\color{black}\parbox{16cm}{
\begin{prop}
\label{prop:minimal_captive}~
\begin{enumerate}
\item The map $F$ is minimally captive (i.e. Eq.(\ref{eq:hyp_minimal_captivity_epsilon})
holds true) if and only if the map $F$ satisfies
\begin{equation}
\exists a,\quad\forall\left(x,\xi\right)\in\mathcal{K}_{a,a},\quad\sharp\left\{ F\left(x,\xi\right)\bigcap\mathcal{K}_{a,a}\right\} \leq1.\label{eq:hyp_minimal_captivity_bis}
\end{equation}

\item If map $F$ is minimally captive then 
\begin{equation}
\exists a,\exists C,\forall n\mbox{ s.t. }\forall\left(x,\xi\right)\in\mathcal{K}_{a,0},\quad\sharp\left\{ F^{n}\left(x,\xi\right)\bigcap\mathcal{K}_{a,0}\right\} \leq C.\label{eq:Hyp_2}
\end{equation}
where $\mathcal{K}_{a,0}:=\left(\pi^{-1}\left(K_{a}\right)\bigcap\left[-R,R\right]\right)$
has been defined in (\ref{hyp:minimal_capt}).
\end{enumerate}
\end{prop}
}}}\end{center}
\begin{proof}
The fact that (\ref{eq:hyp_minimal_captivity_bis}) is equivalent
to (\ref{eq:hyp_minimal_captivity_epsilon}) is because
\[
\forall\varepsilon>0,\exists a\quad\mbox{s.t. }\mathcal{K}_{a,a}\subset\mathcal{K}_{\varepsilon}
\]
\[
\forall a,\exists\varepsilon>0\mbox{ s.t. }\mathcal{K}_{\varepsilon}\subset\mathcal{K}_{a,a}
\]

\end{proof}

\subsection{Symbolic dynamics}

The purpose of this section is to describe precisely the dynamics
of $\phi$ and $F$ using ``symbolic dynamics''. This is very standard
for expanding maps \cite{brin-02}. This somehow refines the structure
of the sets $\mathcal{K}_{a,b}$ introduced before.

\subsubsection{Symbolic dynamics on the trapped set $K\subset I$}

Let

\begin{equation}
\mathcal{W}_{-}:=\left\{ \left(\ldots,w_{-2},w_{-1},w_{0}\right)\in\left\{ 1,\ldots,N\right\} ^{-\mathbb{N}},w_{l-1}\rightsquigarrow w_{l},\forall l\leq0\right\} \label{eq:def_W_-}
\end{equation}
be the set of \textbf{admissible left semi-infinite sequences}. For
$w\in\mathcal{W}_{-}$ and $i<j$ we write $w_{i,j}:=\left(w_{i},w_{i+1},\ldots w_{j}\right)$
for an extracted sequence. For simplicity we will use the notation
\begin{equation}
\phi_{w_{i,j}}:=\phi_{w_{j-1},w_{j}}\circ\ldots\circ\phi_{w_{i},w_{i+1}}:I_{w_{i}}\rightarrow I_{w_{j}}\label{eq:def_phi_w_ij}
\end{equation}
 for the composition of maps. For $n\geq0$, let
\begin{equation}
I_{w_{-n,0}}:=\phi_{w_{-n,0}}\left(I_{w_{-n}}\right)\subset I_{w_{0}}.\label{eq:def_Iwn}
\end{equation}
For any $0<m<n$ we have the strict inclusions 
\[
I_{w_{-n,0}}\subset I_{w_{-m,0}}\subset I_{w_{0}}.
\]
From (\ref{eq:theta_contraction}), the size of $I_{w_{-n,0}}$ is
bounded by
\[
\left|I_{w_{-n,0}}\right|\leq\theta^{n}\left|I_{w_{0}}\right|,
\]
hence the sequence of sets $\left(I_{w_{-n,0}}\right)_{n\geq1}$ is
a sequence of non empty and decreasing closed intervals and $\bigcap_{n=1}^{\infty}I_{w_{-n,0}}$
is a point in $K$. We define

\begin{center}{\color{red}\fbox{\color{black}\parbox{16cm}{
\begin{defn}
The ``\textbf{symbolic coding map}'' is 
\begin{equation}
S:\begin{cases}
\mathcal{W}_{-} & \rightarrow K\\
w & \rightarrow S\left(w\right):=\bigcap_{n=1}^{\infty}I_{w_{-n,0}}
\end{cases}\label{eq:def_coding_S}
\end{equation}

\end{defn}
}}}\end{center}

In some sense we have decomposed the sets $K_{n}$, Eq.(\ref{eq:def_Kn}),
into individual components:

\begin{equation}
K_{n}=\bigcup_{w_{-n,0}\in\mathcal{W}_{-}}I_{w_{-n,0}}\label{eq:def_K_n}
\end{equation}
\[
K=\bigcup_{w\in\mathcal{W}_{-}}S\left(w\right)
\]

Let us introduce the \textbf{left shift,} a multivalued map, defined
by
\[
L:\begin{cases}
\mathcal{W}_{-} & \rightarrow\mathcal{W}_{-}\\
\left(\ldots,w_{-2},w_{-1},w_{0}\right) & \rightarrow\left(\ldots,w_{-2},w_{-1},w_{0},w_{1}\right)
\end{cases}
\]
with $w_{1}\in\left\{ 1,\ldots,N\right\} $ such that $w_{0}\rightsquigarrow w_{1}$.
Let the \textbf{right shift} be the univalued map defined by
\[
R:\begin{cases}
\mathcal{W}_{-} & \rightarrow\mathcal{W}_{-}\\
\left(\ldots,w_{-2},w_{-1},w_{0}\right) & \rightarrow\left(\ldots,w_{-2},w_{-1}\right)
\end{cases}.
\]
\begin{framed}%
\begin{prop}
\label{prop:symbolic dynamics}The following diagram is commutative
\begin{eqnarray}
\mathcal{W}_{-} & \overset{S}{\longrightarrow} & \qquad K\label{eq:diagram_S}\\
R\uparrow\downarrow L &  & \phi^{-1}\uparrow\downarrow\phi\nonumber \\
\mathcal{W}_{-} & \overset{S}{\longrightarrow} & \qquad K\nonumber 
\end{eqnarray}
and the map $S:\mathcal{W}_{-}\rightarrow K$ is one to one. This
means that the dynamics of points on the trapped set $K$ under the
maps $\phi^{-1},\phi$ is equivalent to the symbolic dynamics of the
shift maps $R,L$ on the set of admissible words $\mathcal{W}_{-}$.
Notice that the maps $R$ and $\phi^{-1}$ are univalued, whereas
the maps $L$ and $\phi$ are (in general) multivalued. \end{prop}
\end{framed}
\begin{proof}
From the definition of $S$ we have 
\begin{equation}
\phi_{w_{0}w_{1}}\left(S\left(\ldots,w_{-2},w_{-1},w_{0}\right)\right)=S\left(\ldots,w_{-2},w_{-1},w_{0},w_{1}\right)\label{eq:e1}
\end{equation}
 and 
\begin{equation}
\phi_{w_{-1}w_{0}}^{-1}\left(S\left(\ldots,w_{-2},w_{-1},w_{0}\right)\right)=S\left(\ldots,w_{-2},w_{-1}\right)\label{eq:e2}
\end{equation}
which gives the diagram (\ref{eq:diagram_S}). The map $S:\mathcal{W}_{-}\rightarrow K$
is surjective by construction. Let us show that the hypothesis (\ref{eq:hyp_non_intersect})
implies that it is also injective. Let $w,w'\in\mathcal{W}_{-}$ and
suppose that $w\neq w'$, i.e. there exists $k\geq0$ such that $w_{-k}\neq w'_{-k}$.
From (\ref{eq:hyp_non_intersect}) we have $\phi_{w_{-k},w_{-k+1},}\left(I_{w_{-k}}\right)\cap\phi_{w'_{-k},w'_{-k+1}}\left(I_{w'_{-k}}\right)=\emptyset$.
We deduce recursively that $\phi_{w_{-k,0}}\left(I_{w_{-k}}\right)\cap\phi_{w'_{-k,0}}\left(I_{w'_{-k}}\right)=\emptyset$.
Since $S\left(w\right)\in\phi_{w_{-k,0}}\left(I_{w_{-k}}\right)$
and $S\left(w'\right)\in\phi_{w'_{-k,0}}\left(I_{w'_{-k}}\right)$
we deduce that $S\left(w\right)\neq S\left(w'\right)$. Hence $S$
is one to one.
\end{proof}

\subsubsection{The ``future trapped set'' $\tilde{K}$ in phase space $T^{*}I$}

Let

\[
\mathcal{W}_{+}:=\left\{ \left(w_{0},w_{1},w_{2}\ldots\right)\in\left\{ 1,\ldots,N\right\} ^{\mathbb{N}},\quad w_{l}\rightsquigarrow w_{l+1},\forall l\geq0\right\} 
\]
be the set of admissible right semi-infinite sequences. We still use
the notation $w_{i,j}:=\left(w_{i},w_{i+1},\ldots w_{j}\right)$ for
an extracted sequence. For any $n\geq0$ let
\begin{equation}
\tilde{I}_{w_{0,n}}:=F^{-n}\left(I_{w_{0,n}}\times\left[-R,R\right]\right)\label{eq:def_set_I_tilde}
\end{equation}
be the image of the rectangle under the univalued map $F^{-n}$. Notice
that $\pi\left(\tilde{I}_{w_{0,n}}\right)=I_{w_{0}}$ where $\pi\left(x,\xi\right)=x$
is the canonical projection map. Since the map $F^{-1}$ contracts
strictly in variable $\xi$ by the factor $\theta<1$ then $\left(\tilde{I}_{w_{0,n}}\right)_{n\in\mathbb{N}}$
is a sequence of decreasing sets: $\tilde{I}_{w_{0,n+1}}\subset\tilde{I}_{w_{0,n}}$
and we can define the limit

\begin{equation}
\tilde{S}:w\in\mathcal{W}_{+}\rightarrow\tilde{S}\left(w\right):=\bigcap_{n\geq0}\tilde{I}_{w_{0,n}}\subset\tilde{K}.\label{eq:def_S_tilde}
\end{equation}

\begin{center}{\color{blue}\fbox{\color{black}\parbox{16cm}{
\begin{prop}
For every $w\in\mathcal{W}_{+}$, the set $\tilde{S}\left(w\right)$
is a smooth curve given by
\[
\tilde{S}\left(w\right)=\left\{ \left(x,\zeta_{w}\left(x\right)\right),\quad x\in I_{w_{0}},w\in\mathcal{W}_{+}\right\} 
\]
with
\begin{equation}
\zeta_{w}\left(x\right)=-\sum_{k\geq1}\phi'_{w{}_{0,k}}(x)\cdot\tau'\left(\phi_{w_{0,k}}(x)\right),.\label{eq:expression_zeta_w}
\end{equation}

We have an estimate of regularity, uniform in $w$: $\forall\alpha\in\mathbb{N}$,
$\exists C_{\alpha}>0$, $\forall w\in\mathcal{W}_{+}$, $\forall x\in I_{w_{0}}$,
\begin{equation}
\left|\left(\partial_{x}^{\alpha}\zeta_{w}\right)\left(x\right)\right|\leq C_{\alpha}.\label{eq:estimate}
\end{equation}

Moreover, with the hypothesis \ref{hyp:minimal_capt} of minimal captivity
there exists $a\geq1$ such that these branches do not intersect on
$\pi^{-1}\left(K_{a}\right)$, 
\begin{equation}
\forall w,w'\in\mathcal{W}_{+},\quad w\neq w'\Rightarrow\pi^{-1}\left(K_{a}\right)\cap\tilde{S}\left(w\right)\cap\tilde{S}\left(w'\right)=\emptyset\label{eq:branchs_disjoints}
\end{equation}
The set (\ref{eq:def_K_tilde-2}) can be expressed as
\[
\tilde{K}=\bigcup_{w\in\mathcal{W}_{+}}\tilde{S}\left(w\right).
\]

\end{prop}
}}}\end{center}
\begin{proof}
From (\ref{eq:def_symplectic_map_Fij-1}) we get 
\begin{equation}
F^{-1}\left(\phi_{i,j}\left(x\right),\xi\right)=\left(x,\phi'_{i,j}\left(x\right)\left(\xi-\tau'\left(\phi_{i,j}\left(x\right)\right)\right)\right).\label{eq:F_inverse}
\end{equation}
Iterating this equation we get, that
\[
\zeta_{w,n}\left(x\right):=-\sum_{k=1}^{n}\phi'_{w{}_{0,k}}\left(x\right)\cdot\tau'\left(\phi_{w_{0,k}}\left(x\right)\right)=F^{-n}\left(\phi_{w_{0,n}}\left(x\right),0\right),
\]
thus $\zeta_{w,n}(x)\in\tilde{S}\left(w\right)$ for all $n\in\mathbb{N}$
and we get (\ref{eq:expression_zeta_w}). In order to prove (\ref{eq:estimate})
we can check, that the series of $\zeta_{w,n}(x)$ and $\partial_{x}^{\alpha}\zeta_{w,n}(x)$
converge with uniform bounds in $w$ which follows after some calculations
from (\ref{eq:theta_contraction}) and the fact that $\phi'_{w_{0,k}}(x)\leq\theta^{k}$
independent of $w$.
\end{proof}

\subsubsection{Symbolic dynamics on the trapped set $\mathcal{K}$ in phase space
$T^{*}I$}

Recall from (\ref{eq:def2_trapped_K}) that $\mathcal{K}=\pi^{-1}\left(K\right)\cap\tilde{K}$.
Let

\[
\mathcal{W}:=\left\{ \left(\ldots w_{-2},w_{-1},w_{0},w_{1},\ldots\right)\in\left\{ 1,\ldots,N\right\} ^{\mathbb{Z}},\quad w_{l}\rightsquigarrow w_{l+1},\forall l\in\mathbb{Z}\right\} 
\]
be the set of bi-infinite admissible sequences. For a given $w\in\mathcal{W}$
and$a,b\in\mathbb{N}$, let
\[
\mathcal{I}_{w_{-a,0},w_{0,b}}:=\left(\pi^{-1}\left(I_{w_{-a,0}}\right)\cap\tilde{I}_{w_{0,b}}\right)\quad\subset\mathcal{K}_{a,b}
\]
where $\mathcal{K}_{a,b}$ has been defined in (\ref{eq:def_K_a_b}).

\begin{center}{\color{red}\fbox{\color{black}\parbox{16cm}{
\begin{defn}
The \textbf{symbolic coding map} is
\begin{equation}
\mathcal{S}:\begin{cases}
\mathcal{W} & \rightarrow\mathcal{K}\\
w & \rightarrow\mathcal{S}\left(w\right):=\bigcap_{n=1}^{\infty}\mathcal{I}_{w_{-n,0},w_{0,n}}=\left(\pi^{-1}\left(S\left(w_{-}\right)\right)\cap\tilde{S}\left(w_{+}\right)\right)
\end{cases}\label{eq:def_S_trapped}
\end{equation}
with $w_{-}=\left(\ldots w_{-1},w_{0}\right)\in\mathcal{W}_{-}$,
$w_{+}=\left(w_{0},w_{1},\ldots\right)\in\mathcal{W}_{+}$. 
\end{defn}
}}}\end{center}

More precisely we can express the point $\mathcal{S}\left(w\right)\in\mathcal{K}$
as
\begin{equation}
\mathcal{S}\left(w\right)=\left(x_{w_{-}},\xi_{w}\right),\quad x_{w_{-}}=S\left(w_{-}\right),\quad\xi_{w}=\zeta_{w_{+}}\left(S\left(w_{-}\right)\right),\label{eq:point_in_trapped_K}
\end{equation}
with $\zeta_{w_{+}}$ given in (\ref{eq:expression_zeta_w}). We also
have 
\[
\mathcal{K}_{a,b}=\bigcup_{w\in\mathcal{W}}\mathcal{I}_{w_{-a,0},w_{0,b}}.
\]

\begin{center}{\color{blue}\fbox{\color{black}\parbox{16cm}{
\begin{prop}
\label{prop:S_tilde_bijective}The following diagram is commutative
\begin{eqnarray}
\mathcal{W} & \overset{\mathcal{S}}{\longrightarrow} & \qquad\mathcal{K}\label{eq:diagram_S-1}\\
R\uparrow\downarrow L &  & F^{-1}\uparrow\downarrow F\nonumber \\
\mathcal{W} & \overset{\mathcal{S}}{\longrightarrow} & \qquad\mathcal{K}.\nonumber 
\end{eqnarray}
If assumption \ref{hyp:minimal_capt} of minimal captivity holds true
then the map $\mathcal{S}:\mathcal{W}\rightarrow\mathcal{K}$ is one
to one. This means that the univalued dynamics of points on the trapped
set $\mathcal{K}$ under the maps $F^{-1},F$ is equivalent to the
symbolic dynamics of the full shift maps $R,L$ on the set of words
$\mathcal{W}$.
\end{prop}
}}}\end{center}
\begin{proof}
Commutativity of the diagram comes from the construction of $\mathcal{S}$.
Also $\mathcal{S}$ is surjective. Let us show that $\mathcal{S}$
is injective. Let $w,w'\in\mathcal{W}$, with $w\neq w'$. There exists
$n\geq0$ such that $\left(L^{n}\left(w\right)\right)_{-}\neq\left(L^{n}\left(w'\right)\right)_{-}$.
So $S\left(\left(L^{n}\left(w\right)\right)_{-}\right)\neq S\left(\left(L^{n}\left(w'\right)\right)_{-}\right)$
because $S:\mathcal{W}_{-}\rightarrow K$ is one to one from Lemma
\ref{prop:symbolic dynamics}. Hence $\mathcal{S}\left(L^{n}\left(w\right)\right)\neq\mathcal{S}\left(L^{n}\left(w'\right)\right)$
and $F^{n}\left(\mathcal{S}\left(w\right)\right)\neq F^{n}\left(\mathcal{S}\left(w'\right)\right)$
from commutativity of the diagram. We apply $F^{-n}$ and deduce that
$\mathcal{S}\left(w\right)\neq\mathcal{S}\left(w'\right)$ because
$F^{-1}$ and $F^{-n}$ are injective on $\mathcal{K}$ from Assumption
\ref{hyp:minimal_capt}.
\end{proof}

\subsection{Dimension of the trapped set $\mathcal{K}$}

\subsubsection{Proof of Theorem \ref{thm:dim_of_K_tilde}}

For $w=\left(w_{k}\right)_{k\in\mathbb{Z}}\in\mathcal{W}$, we note
$w_{-}=\left(\ldots,w_{-2},w_{-1},w_{0}\right)\in\mathcal{W}_{-}$
and $w_{+}=\left(w_{0},w_{1},\ldots\right)\in\mathcal{W}_{+}$. Let
\[
\mathrm{Inv}\left(w_{+}\right):=\left(\ldots w_{2},w_{1},w_{0}\right)
\]
be the reversed word. Since the adjency matrix $A$ is supposed to
be symmetric we have that $\mathrm{Inv}\left(w_{+}\right)\in\mathcal{W}_{-}$.
Then, let us consider the following one to one map

\[
D:\begin{cases}
\mathcal{W} & \rightarrow\left(\mathcal{W}_{-}\times\mathcal{W}_{-}\right)_{l}\\
w & \rightarrow\left(w_{-},\mathrm{Inv}\left(w_{+}\right)\right)
\end{cases}
\]
where 
\begin{equation}
\left(\mathcal{W}_{-}\times\mathcal{W}_{-}\right)_{l}:=\left\{ \left(w,w'\right)\in\mathcal{W}_{-}\times\mathcal{W}_{-},\quad w_{0}=w'_{0}\right\} \label{eq:def_W_W_l}
\end{equation}
is a subset of $\mathcal{W}_{-}\times\mathcal{W}_{-}$. The index
$l$ stands for ``linked''. Let
\[
\Phi:=\left(S\otimes S\right)\circ D\circ\mathcal{S}^{-1}\quad:\mathcal{K}\rightarrow K\times K
\]
where $\mathcal{S}:\mathcal{W}\rightarrow\mathcal{K}$ has been defined
in (\ref{eq:def_S_trapped}) and is shown in Proposition \ref{prop:S_tilde_bijective}
to be one to one under assumption \ref{hyp:minimal_capt}. The map
$S:\mathcal{W}_{+}\rightarrow K$ has been defined in (\ref{eq:def_coding_S})
and is also one to one. Consider 
\begin{equation}
\left(K\times K\right)_{l}:=\left(S\otimes S\right)\left(\left(\mathcal{W}_{-}\times\mathcal{W}_{-}\right)_{l}\right)\quad\subset K\times K\label{eq:def_K_K_l}
\end{equation}
the image of (\ref{eq:def_W_W_l}) under the map $S\otimes S$. From
the previous remarks, the map $\Phi:\mathcal{K}\rightarrow\left(K\times K\right)_{l}$
is one to one.

\begin{center}{\color{blue}\fbox{\color{black}\parbox{16cm}{
\begin{lem}
\label{lem:25}The map $\Phi:\mathcal{K}\rightarrow\left(K\times K\right)_{l}$
is bi-Lipschitz.
\end{lem}
}}}\end{center}

As a consequence of this Lemma, since the Hausdorff and Minkowski
dimension is invariant under bi-Lipschitz maps \cite[p.24]{Falconer_97},
we deduce that
\begin{equation}
\dim_{M}\left(\mathcal{K}\right)=\dim_{M}\left(K\times K\right)_{l}\label{eq:dimdim}
\end{equation}

Before proving Lemma \ref{lem:25}, let us show how to deduce Theorem
\ref{thm:dim_of_K_tilde} from it. Let us temporally write $K_{i}:=K\cap I_{i}$.
From (\ref{eq:def_K_K_l}) we have that
\[
\left(K\times K\right)_{l}=\bigcup_{i}K_{i}\times K_{i}
\]
hence 
\begin{equation}
\mbox{dim}_{M}\left(K\times K\right)_{l}=\sup_{i}\left(2\dim_{M}K_{i}\right)=2\dim_{M}K\label{eq:2dimK}
\end{equation}
Eq.(\ref{eq:dimdim}) and (\ref{eq:2dimK}) give Theorem \ref{thm:dim_of_K_tilde}.
\begin{proof}
of Lemma \ref{lem:25}. Let $w\in\mathcal{W}$. We write $w=\left(w_{-},w_{+}\right)$
as before and $x_{w_{-}}:=S\left(w_{-}\right)\in K$, $\rho=\left(x_{w_{-}},\xi_{w}\right)=\mathcal{S}\left(w\right)\in\mathcal{K}$.
Similarly for another $w'\in\mathcal{W}$ we get another point $\rho'=\left(x_{w'_{-}},\xi_{w'}\right)\in\mathcal{K}$.
We have that 
\[
\Phi\left(\rho\right)=\left(S\left(w_{-}\right),S\left(\mathrm{Inv}\left(w_{+}\right)\right)\right)=\left(x_{w_{-}},x_{\mathrm{Inv}\left(w_{+}\right)}\right)\in K\times K.
\]
That the map $\Phi$ is bi-Lipschitz means that
\[
\left|\Phi\left(\rho\right)-\Phi\left(\rho'\right)\right|\asymp\left|\rho-\rho'\right|
\]
uniformly%
\footnote{The notation $\left|\Phi\left(\rho\right)-\Phi\left(\rho'\right)\right|\asymp\left|\rho-\rho'\right|$
means precisely that there exist $C>0$ such that for every $\rho,\rho'$,
$C^{-1}\left|\rho-\rho'\right|\leq\left|\Phi\left(\rho\right)-\Phi\left(\rho'\right)\right|\leq C\left|\rho-\rho'\right|$.%
} over $\rho,\rho'$. Equivalently this is
\begin{equation}
\left|x_{w_{-}}-x_{w'_{-}}\right|+\left|x_{\mathrm{Inv}\left(w_{+}\right)}-x_{\mathrm{Inv}\left(w'_{+}\right)}\right|\asymp\left|x_{w_{-}}-x_{w'_{-}}\right|+\left|\xi_{w}-\xi_{w'}\right|\label{eq:eq}
\end{equation}
uniformly over $w,w'\in\mathcal{W}$. Let us show (\ref{eq:eq}).
Let $w,w'\in\mathcal{W}$ and let $n\geq0$ be the integer such that
that $\left(w_{+}\right)_{j}=\left(w'_{+}\right)_{j}$ for $-n\leq j\leq0$
but $\left(w_{+}\right)_{-n-1}\neq\left(w'_{+}\right)_{-n-1}$. From
the definition (\ref{eq:def_Iwn}) of the intervals $I_{w_{-n,0}}$,
we see that the two points $x_{\mathrm{Inv}\left(w_{+}\right)},x_{\mathrm{Inv}\left(w'_{+}\right)}$
belong both to the interval $I_{\left(\mathrm{Inv}\left(w_{+}\right)\right)_{-n,0}}$
but inside it, they belong to the disjoint sub-intervals $I_{\left(\mathrm{Inv}\left(w_{+}\right)\right)_{-n-1,0}}$
and $I_{\left(\mathrm{Inv}\left(w'_{+}\right)\right)_{-n-1,0}}$ respectively.
Hence
\[
\left|x_{\mathrm{Inv}\left(w_{+}\right)}-x_{\mathrm{Inv}\left(w'_{+}\right)}\right|\asymp\left|I_{\left(\mathrm{Inv}\left(w_{+}\right)\right)_{-n,0}}\right|
\]
uniformly over $w,w'\in\mathcal{W}$, where $\left|I\right|$ is the
length of the interval $I$. From the definition (\ref{eq:def_set_I_tilde})
of the sets $\tilde{I}_{w_{0,n}}$ we observe that the points $\rho=\left(x_{w_{-}},\xi_{w}\right)$
and $\rho'=\left(x_{w'_{-}},\xi_{w'}\right)$ belong respectively
to the sets $\tilde{I}_{w_{0,n}}$ and $\tilde{I}_{w'_{0,n}}$. Let
$\tilde{w}':=\left(w'_{-},w_{+}\right).$We have
\begin{eqnarray}
\left|\rho-\rho'\right| & = & \left|\left(x_{w_{-}},\xi_{w}\right)-\left(x_{w'_{-}},\xi_{w'}\right)\right|\label{eq:e1-2}\\
 & \asymp & \left|\left(x_{w_{-}},\xi_{w}\right)-\left(x_{w{}_{-}},\xi_{\tilde{w}'}\right)\right|+\left|\left(x_{w{}_{-}},\xi_{\tilde{w}'}\right)-\left(x_{w'_{-}},\xi_{w'}\right)\right|\\
 & \asymp & \left|x_{w_{-}}-x_{w'_{-}}\right|+\left|\xi_{w}-\xi_{\tilde{w}'}\right|\nonumber 
\end{eqnarray}
The points $\xi_{w},\xi_{\tilde{w}'}$ belong to the same set $\tilde{I}_{w_{0,n}}$.
However if assumption of ``minimal captivity'' holds, they belong
to disjoint sub-sets $\tilde{I}_{w_{0,n+1}}$ and $\tilde{I}_{w'_{0,n+1}}$
respectively. Hence

\begin{equation}
\left|\xi_{w}-\xi_{\tilde{w}'}\right|\asymp\left|J_{w,n}\right|\label{eq:e4}
\end{equation}
with the interval $J_{w,n}:=\tilde{I}_{w_{0,n}}\cap\pi^{-1}\left(x_{w_{-}}\right)$.
From the bounded distortion principle \cite{Falconer_97} we have
that
\[
\forall x,y\in I_{w_{-n,0}},\quad\left|\left(D\phi_{w_{-n,0}}\right)\left(x\right)\right|\asymp\left|\left(D\phi_{w_{-n,0}}\right)\left(y\right)\right|\asymp\left|I_{w_{-n,0}}\right|
\]
uniformly with respect to $w,n,x,y$. From the expression of the canonical
map $F$ in (\ref{eq:def_symplectic_map_Fij-1}) and the bounded distortion
principle, we have that
\[
\left|J_{w,n}\right|\asymp\left|\left(D\phi_{w_{-n,0}}\right)\left(x\right)\right|,\quad\forall x\in I_{w_{0}},
\]
uniformly with respect to $w,n,x$. Using the previous results we
get
\begin{eqnarray*}
\left|x_{w_{-}}-x_{w'_{-}}\right|+\left|\xi_{w}-\xi_{w'}\right| & \asymp & \left|x_{w_{-}}-x_{w'_{-}}\right|+\left|\xi_{w}-\xi_{\tilde{w}'}\right|\\
 & \asymp & \left|x_{w_{-}}-x_{w'_{-}}\right|+\left|J_{w,n}\right|\\
 & \asymp & \left|x_{w_{-}}-x_{w'_{-}}\right|+\left(D\phi_{w_{0,n}}\right)\left(x\right),\quad\forall x\in I_{w_{0}},\\
 & \asymp & \left|x_{w_{-}}-x_{w'_{-}}\right|+\left|I_{\mathrm{Inv}\left(w_{0,n}\right)}\right|\\
 & \asymp & \left|x_{w_{-}}-x_{w'_{-}}\right|+\left|x_{\mathrm{Inv}\left(w_{+}\right)}-x_{\mathrm{Inv}\left(w'_{+}\right)}\right|.
\end{eqnarray*}
We have obtained (\ref{eq:eq}) and finished the proof of Lemma \ref{lem:25}
and Theorem \ref{thm:dim_of_K_tilde}.\end{proof}

\section{\label{sec:Proof-of-the_spectral_gap}Proof of Theorem \ref{th:gap_spectral}
for the spectral gap in the semiclassical limit}

For the proof of Theorem \ref{th:gap_spectral}, we will follow step
by step the same analysis as in Section \ref{sec:Proof-of-Theorem_discrete_spectrum}
(and also follow closely the proof of Theorem 2 in \cite{fred_expanding_09}).
The main difference now is that $\hbar\ll1$ is a semi-classical parameter
(no fixed anymore). In other words, we just perform a linear rescaling
in cotangent space: $\xi_{h}:=\hbar\xi$. Our quantization rule for
a symbol $A\left(x,\xi_{h}\right)\in S^{-m}\left(\mathbb{R}\right)$,
Eq.(\ref{eq:quantiz_rule}) writes now (see \cite{martinez-01} p.22),
for $\varphi\in\mathcal{S}\left(\mathbb{R}\right)$:
\begin{equation}
\left(\hat{A}\varphi\right)\left(x\right):=\frac{1}{2\pi\hbar}\int A\left(x,\xi_{h}\right)e^{i\left(x-y\right)\xi_{h}/\hbar}\varphi\left(y\right)dyd\xi_{h}\label{eq:quantiz_rule_n}
\end{equation}
For simplicity we will still write $\xi$ instead of $\xi_{h}$ below.

\subsection{The escape function}

Let $1<\kappa<1/\theta$ and $R>0$ given in Lemma \ref{lem:escape_F}.
Let $m>0$, $\eta>0$ (small) and consider a $C^{\infty}$ function
$A_{m}\left(x,\xi\right)$ on $T^{*}\mathbb{R}$ so that:
\begin{eqnarray*}
A_{m}\left(x,\xi\right) & := & \left\langle \xi\right\rangle ^{-m}\qquad\mbox{for }\left|\xi\right|>R+\eta\\
 & := & 1\qquad\mbox{for }\xi\leq R
\end{eqnarray*}
where $\left\langle \xi\right\rangle :=\left(1+\xi^{2}\right)^{1/2}$.
$A_{m}$ belongs to the symbol class $S^{-m}\left(\mathbb{R}\right)$
defined in (\ref{eq:def_symbol_class_S-m}).

From Eq. (\ref{eq:expand_nu}) we can deduce, similarly to Eq.(\ref{eq:A_decreases})
and if $\eta$ is small enough, that:
\begin{equation}
\forall x\in I,\forall\left|\xi\right|>R,\forall i\rightsquigarrow j\qquad\frac{A_{m}\left(F_{i,j}\left(x,\xi\right)\right)}{A_{m}\left(x,\xi\right)}\leq C^{m}<1,\quad\mbox{with }C=\sqrt{\frac{R^{2}+1}{\kappa^{2}R^{2}+1}}<1.\label{eq:A_decreases_2}
\end{equation}
This means that the function $A_{m}$ is an \textbf{escape function}
since it decreases strictly along the trajectories of $F$ outside
the zone $\mathcal{Z}_{0}:=I\times\left[-R,R\right]$. For any point
$\left(x,\xi\right)\in T^{*}I$ we have the more general bound:
\begin{equation}
\forall x\in I,\forall\xi\in\mathbb{R},\forall i\rightsquigarrow j\qquad\frac{A_{m}\left(F_{i,j}\left(x,\xi\right)\right)}{A_{m}\left(x,\xi\right)}\leq1.\label{eq:general_bound_A}
\end{equation}
Let $\hbar>0$. Using the quantization rule (\ref{eq:quantiz_rule_n}),
the symbol $A_{m}$ can be quantized giving a $\hbar$-pseudodifferential
operator $\hat{A}_{m}$ which is self-adjoint and invertible on $C^{\infty}\left(I\right)$.
In our case $\hat{A}_{m}$ is simply a multiplication operator by
$A_{m}\left(\xi\right)$ in $\hbar-$Fourier space.

\subsection{Using the Egorov Theorem}

Let us consider the Sobolev space
\[
H^{-m}\left(\mathbb{R}\right):=\hat{A}_{m}^{-1}\left(L^{2}\left(\mathbb{R}\right)\right)
\]
 which is the usual Sobolev space as a linear space, except for the
norm which depends on $\hbar$. Then $\hat{F}:H^{-m}\left(\mathbb{R}\right)\rightarrow H^{-m}\left(\mathbb{R}\right)$
is unitary equivalent to 
\[
\hat{Q}:=\hat{A}_{m}\hat{F}\hat{A}_{m}^{-1}\quad:L^{2}\left(\mathbb{R}\right)\rightarrow L^{2}\left(\mathbb{R}\right).
\]
Let $n\in\mathbb{N}^{*}$, a fixed time which will be made large at
the end of the proof, and define

\begin{equation}
\hat{P}^{\left(n\right)}:=\hat{Q}^{*n}\hat{Q}^{n}=\hat{A}_{m}^{-1}\hat{F}^{*n}\hat{A}_{m}^{2}\hat{F}^{n}\hat{A}_{m}^{-1}.\label{eq:def_Pn}
\end{equation}
From Egorov Theorem, as in Lemma \ref{lem:Egorov-theorem}), we have
that $\hat{B}:=\hat{F}^{*}\hat{A}_{m}^{2}\hat{F}$ is a PDO with principal
symbol

\begin{eqnarray*}
B\left(x,\xi\right) & = & \chi^{2}\left(x\right)\sum_{j\mbox{ s.t.}i\rightsquigarrow j}\left|\phi_{i,j}'\left(x\right)\right|e^{2\mathrm{Re}\left(V\left(\phi_{i,j}\left(x\right)\right)\right)}A_{m}^{2}\left(F_{i,j}\left(x,\xi\right)\right),\qquad\left(x,\xi\right)\in T^{*}I\\
 & = & \chi^{2}\left(x\right)\sum_{j\mbox{ s.t.}i\rightsquigarrow j}e^{2D\left(\left(\phi_{i,j}\left(x\right)\right)\right)}A_{m}^{2}\left(F_{i,j}\left(x,\xi\right)\right)
\end{eqnarray*}
where we have used the ``damping function'' $D\left(x\right):=\mathrm{Re}\left(V\left(x\right)\right)-\frac{1}{2}\log\left(\left|\left(\phi^{-1}\right)'\left(x\right)\right|\right)$
already defined in (\ref{eq:def_damping_function}). Iteratively for
every $n\geq1$, Egorov's Theorem gives that $\hat{F}^{*n}\hat{A}_{m}^{2}\hat{F}^{n}$
is a PDO with principal symbol
\begin{eqnarray*}
B_{n}\left(x,\xi\right) & = & \chi^{2}\left(x\right)\sum_{w_{-n,0}\in\mathcal{W}_{-}}e^{2D_{w_{-n,0}}\left(x\right)}A_{m}^{2}\left(F_{w_{-n,0}}\left(x,\xi\right)\right)
\end{eqnarray*}
where $\mathcal{W}_{+}$ is the set of admissible sequences, defined
in (\ref{eq:def_W_-}), with the Birkhoff sum $D_{w_{-n,0}}\left(x\right):=\sum_{k=1}^{n}D\left(\phi_{w_{-n,-k}}\left(x\right)\right)$
and 
\[
F_{w_{-n,0}}:=F_{w_{-1},w_{0}}\circ\ldots\circ F_{w_{-n},w_{-n+1}}.
\]
With the theorem of composition of PDO \cite[chap.4]{zworski-03}
we obtain that $\hat{P}^{\left(n\right)}$ is a PDO of order $0$
with principal symbol given by

\begin{eqnarray}
P^{\left(n\right)}\left(x,\xi\right) & = & \left(\chi^{2}\left(x\right)\sum_{w_{-n,0}\in\mathcal{W}_{-}}e^{2D_{w_{-n,0}}\left(x\right)}\frac{A_{m}^{2}\left(F_{w_{-n,0}}\left(x,\xi\right)\right)}{A_{m}^{2}\left(x,\xi\right)}\right).\label{eq:e1-1}
\end{eqnarray}
We define
\[
\gamma_{\left(n\right)}:=\sup_{x\in I,w_{-n,0}\in\mathcal{W}_{-}}\frac{1}{n}D_{w_{-n,0}}\left(x\right)
\]
hence $e^{2D_{w_{-n,0}}\left(x\right)}\leq e^{2n\gamma_{\left(n\right)}}$.

From Theorem \ref{thm:discrete-spectrum}, the spectrum of $\hat{F}_{\hbar}$
does not depend on the choice of $\chi$. Here we take $a\geq0$ as
given in Assumption \ref{hyp:minimal_capt} and we choose $\chi$
such that $\chi\equiv1$ on $K_{a+1}$, $\chi\equiv0$ on $\mathbb{R}\backslash K_{a}$.
We have $P\left(x,\xi\right)=0$ if $x\in\mathbb{R}\backslash K_{a}$.

Now we will bound the positive symbol $P^{\left(n\right)}\left(x,\xi\right)$
from above, considering $x\in K_{a}$ and different possibilities
for the trajectory $F_{w_{-n,0}}\left(x,\xi\right)$:
\begin{enumerate}
\item If $\left|\xi\right|>R$, Eq.(\ref{eq:A_decreases_2}) gives 
\begin{align}
\frac{A_{m}^{2}\left(F_{w_{-n,0}}\left(x,\xi\right)\right)}{A_{m}^{2}\left(x,\xi\right)} & =\frac{A_{m}^{2}\left(F_{w_{-n,0}}\left(x,\xi\right)\right)}{A_{m}^{2}\left(F_{w_{-n,-1}}\left(x,\xi\right)\right)}\frac{A_{m}^{2}\left(F_{w_{-n,-1}}\left(x,\xi\right)\right)}{A_{m}^{2}\left(F_{w_{-n,-2}}\left(x,\xi\right)\right)}\ldots\frac{A^{2}\left(F_{w_{-n,-n+1}}\left(x,\xi\right)\right)}{A^{2}\left(x,\xi\right)}\label{eq:bound1}\\
 & \leq\left(C^{2m}\right)^{n}
\end{align}
therefore 
\[
P^{\left(n\right)}\left(x,\xi\right)\leq\left(\sharp\mathcal{W}_{n}\right)e^{2n\gamma_{\left(n\right)}}\left(C^{2m}\right)^{n}\leq\left(Ne^{2\gamma_{\left(n\right)}}C^{2m}\right)^{n}.
\]
We have used that $\sharp\mathcal{W}_{n}\leq N^{n}$. Notice that
$C^{2m}$ can be made arbitrarily small if $m$ is large.
\item If $\left|\xi\right|\leq R$, we have from the hypothesis of minimal
captivity \ref{hyp:minimal_capt} and Proposition \ref{prop:minimal_captive}
that at time $(n-1)$ every point $\left(x',\xi'\right)$ of the set
$F^{n-1}\left(x,\xi\right)$ except finitely many points, satisfy
$\left|\xi'\right|>R$. Using (\ref{eq:general_bound_A}) and (\ref{eq:A_decreases_2}),
for all these points one has $\frac{A_{m}^{2}\left(F_{w_{-n,0}}\left(x,\xi\right)\right)}{A_{m}^{2}\left(x,\xi\right)}\leq C^{2m}$
and for the exceptional point one can only write $\frac{A_{m}^{2}\left(F_{w_{-n,0}}\left(x,\xi\right)\right)}{A_{m}^{2}\left(x,\xi\right)}\leq1$.
This gives
\[
P^{\left(n\right)}\left(x,\xi\right)\leq e^{2n\gamma_{\left(n\right)}}\left(\left(\sharp\mathcal{W}_{n}-1\right)C^{2m}+C'\right)\leq\mathcal{B}
\]
with the bound
\begin{equation}
\mathcal{B}:=e^{2n\gamma_{\left(n\right)}}\left(N^{n}C^{2m}+C'\right)\label{eq:bound2}
\end{equation}

\end{enumerate}
With the \textbf{$L^{2}$ continuity theorem} for pseudodifferential
operators \cite{martinez-01,dimassi-99} this implies that in the
limit $\hbar\rightarrow0$
\begin{equation}
\left\Vert \hat{P}^{\left(n\right)}\right\Vert \leq\mathcal{B}+\mathcal{O}_{n}\left(\hbar\right)
\end{equation}

Polar decomposition of $\hat{Q}^{n}$ gives
\begin{equation}
\left\Vert \hat{Q}^{n}\right\Vert \leq\left\Vert \left|\hat{Q}^{n}\right|\right\Vert =\sqrt{\left\Vert \hat{P}^{\left(n\right)}\right\Vert }\leq\left(\mathcal{B}+\mathcal{O}_{n}\left(\hbar\right)\right)^{1/2}\label{eq:norm_Q^n}
\end{equation}

Let $\gamma_{+}=\limsup_{n\rightarrow\infty}\gamma_{\left(n\right)}$.
If we let $\hbar\rightarrow0$ first, and $m\rightarrow+\infty$ giving
$C^{2m}\rightarrow0$, and $n\rightarrow\infty$, we obtain $\left(\mathcal{B}+\mathcal{O}_{n}\left(\hbar\right)\right)^{1/\left(2n\right)}\rightarrow e^{\gamma_{+}}$.
Therefore for any $\rho>e^{\gamma_{+}}$, there exists $n_{0}\in\mathbb{N}$,
$\hbar_{0}>0$, $m_{0}>0$ such that for any $\hbar\leq\hbar_{0}$,
$m>m_{0}$, 
\begin{equation}
\left\Vert \hat{F}_{\hbar}^{n_{0}}\right\Vert _{H^{-m}}=\left\Vert \hat{Q}^{n_{0}}\right\Vert _{L^{2}}\leq\rho^{n_{0}}.\label{eq:norm_F^n-1}
\end{equation}

Also, there exists $c>0$ independent of $\hbar\leq\hbar_{0}$, such
that for any $r$ such that $0\leq r<n_{0}$ we have $\left\Vert \hat{Q}^{r}\right\Vert _{L^{2}}<c$.
As a consequence for any $n\in\mathbb{N}$ we write $n=kn_{0}+r$
with $0\leq r<n_{0}$ and
\[
\left\Vert \hat{F}_{\hbar}^{n}\right\Vert _{H^{-m}}=\left\Vert \hat{Q}^{n}\right\Vert _{L^{2}}\leq\left\Vert \hat{Q}^{n_{0}}\right\Vert _{L^{2}}^{k}\left\Vert \hat{Q}^{r}\right\Vert _{L^{2}}\leq\rho^{n}\frac{\left\Vert \hat{Q}^{r}\right\Vert _{L^{2}}}{\rho^{r}}\leq c\rho^{n}
\]
We have obtained (\ref{eq:bound_for_F^n}). Equivalently this gives
(\ref{eq:bound_on_resolvent}).

For any $n$ the spectral radius of $\hat{Q}$ satisfies \cite[p.192]{reed-simon1}
\[
r_{s}\left(\hat{Q}\right)\leq\left\Vert \hat{Q}^{n}\right\Vert ^{1/n}\leq c^{1/n}\rho.
\]
So we get that for $\hbar\rightarrow0$,
\begin{equation}
r_{s}\left(\hat{F}_{\hbar}\right)=r_{s}\left(\hat{Q}\right)\leq e^{\gamma_{+}}+o\left(1\right)\label{eq:bound_rs}
\end{equation}
which finishes the proof of Theorem \ref{th:gap_spectral}.

\section{\label{sec:Fractal-Weyl-law}Proof of Theorem \ref{thm:fractal-weyl-upper}
about the fractal Weyl law}

We will prove this result once more by conjugating the transfer operator
by an escape function as in previous Section \ref{sec:Proof-of-the_spectral_gap}.
However we first have to improve the properties of the escape function.

\subsection{A refined escape function\label{sub:Refined-escape-function}}

\subsubsection{Distance function}

The escape function $A$ will be constructed from a distance function
$\delta$. For $x\in I$, let
\begin{equation}
\tilde{K}\left(x\right):=\tilde{K}\bigcap\left(\left\{ x\right\} \times\mathbb{R}\right)\label{eq:def_Ktilde_call}
\end{equation}
where $\tilde{K}$ has been defined in (\ref{eq:def_K_tilde-2}).
With this notation we can define the following distance function.

\begin{center}{\color{red}\fbox{\color{black}\parbox{16cm}{
\begin{defn}
Let $x\in I_{w_{0}}$ and $\xi\in\mathbb{R}$, we define the distance
of $\left(x,\xi\right)$ to the set $\tilde{K}$ given in (\ref{eq:def_K_tilde-2})
by
\end{defn}
\begin{equation}
\delta\left(x,\xi\right):=\mathrm{dist}\left(\xi,\tilde{K}\left(x\right)\right)=\min_{w\in\mathcal{W}_{+}}\left|\xi-\zeta_{w}\left(x\right)\right|.\label{eq:dist_func_delta}
\end{equation}

}}}\end{center}

We will show that the distance function $\delta\left(x,\xi\right)$
decreases along the trajectories of $F$. First, the next Lemma shows
how the branches $\zeta_{w}$ are transformed under the canonical
map $F$. This formula follows from straightforward calculations.

\begin{center}{\color{blue}\fbox{\color{black}\parbox{16cm}{
\begin{lem}
\label{lem:36-1}For every $w_{+}=\left(w_{0},w_{1},\ldots\right)\in\mathcal{W}_{+}$,
$x\in I_{w_{0}}$ we have
\begin{equation}
F_{w_{0},w_{1}}\left(x,\zeta_{w_{+}}\left(x\right)\right)=\left(x',\zeta_{L\left(w_{+}\right)}\left(x'\right)\right)\label{eq:zeta_transform}
\end{equation}
with $L\left(w_{+}\right):=\left(w_{1},w_{2},\ldots\right)$and$x'=\phi_{w_{0},w_{1}}\left(x\right)$.
\end{lem}
}}}\end{center}

\begin{center}{\color{blue}\fbox{\color{black}\parbox{16cm}{
\begin{lem}
\label{lem:delta_decay}$\forall i,j,i\rightsquigarrow j,\forall x\in I_{i},\forall\xi\in\mathbb{R},$
\[
\delta\left(F_{i,j}\left(x,\xi\right)\right)\geq\frac{1}{\theta}\delta\left(x,\xi\right)
\]
where $\theta<1$ is given by (\ref{eq:theta_contraction}).
\end{lem}
}}}\end{center}
\begin{proof}
Let $i=w_{0}\rightsquigarrow j=w_{1}$, $x\in I_{w_{0}}$. Let $\left(x',\xi'\right):=F_{w_{0},w_{1}}\left(x,\xi\right)$
with $x'\in I_{w_{1}}$. We use (\ref{eq:zeta_transform}) and also
that $F_{w_{0},w_{1}}$ is expansive in $\xi$ by a factor larger
than $\theta^{-1}>1$ (Eq.(\ref{eq:def_symplectic_map_Fij-1})), and
get 
\[
\left|\xi'-\zeta_{L\left(w_{+}\right)}\left(x'\right)\right|=\left|\left(F_{w_{0},w_{1}}\left(x,\xi\right)-F_{w_{0},w_{1}}\left(x,\zeta_{w_{+}}\left(x\right)\right)\right)_{\xi}\right|\geq\frac{1}{\theta}\left|\xi-\zeta_{w_{+}}\left(x\right)\right|.
\]
Thus 
\begin{eqnarray*}
\delta\left(F_{w_{0},w_{1}}\left(x,\xi\right)\right) & = & \min_{w\in\mathcal{W}_{+}}\left|\xi'-\zeta_{w_{+}}\left(x'\right)\right|=\min_{w_{+}\in\mathcal{W}_{+}}\left|\xi'-\zeta_{L\left(w_{+}\right)}\left(x'\right)\right|\\
 & \geq & \frac{1}{\theta}\min_{w_{+}\in\mathcal{W}_{+}}\left|\xi-\zeta_{w_{+}}\left(x\right)\right|=\frac{1}{\theta}\delta\left(x,\xi\right).
\end{eqnarray*}

\end{proof}

\subsubsection{Escape function}

The aim of this section is to prove the existence of an escape function
with the following properties:

\begin{framed}%
\begin{prop}
\textbf{\label{lem:escape_function-1}} $\forall1<\kappa<\theta^{-1}$,
$\mathsf{\exists C}_{0}>0,$ $\forall\mu,\mbox{ s.t. }0\leq\mu<\frac{1}{2}$,
$\forall m>0$, there exists an $\hbar$-dependent order function
$A_{m,\mu}\in\mathcal{OF}^{m\mu}(\langle\xi\rangle^{-m})$ (as defined
in Definition \ref{def:h-OrderFunction}) which fulfills the following
``decay condition'':

$\forall i,j,\quad\mbox{s.t. }i\rightsquigarrow j$ and $\forall\left(x,\xi\right)\in I_{i}\times\mathbb{R}$
s.t. $\delta\left(x,\xi\right)>\mathsf{C}_{0}\hbar^{\mu}$ the following
estimate holds:

\begin{equation}
\left(\frac{A_{m,\mu}\circ F_{i,j}}{A_{m,\mu}}\right)(x,\xi)\leq\kappa^{-m}.\label{eq:decroissance_fct_fuite-1}
\end{equation}
 \end{prop}
\end{framed}In order to prove the above proposition we first remark that the distance
function (\ref{eq:dist_func_delta}) is not differentiable, however
Lipschitz.

\begin{framed}%
\begin{lem}
\label{lem:delta_Lipschitz}Let $C_{1}:=\sup_{x\in I,\omega\in\mathcal{W}_{+}}|(\partial_{x}\zeta_{\omega})(x)|$.
Then $\delta:T^{*}I\to\mathbb{R}^{+}$is a Lipschitz function with
constant $C_{1}+1$\end{lem}
\end{framed}
\begin{proof}
Let $x,y\in I_{i}$, then from the fact, that $|(\partial_{x}\zeta_{\omega})(x)|$
is uniformly bounded by $C_{1}$ we have

\[
|\delta(x,\xi)-\delta(y,\xi)|\leq C_{1}|x-y|.
\]
On the other hand clearly

\[
|\delta(y,\xi)-\delta(y,\zeta)|\leq|\xi-\zeta|
\]
thus

\[
|\delta(x,\xi)-\delta(y,\zeta)|\leq C_{1}|x-y|+|\xi-\zeta|\leq(C_{1}+1)\textup{dist}((x,\xi),(y,\zeta)).
\]

\end{proof}
Next we choose $0\leq\mu<1/2$ and regularize the function $\delta$
at the scale $\hbar^{\mu}$. For this we choose $\chi\in C_{0}^{\infty}\left(\mathbb{R}^{2}\right)$
with support in the unit ball $B_{1}(0)$ of $\mathbb{R}^{2}$ and
$\chi\left(x,\xi\right)>0$ for $\|(x,\xi)\|<1$. This function can
be rescaled to 
\[
\chi_{\hbar^{\mu}}\left(x,\xi\right):=\frac{1}{\hbar^{2\mu}\left\Vert \chi\right\Vert _{L^{1}}}\chi\left(\frac{x}{\hbar^{\mu}},\frac{\xi}{\hbar^{\mu}}\right)
\]
 such that $\textrm{supp}\chi_{\hbar^{\mu}}\subset B_{\hbar^{\mu}}(0)$
and $\int\chi_{\hbar^{\mu}}\left(x\right)dx=1$. Now we can define
the regularized distance function by
\[
\tilde{\delta}\left(x,\xi\right):=\int_{T^{*}I}\delta\left(x',\xi'\right)\chi_{\hbar^{\mu}}\left(x-x',\xi-\xi'\right)dx'd\xi'.
\]
 This smoothed distance function $\tilde{\delta}$ differs only at
order $\hbar^{\mu}$ from the original one because 
\begin{eqnarray}
\left|\tilde{\delta}(x,\xi)-\delta(x,\xi)\right| & = & \left|\int_{\mathbb{R}^{2}}(\delta(x,\xi)-\tilde{\delta(x-x',\xi-\xi'))\chi_{\hbar^{\mu}}(x',\xi')}dx'd\xi'\right|\nonumber \\
 & \leq & \sup_{(x',\xi')\in B_{\hbar^{\mu}}(0)}|(\delta(x,\xi)-\delta(x-x',\xi-\xi')|\nonumber \\
 & \leq & (C_{1}+1)\hbar^{\mu}.\label{eq:delta_hmu}
\end{eqnarray}
 Furthermore we get the following estimates for its derivatives:

\begin{center}{\color{blue}\fbox{\color{black}\parbox{16cm}{
\begin{lem}
\label{lem:DerivDelta}For all $\alpha,\beta\in\mathbb{N}$ the estimate
\[
|\partial_{x}^{\alpha}\partial_{\xi}^{\beta}\tilde{\delta}(x,\xi)|\leq C_{\alpha,\beta}\hbar^{-\mu(\alpha+\beta)}(\delta(x,\xi)+C\hbar^{\mu})
\]

holds
\end{lem}
}}}\end{center}
\begin{proof}
From the definition of $\chi_{\hbar^{\mu}}$ we have $\|\partial_{x}^{\alpha}\partial_{\xi}^{\beta}\chi_{\hbar^{\mu}}\|_{\infty}\leq C_{\alpha,\beta}\hbar^{-2-(\alpha+\beta)\mu}$
and thus: 
\begin{eqnarray*}
\left|\left(\partial_{x}^{\alpha}\partial_{\xi}^{\beta}\tilde{\delta}(x,\xi)\right)\right| & = & \int_{T^{*}I}\delta\left(x',\xi'\right)\partial_{x}^{\alpha}\partial_{\xi}^{\beta}\chi_{\hbar^{\mu}}(x-x',\xi-\xi')dx'd\xi'\\
 & \leq & \pi\hbar^{2\mu}\|\delta\|_{\infty,B_{\hbar^{\mu}(x,\xi)}}C_{\alpha,\beta}\hbar^{-(2+\alpha+\beta)\mu}\\
 & \leq & \pi C_{\alpha,\beta}\hbar^{-(\alpha+\beta)\mu}(\delta(x,\xi)+(C_{1}+1)\hbar^{\mu})
\end{eqnarray*}
where we used the Lipschitz property of $\delta$ in the last inequality. 
\end{proof}
As $|\delta(x,\xi)|\leq\left|\xi\right|+C$ the above lemma gives
us directly that $\tilde{\delta}\in S_{\mu}^{1}\left(T^{*}I\right)$.
Now we define the escape function as:

\begin{equation}
A_{m,\mu}\left(x,\xi\right):=\hbar^{m\mu}\left(\hbar^{2\mu}+\left(\tilde{\delta}\left(x,\xi\right)\right)^{2}\right){}^{-\frac{m}{2}}.\label{eq:EscFuncWeyl}
\end{equation}
This is obviously a smooth function and it obeys the following estimates:

\begin{center}{\color{blue}\fbox{\color{black}\parbox{16cm}{
\begin{lem}
\label{lem:derivAm}The function $A_{m,\mu}$ defined in (\ref{eq:EscFuncWeyl})
is an $\hbar$-dependent order function: $A_{m,\mu}\in\mathcal{OF}^{m\mu}(\langle\xi\rangle^{-m})$.
\end{lem}
}}}\end{center}
\begin{proof}
As $\min(0,|\xi|-C)\leq\tilde{\delta}(x,\xi)\leq|\xi|+C$ it follows,
that $A_{m,\mu}(x,\xi)\leq\tilde{C}\langle\xi\rangle^{-m}$ and that
$A_{m,\mu}(x,\xi)\geq C^{\prime}\hbar^{m\mu}\langle\xi\rangle^{-m}$.
It remains thus to show, that for arbitrary $\alpha,\beta\in\mathbb{N}$
one has: 
\begin{equation}
|\partial_{x}^{\alpha}\partial_{\xi}^{\beta}A_{m,\mu}(x,\xi)|\leq C_{\alpha,\beta}\hbar^{-\mu(\alpha+\beta)}A_{m,\mu}(x,\xi)\label{eq:A_mmu_estimates}
\end{equation}
where $C_{\alpha,\beta}$ depends only on $\alpha$ and $\beta$.
First consider the case $\alpha=1$, $\beta=0$:

\[
|\partial_{x}A_{m,\mu}(x,\xi)|=\left|\hbar^{m\mu}m\frac{(\partial_{x}\tilde{\delta}(x,\xi))\tilde{\delta}(x,\xi)}{\left(\hbar^{2\mu}+\left(\tilde{\delta}\left(x,\xi\right)\right)^{2}\right){}^{\frac{m+2}{2}}}\right|\leq C\hbar^{-\mu}A_{m,\mu}\left(x,\xi\right)
\]
where we used $\tilde{\delta}\leq\sqrt{\hbar^{2\mu}+\tilde{\delta^{2}}}$
and $|\partial_{x}\tilde{\delta}|\leq C\hbar^{-\mu}\sqrt{\hbar^{2\mu}+\tilde{\delta^{2}}}$
which follows from Lemma \ref{lem:DerivDelta} together with (\ref{eq:delta_hmu}).
Inductively one obtains the estimate for arbitrary $\alpha,\beta\in\mathbb{N}$
by repeated use of Lemma \ref{lem:DerivDelta} and (\ref{eq:delta_hmu}).
\end{proof}
Finally it remains to show the decay estimates for $\left(\frac{A_{m,\mu}\circ F_{i,j}}{A_{m,\mu}}\right)\left(x,\xi\right)$.
Combining (\ref{eq:delta_hmu}) with lemma \ref{lem:delta_decay}
we then get

\begin{align*}
\tilde{\delta}(F_{i,j}(x,\xi)) & \geq\delta(F_{i,j}(x,\xi))-(C_{1}+1)\hbar^{\mu}\geq\frac{1}{\theta}\delta(x,\xi)-(C_{1}+1)\hbar^{\mu}\\
 & \geq\frac{1}{\theta}\tilde{\delta}(x,\xi)-(\frac{1}{\theta}+1)(C_{1}+1)\hbar^{\mu}
\end{align*}
and thus
\begin{equation}
\frac{A_{m,\mu}(F_{i,j}(x,\xi))}{A_{m,\mu}(x,\xi)}\leq\left(\frac{1+(\frac{1}{\theta}\cdot\frac{\tilde{\delta}(x,\xi)}{\hbar^{\mu}}-\tilde{C})^{2}}{1+(\frac{\tilde{\delta}(x,\xi)}{\hbar^{\mu}})^{2}}\right)^{\frac{m}{2}}\label{eq:Ammu_decay}
\end{equation}
where $\tilde{C}=(\frac{1}{\theta}+1)(C_{1}+1)$. Clearly the right
side of (\ref{eq:Ammu_decay}) converges to $(\frac{1}{\theta})^{-m}$
for $\text{\ensuremath{\frac{\tilde{\delta}\left(x,\xi\right)}{\hbar^{\mu}}}}\to\infty$
which proves the existence of a desired $C_{0}$ and finishes the
proof of Proposition \ref{lem:escape_function-1}.

\subsubsection{Truncation in $x$}

Here we choose a similar truncation operator $\hat{\chi}$ as in Eq.(\ref{eq:def_F_Chi})
but in a finer vicinity of the trapped set $K$. First notice that
$K_{\hbar^{\mu}}\Subset\phi^{-1}\left(K_{\hbar^{\mu}}\right)$ where
$K_{\hbar^{\mu}}$ has been defined in Definition \ref{Minkowski}.
For $\hbar$ small enough we have $\phi^{-1}\left(K_{\hbar^{\mu}}\right)\Subset I$.
Let $\chi\in C_{\phi^{-1}\left(K_{\hbar^{\mu}}\right)}^{\infty}$
such that $\chi\left(x\right)=1$ for $x\in K_{\hbar^{\mu}}$. $\chi$
can be considered as a function $\chi\left(x,\xi\right):=\chi\left(x\right)$
(independent of $\xi$) and we have that $\chi_{\mu}\in S_{\mu}^{0}\left(T^{*}\mathbb{R}\right)$.
As in Eq.(\ref{eq:def_F_Chi}) we define $\hat{\chi}:=\mbox{Op}_{\hbar}^{w}(\chi)$
which is the multiplication operator by $\chi$ and 
\[
\hat{F}_{i,j,\chi}:=\hat{F}_{i,j}\hat{\chi},\qquad\hat{F}_{\chi}:=\hat{F}\hat{\chi}.
\]
We will again omit the $\chi$ in the notation and write $\hat{F}_{\hbar}$
for $\hat{F}_{\chi}$ in the sequel.

\subsection{Weyl law\label{sub:Weyl-law}}

The Weyl law will give an upper bound on the number of eigenvalues
of $\hat{F}_{\hbar}$ in the Sobolev spaces $H^{m}$. These estimates
will be obtained by conjugating $\hat{F}_{\hbar}$ with $Op_{\hbar}^{w}(A_{m,\mu})$
in the same way as for the discrete spectrum or the spectral gap.
Note that we use the Weyl quantization (see Definition \ref{def:Quantization})
in this section, because we want to obtain self adjoint operators.
In order to be able to conjugate we have to show, that $Op_{\hbar}^{w}(A_{m,\mu}):H^{-m}\to L^{2}$
is an isomorphism. We already know that $Op_{\hbar}^{w}(\langle\xi\rangle^{m}):L^{2}\to H^{-m}$
is an isomorphism, thus it suffices to show, that $\hat{B}:=Op_{\hbar}^{w}(A_{m,\mu})Op_{\hbar}^{w}(\langle\xi\rangle^{m}):L^{2}\to L^{2}$
is invertible. From the $\hbar$-local symbol calculus (Theorem \ref{thm:Comp})
it follows that $\hat{B}$ is an elliptic operator in the $\hbar$-local
symbol class $S_{\mu}(A_{m,\mu}\langle\xi\rangle^{m})$ and thus the
invertibility follows from proposition \ref{prop:L2inverse}. Note
that it is necessary to work in the $\hbar$-local symbol classes
as $\hat{B}$ would not be an elliptic operator in $S_{\mu}(1)$.
Proposition \ref{prop:L2inverse} also gives us the leading order
of our inverse $\hat{B}^{-1}$ which is $A_{m,\mu}^{-1}\langle\xi\rangle^{-m}$.
So the inverse of $Op_{\hbar}^{w}(A_{m,\mu})$ is  again a PDO with
leading symbol $A_{m,\mu}^{-1}$.

With the isomorphism $Op_{\hbar}^{w}(A_{m,\mu}):H^{m}\to L^{2}$ we
can thus define a different scalar product on the Sobolev spaces which
turns $Op_{\hbar}^{w}(A_{m,\mu})$ into a unitary operator. The Sobolev
space equipped with this scalar product will be denoted by $\mathcal{H}_{\hbar,\mu}^{m}$
and the study of $\hat{F}_{\hbar}$ is thus unitary equivalent to
the study of $\hat{Q}_{m}$ defined by the following commutative diagram
(where we noted $\hat{A}_{m,\mu}:=Op_{\hbar}^{w}(A_{m,\mu})$:

\begin{equation}
\begin{array}{ccc}
L^{2}\left(\mathbb{R}\right) & \overset{\hat{Q}_{m}}{\rightarrow} & L^{2}\left(\mathbb{R}\right)\\
\downarrow\hat{A}_{m,\mu}^{-1} &  & \downarrow\hat{A}_{m,\mu}^{-1}\\
\mathcal{H}_{\hbar,\mu}^{-m} & \overset{\hat{F}_{\hbar}}{\rightarrow} & \mathcal{H}_{\hbar,\mu}^{-m}
\end{array}\label{eq:ComDiagWeyl}
\end{equation}
In the next Lemma, $\mathsf{C}_{0}$ and $\kappa$ are as in lemma
\ref{lem:escape_function-1}.

\begin{center}{\color{blue}\fbox{\color{black}\parbox{16cm}{
\begin{lem}
\label{lem:weyl_Q_m_mu}$\exists\mathsf{C}>\mathsf{C}_{0}$, $\forall\epsilon>0$,
$\forall\mu$ s.t. $0\leq\mu<\frac{1}{2}$, $\forall m>0$ sufficiently
large, as $\hbar\rightarrow0$ we have: 
\begin{equation}
\sharp\left\{ \lambda_{i}^{\hbar}\in\sigma\left(\hat{F}_{\hbar}|_{\mathcal{H}_{\hbar,\mu}^{m}}\right)\,|\,\left|\lambda_{i}^{\hbar}\right|\geq\epsilon\right\} \leq\frac{1}{2\pi\hbar}\left(\tilde{C}_{1}\mathrm{Leb}\left\{ \mathcal{K}_{C_{1}\hbar^{\mu}}\right\} +\tilde{C}_{2}\hbar\right).\label{eq:lem_nbre_resonances}
\end{equation}

\end{lem}
}}}\end{center}

Before proving Lemma \ref{lem:weyl_Q_m_mu}, let us show that it implies
Theorem \ref{thm:fractal-weyl-upper}. From Theorem \ref{thm:discrete-spectrum},
the discrete spectrum of $\hat{F}_{\hbar}|_{\mathcal{H}_{\hbar,\mu}^{m}}$
is the Ruelle spectrum of resonances $\mathrm{Res}\left(\hat{F}_{\hbar}\right)$,
independent of $\mu$ and $m$. With assumption \ref{hyp:minimal_capt}
we can use Eq.(\ref{eq:dim_Ktilde}) and that $\mathcal{K}$ has pure
dimension, thus equation (\ref{eq:pure_dim}) gives $\mathrm{Leb}\left\{ \mathcal{K}_{C_{1}\hbar^{\mu}}\right\} =\mathcal{O}\left(\left(\hbar^{\mu}\right)^{\mathrm{codim}_{M}\left(\mathcal{K}\right)}\right)$.
As $\mathrm{codim}_{M}\left(\mathcal{K}\right)<2$ and $\mu<\frac{1}{2}$
equation(\ref{eq:lem_nbre_resonances}) and gives
\begin{eqnarray*}
\sharp\left\{ \lambda_{i}^{\hbar}\in\mathrm{Res}\left(\hat{F}_{\hbar}\right)\,|\,\left|\lambda_{i}^{\hbar}\right|\geq\epsilon\right\}  & = & \mathcal{O}\left(\hbar^{-1}\left(\hbar^{\mu}\right)^{\mathrm{codim}_{M}\left(\mathcal{K}\right)}\right)\\
 & = & \mathcal{O}\left(\hbar^{-1}\left(\hbar^{\mu}\right)^{2-2\mathrm{dim}_{H}\left(K\right)}\right)=\mathcal{O}\left(\hbar^{2\mu-1-2\mu\mathrm{dim}_{H}\left(K\right)}\right)
\end{eqnarray*}
for any fixed $0\leq\mu<1/2$. This gives Theorem \ref{thm:fractal-weyl-upper}
with $\eta=(1-2\mu)(1-\dim_{\textup{H}}(K))$.
\begin{proof}
\emph{of Lemma \ref{lem:weyl_Q_m_mu}}. From (\ref{eq:ComDiagWeyl}),
$\hat{F}_{\chi,\hbar}:\mathcal{H}_{\hbar,\mu}^{m}\rightarrow\mathcal{H}_{\hbar,\mu}^{m}$
is unitary equivalent to 
\[
\hat{Q}_{m,\mu}:=\mbox{Op}_{\hbar}^{w}\left(A_{m,\mu}\right)\hat{F}_{\hbar}\hat{\chi}\mbox{Op}_{\hbar}^{w}\left(A_{m,\mu}\right)^{-1}:L^{2}\left(\mathbb{R}\right)\rightarrow L^{2}\left(\mathbb{R}\right).
\]
 Consider 
\[
\hat{P}_{\mu}:=\hat{Q}_{m,\mu}^{*}\hat{Q}_{m,\mu}=\mbox{Op}_{\hbar}^{w}\left(A_{m,\mu}\right)^{-1}\hat{\chi}\hat{F}_{\hbar}^{*}\mbox{Op}_{\hbar}^{w}\left(A_{m,\mu}\right)^{2}\hat{F}_{\hbar}\hat{\chi}\mbox{Op}_{\hbar}^{w}\left(A_{m,\mu}\right)^{-1}.
\]
By the composition Theorem \ref{thm:Comp} and the Egorov theorem
\ref{thm:Ego} for $\hbar$-local symbols, $\hat{P}_{\mu}$ is a PDO
with leading symbol $P_{\mu}\left(x,\xi\right)\in S_{\mu}\left(1\right)$,
for $x\in I_{i}$, $\xi\in\mathbb{R}$, given by the same expression
as in (\ref{eq:symb_P}):%
\footnote{Also for this calculation it is crucial to work with the $\hbar$-local
calculus in order to obtain sufficient remainder estimates.%
} 
\begin{equation}
P_{\mu}\left(x,\xi\right)=\chi^{2}\left(x\right)\sum_{j\mbox{ s.t. }i\rightsquigarrow j}\left|\phi_{i,j}'\left(x\right)\right|e^{2\Re\left(V\left(\phi_{i,j}\left(x\right)\right)\right)}\frac{A_{m,\mu}^{2}\left(F_{i,j}\left(x,\xi\right)\right)}{A_{m,\mu}^{2}\left(x,\xi\right)}\mbox{ mod }\hbar^{1-2\mu}S_{\mu}^{-1}\left(T^{*}\mathbb{R}\right).\label{eq:P_mu}
\end{equation}
From the definition of $\chi$ and Eq.(\ref{eq:decroissance_fct_fuite-1}),
the operator $\hat{P}_{\mu}$ can be decomposed into self-adjoint
operators 
\[
\hat{P}_{\mu}=\hat{k}_{\mu}+\hat{r}_{\mu}
\]
where $\hat{k}_{\mu}$ is a PDO with symbol $k_{\mu}\in S_{\mu}^{0}\left(T^{*}\mathbb{R}\right)$
supported on $\mathcal{K}_{C_{1}\hbar^{\mu}}$ for some $C_{1}>0$.
Hence $\hat{k}_{\mu}$ is a trace-class operator. The operator $\hat{r}_{\mu}$
is a PDO with symbol $r_{\mu}\in S_{\mu}^{0}\left(T^{*}\mathbb{R}\right)$
such that 
\[
\left\Vert r_{\mu}\right\Vert _{\infty}\leq\theta e^{2\left\Vert \Re(V)\right\Vert _{\infty}}\kappa^{-2m}+\mathcal{O}\left(\hbar^{1-2\mu}\right),
\]
hence $\left\Vert \hat{r}_{\mu}\right\Vert \leq C\kappa^{-2m}+\mathcal{O}(\hbar^{1-2\mu})$.

Using Lemma \ref{lem:weyl_law_compact} in Appendix A we have that
for every $\epsilon>0$, in the limit $\hbar\rightarrow0$,
\begin{equation}
\sharp\left\{ \mu_{i}^{\hbar}\in\sigma\left(\hat{k}_{\mu}\right)\,|\,\left|\mu_{i}^{\hbar}\right|\geq\epsilon\right\} \leq\left(2\pi\hbar\right)^{-1}\left(\tilde{C}_{1}\mathrm{Leb}\left\{ \mathcal{K}_{C_{1}\hbar^{\mu}}\right\} +\tilde{C}_{2}\hbar\right).\label{eq:sigma(k_mu)}
\end{equation}
By a standard perturbation argument the same estimates holds for the
operator $\hat{P}_{\mu}$ (for $m$ sufficiently large): for every
$\epsilon>0$, in the limit $\hbar\rightarrow0$,
\begin{equation}
\sharp\left\{ \mu_{i}^{\hbar}\in\sigma\left(\hat{P}_{\mu}\right)\,|\,\left|\mu_{i}^{\hbar}\right|\geq\epsilon+\left\Vert \hat{r}_{\mu}\right\Vert \right\} \leq\left(2\pi\hbar\right)^{-1}\left(\tilde{C}_{1}\mathrm{Leb}\left\{ \mathcal{K}_{C_{1}\hbar^{\mu}}\right\} +\tilde{C}_{2}\hbar\right).\label{eq:sigma(k_mu)-1}
\end{equation}
From the definition $\hat{P}_{\mu}:=\hat{Q}_{m,\mu}^{*}\hat{Q}_{m,\mu}$,
the $\sqrt{\mu_{i}^{\hbar}}$ are singular values of $\hat{Q}_{m,\mu}$.
Then corollary \ref{cor:singular} from Appendix B shows that the
same estimate holds true for the eigenvalues of $\hat{Q}_{m,\mu}$,
hence of $\hat{F}_{\hbar}$, yielding the result (\ref{eq:lem_nbre_resonances}).\end{proof}

\appendix

\section{Adapted Weyl type estimates}

\begin{framed}%
\begin{lem}
\label{lem:weyl_law_compact}Let $a_{\hbar}\in S_{\mu}\left(\langle x\rangle^{-2}\langle\xi\rangle^{-2}\right)$
with $0\leq\mu<\frac{1}{2}$ be a real compactly supported symbol.
$\forall\hbar>0,$ $\hat{A}:=\mbox{\emph{Op}}_{\hbar}^{w}(a_{\hbar})$
is self-adjoint and trace class on $L^{2}\left(\mathbb{R}\right)$
and for any $\epsilon>0$, as $\hbar\rightarrow0$ : 
\begin{equation}
\left(2\pi\hbar\right)\sharp\left\{ \lambda_{i}^{\hbar}\in\sigma\left(\hat{A}\right)\,|\,\left|\lambda_{i}^{\hbar}\right|\geq\epsilon\right\} \leq C_{1}\mathrm{Leb}\left\{ (x,\xi)\,;\,|a|>0\right\} +C_{2}\hbar\label{eq:loi_weyl}
\end{equation}
where $C_{1}$ and $C_{2}$ depend only on $\mu$ and $\epsilon$.\end{lem}
\end{framed}
\begin{proof}
As $a_{\hbar}$ is compactly supported $\hat{A}$ is trace class for
every $\hbar$ (see theorem C.17 \cite{zworski-03}). Consequently
also $\frac{1}{\epsilon^{2}}\hat{A}^{2}$ is trace class and its trace
is given by Lidskii's theorem by $Tr(\frac{1}{\epsilon^{2}}\hat{A}^{2})=\sum{}_{i}\left(\frac{{\lambda_{i}^{\hbar}}}{\epsilon}\right)^{2}$.
As $\hat{A}$ is self adjoint all $\lambda_{i}^{\hbar}$ are real
and one clearly has
\[
\sharp\left\{ \lambda_{i}^{\hbar}\in\sigma\left(\hat{A}\right)\,|\,\left|\lambda_{i}^{\hbar}\right|\geq\epsilon\right\} \leq Tr\left(\frac{1}{\epsilon^{2}}\hat{A}^{2}\right).
\]
If we denote by $b_{\hbar}(x,\xi)$ the complete symbol of $\hat{A^{2}}$
we can calculate the trace by the following exact formula
\[
Tr(\hat{A}^{2})=\frac{1}{2\pi\hbar}\int b_{\hbar}(x,\xi)dxd\xi
\]
According to the theorem of composition of PDOs $b_{\hbar}$ can be
written as $b_{\hbar}=b_{\hbar}^{(1)}+\hbar^{1}b_{\hbar}^{(2)}$ where
$\mathrm{supp}b_{\hbar}^{(1)}=\mathrm{supp}a_{\hbar}$ and $b_{\hbar}^{(2)}\in\text{ \ensuremath{S_{\mu}\left(\langle x\rangle^{-2}\langle\xi\rangle^{-2}\right)}}$(note
that this decomposition depends on $\mu$. Thus
\[
\frac{1}{\epsilon\text{\texttwosuperior\ }}Tr(\hat{A}^{2})=\frac{1}{2\pi\hbar\epsilon^{2}}\left(\int b_{\hbar}^{(1)}(x,\xi)dxd\xi+\hbar^{1}\int b_{\hbar}^{(1)}(x,\xi)dxd\xi\right)\leq\left(\frac{1}{2\pi\hbar}C_{1}\textup{Leb}(\textup{supp}(a_{\hbar}))+C_{2}\hbar\right)
\]

\end{proof}

\section{General lemmas on singular values of compact operators}

Let $\left(P_{n}\right)_{n\in\mathbb{N}}$ be a family of compact
operators on some Hilbert space. For every $n\in\mathbb{N}$ let $(\lambda_{j,n})_{j\in\mathbb{N}}\in\mathbb{C}$
be the sequence of eigenvalues of $P_{n}$ counted with multiplicity
and ordered decreasingly:
\[
|\lambda_{0,n}|\geq|\lambda_{1,n}|\geq...
\]
In the same manner, define $(\mu_{j,n})_{j\in\mathbb{N}}\in\mathbb{R}^{+}$,
the decreasing sequence of singular values of $P_{n}$, i.e. the eigenvalues
of $\sqrt{P_{n}^{*}P_{n}}$ . 

\begin{framed}%
\begin{lem}
\label{lem:``Singular-and-eigenvalues} Suppose there exits a map
$N:\mathbb{N}\rightarrow\mathbb{N}$ s.t. $N(n)\underset{n\rightarrow\infty}{\rightarrow}\infty$
and \textup{$\mu_{N(n),n}\underset{n\rightarrow\infty}{\rightarrow}0$,
}then $\forall C>1,$ $|\lambda_{\left[C\cdot N(n)\right],n}|\underset{n\rightarrow\infty}{\rightarrow}0$
where $\left[.\right]$ stands for the integer part. \end{lem}
\begin{cor}
\label{cor:singular} Suppose there exits a map $N:\mathbb{N}\rightarrow\mathbb{N}$
s.t. $\forall\varepsilon>0,$ $\exists A_{\varepsilon}\geq0$ s.t.
$\forall n\geq A_{\varepsilon}$, 
\[
\#\left\{ j\in\mathbb{N}\;\mbox{s.t.}\;\mu_{j,n}>\varepsilon\right\} <N\left(n\right),
\]
then $\forall C>1,$ $\forall\varepsilon>0$, $\exists B_{C,\varepsilon}\geq0$
s.t. $\forall n\geq B_{C,\varepsilon}$ 
\begin{equation}
\#\left\{ j\in\mathbb{N}\;\mbox{s.t.}\;|\lambda_{j,n}|>\varepsilon\right\} \leq C\cdot N\left(n\right).\label{eq:cor2}
\end{equation}
\end{cor}
\end{framed}
\begin{proof}
\emph{(Of corollary \ref{cor:singular})}. Suppose that for any $\varepsilon>0$,
there exists $A_{\varepsilon}$ s.t. for all $n\geq A_{\varepsilon}$,
$\#\left\{ j\in\mathbb{N}\;\mbox{s.t.}\;\mu_{j,n}>\varepsilon\right\} <N\left(n\right).$
Then $\mu_{N(n),n}\rightarrow_{n\rightarrow\infty}0$ and from Lemma
\ref{lem:``Singular-and-eigenvalues}, $\forall C>1$, $|\lambda_{[C\cdot N(n)],n}|\rightarrow_{n\rightarrow\infty}0,$
which can be directly restated as (\ref{eq:cor2}).
\end{proof}
\medskip{}

\begin{proof}
\emph{(Of lemma \ref{lem:``Singular-and-eigenvalues}).}Let $m_{j,n}:=-\log\mu_{j,n}$
and $l_{j,n}:=-\log|\lambda_{j,n}|$, $M_{k,n}:=\sum_{j=0}^{k}m_{j,n}$
and $L_{k,n}:=\sum_{j=0}^{k}l_{j,n}$. Weyl inequalities relate singular
values and eigenvalues by (see \cite{gohberg-00} p. 50 for a proof)
:
\begin{equation}
\prod_{j=1}^{k}\mu_{j,n}\leq\prod_{j=1}^{k}|\lambda_{j,n}|,\;\forall k\geq1.\label{eq:weyl_inequalities}
\end{equation}
This rewrites: 
\begin{equation}
M_{k,n}\leq L_{k,n},\quad\forall k,n\label{eq:Weyl_ineq_2}
\end{equation}
The sequence $\left(l_{j,n}\right)_{j\geq0}$ is increasing in $j$
so, $\forall n,\forall k$ we have 
\begin{equation}
k\cdot l_{k,n}\geq L_{k,n}.\label{eq:16}
\end{equation}
 Suppose that $\mu_{N(n),n}\rightarrow0$ as $n\rightarrow\infty$
hence 
\begin{equation}
m_{N(n),n}\underset{n\rightarrow\infty}{\rightarrow}\infty\label{eq:17}
\end{equation}
Let $C>1$. The sequence $\left(m_{j,n}\right)_{j\geq0}$ is increasing
in $j$ hence 
\begin{equation}
M_{[C\cdot N(n)],n}\geq\left([C\cdot N(n)]-N(n)\right)\cdot m_{N(n),n},\label{eq:S_vs_m}
\end{equation}
hence
\begin{eqnarray*}
l_{[C\cdot N(n)],n} & \underset{(\ref{eq:16})}{\geq} & \frac{1}{[CN(n)]}\cdot L_{[C\cdot N(n)],n}\underset{(\ref{eq:Weyl_ineq_2})}{\geq}\frac{1}{[C\cdot N(n)]}M_{[C\cdot N(n)],n}\\
 & \underset{(\ref{eq:S_vs_m})}{\geq} & \frac{[C\cdot N(n)]-N(n)}{[C\cdot N(n)]}\cdot m_{N(n),n}\underset{(\ref{eq:17})}{\longrightarrow}\infty
\end{eqnarray*}
Thus $l_{[C\cdot N(n)],n}\underset{n\rightarrow\infty}{\rightarrow}\infty$
and $|\lambda_{\left[C\cdot N(n)\right],n}|\underset{n\rightarrow\infty}{\rightarrow}0$.
\end{proof}

\section{Symbol classes of local $\hbar$-order\label{sec:Appendix-Symbol-classes}}

In this Appendix we will first repeat the definitions of the standard
symbol classes which are used in this article as well as their well
known quantization rules. Then we will introduce a new symbol class
which allows $\hbar$-dependent order functions and will prove some
of the classical results which are known in the usual case for these
new symbol classes.

\subsection{Standard semiclassical Symbol classes and their quantization}

The standard symbol classes (see e.g. \cite{zworski-03} chapter 4
or \cite{dimassi-99} ch 7) of $\hbar$PDO's are defined with respect
to an order function $f(x,\xi)$. This order function is required
to be a smooth positive valued function on $\mathbb{R}^{2n}$ such
that there are constants $C_{0}$ and $N_{0}$ fullfilling
\begin{equation}
f(x,\xi)\leq C_{0}\langle(x,\xi)-(x',\xi')\rangle^{N_{0}}f(x',\xi').\label{eq:OrderFunctionClassical}
\end{equation}
An important example of such an order function is given by $f(x,\xi)=\langle\xi\rangle^{m}$
with $k\in\mathbb{R}$. 

\begin{framed}%
\begin{defn}
\label{def:SymbolClass_S_mu}For $0\leq\mu\leq\frac{1}{2}$ the symbol
classes $\hbar^{k}S_{\mu}(m)$ contain all families of functions $a_{\hbar}(x,\xi)\in C^{\infty}(\mathbb{R}^{2n})$
parametrized by a parameter $\hbar\in]0,\hbar_{0}]$ such that 
\[
|\partial_{x}^{\alpha}\partial_{\xi}^{\beta}a_{\hbar}(x,\xi)|\leq C\hbar^{k-\mu(|\alpha|+|\beta|)}f(x,\xi)
\]
where $C$ depends only on $\alpha$ and $\beta$.\end{defn}
\end{framed}

Unless we want to emphasize the dependence of the symbol $a_{\hbar}$
on $\hbar$ we will drop the index in the following. For the special
case of order function $f(x,\xi)=\langle\xi\rangle^{m}$ we also write
$S_{\mu}^{m}=S_{\mu}(\langle\xi\rangle^{m})$, if $\mu=0$ we write
$S(f):=S_{0}(f)$. 

As quantization we use two different quantization rules in this article
which are called standard quantization respectively Weyl quantization.

\begin{framed}%
\begin{defn}
\label{def:Quantization}Let $a_{\hbar}\in S_{\mu}\left(f\right)$
the Weyl quantization is a family of operators $\mbox{Op}_{\hbar}^{w}(a):\mathcal{S}\left(\mathbb{R}^{n}\right)\rightarrow\mathcal{S}\left(\mathbb{R}^{n}\right)$,
defined by 
\begin{equation}
\left(\mbox{Op}_{\hbar}^{w}(a_{\hbar})\varphi\right)(x)=\left(2\pi\hbar\right)^{-n}\int e^{\frac{i}{\hbar}\xi(x-y)}a_{\hbar}\left(\frac{x+y}{2},\xi\right)\varphi(y)dyd\xi,\qquad\varphi\in\mathcal{S}\left(\mathbb{R}^{n}\right).\label{eq:weyl_quant}
\end{equation}
while the standard quantization $\mbox{Op}_{\hbar}(a):\mathcal{S}\left(\mathbb{R}^{n}\right)\rightarrow\mathcal{S}\left(\mathbb{R}^{n}\right)$
is given by 

\begin{equation}
\left(\mbox{Op}_{\hbar}(a_{\hbar})\varphi\right)(x)=\left(2\pi\hbar\right)^{-n}\int e^{\frac{i}{\hbar}\xi(x-y)}a_{\hbar}\left(x,\xi\right)\varphi(y)dyd\xi,\qquad\varphi\in\mathcal{S}\left(\mathbb{R}^{n}\right).\label{eq:stand_quant}
\end{equation}
\end{defn}
\end{framed}

Both quantization extend continuously to operators on $\mathcal{S}^{\prime}(\mathbb{R}^{n})$.
While the standard quantization is slightly easier to define, the
Weyl quantization has the advantage, that real symbols are mapped
to formally self adjoint operators.

\subsection{Definition of the Symbol classes $S_{\mu}(A_{\hbar})$}

In this standard $\hbar$-PDO calculus the symbols are ordered by
there asymptotic behavior for $\hbar\to0$. If we take for example
a symbol $a\in\hbar^{k}S_{\mu}(f)$ then $a(x,\xi)$ is of order $\hbar^{k}$
for all $(x,\xi)\in\mathbb{R}^{2n}$. The symbol classes that we will
now introduce will also allow $\hbar-$dependent order function which
will allow to control the $\hbar$-order of a symbol locally, i.e.
in dependence of $(x,\xi)$. First we define these $\hbar$-dependent
order functions:

\begin{framed}%
\begin{defn}
\label{def:h-OrderFunction}Let $f$ be an order function on $\mathbb{R}^{2n}$
and $0\leq\mu\leq\frac{1}{2}$. Let $A_{\hbar}\in S_{\mu}(f)$ a (possibly
$\hbar$-dependent) positive symbol such that for some $c\geq0$ there
is a constant $C$ that fulfills 
\begin{equation}
A_{\hbar}(x,\xi)\geq C\hbar^{c}f(x,\xi)\label{eq:AlowerBd}
\end{equation}

\end{defn}
and that for all multiindices $\alpha,\beta\in\mathbb{N}^{n}$: 
\begin{equation}
\left|\partial_{x}^{\alpha}\partial_{\xi}^{\beta}A_{\hbar}(x,\xi)\right|\leq C_{\alpha,\beta}\hbar^{-\mu(|\alpha|+|\beta|)}A_{\hbar}(x,\xi)\label{eq:dxxiEstimate_for_OF}
\end{equation}
holds. Then we call $A_{\hbar}$an $\hbar$-dependent order function
and say $A_{\hbar}\in\mathcal{OF}^{c}(f)$\end{framed}

\begin{framed}%
\begin{defn}
\label{def:hSymbolClassNew}The symbol class $S_{\mu}(A_{\hbar})$
is then defined to be the space of smooth functions $a_{\hbar}(x,\xi)$
defined on $R^{2n}$ and parametrized by $\hbar>0$ such that 
\begin{equation}
\left|\partial_{x}^{\alpha}\partial_{\xi}^{\beta}a_{\hbar}(x,\xi)\right|\leq C_{\alpha,\beta}\hbar^{-\mu(|\alpha|+|\beta|)}A_{\hbar}(x,\xi)\label{eq:dxxiEstimate}
\end{equation}
By $\hbar^{k}S_{\mu}(A_{\hbar})$ we will as usual denote the symbols
$a_{\hbar}$ for which $\hbar^{-k}a_{\hbar}\in S_{\mu}(A_{\hbar})$\end{defn}
\end{framed}

As $A_{h}(x,\xi)\leq C_{0}f(x,\xi)$ and from (\ref{eq:dxxiEstimate_for_OF})
it is obvious, that 
\begin{equation}
S_{\mu}(A_{\hbar})\subset S_{\mu}(f)\label{eq:IncSymbClass}
\end{equation}
and via this inclusion for $a_{\hbar}\in S_{\mu}(A_{\hbar})$ the
standard Quantization $Op_{\hbar}(a)$ and the Weyl quantization $Op_{\hbar}^{w}(a_{\hbar})$
are well defined and give continuous operators on $\mathcal{S}(\mathbb{R}^{n})$
respectively on $\mathcal{S}'(\mathbb{R}^{n})$. Furthermore equation
(\ref{eq:AlowerBd}) gives us a second inclusion 
\begin{equation}
S_{\mu}(f)\subset\hbar^{-c}S_{\mu}(A_{\hbar})\label{eq:IncSymbClass2}
\end{equation}
thus combining these two inclusions we have:
\[
\hbar^{c}S_{\mu}(f)\subset S_{\mu}(A_{\hbar})\subset S_{\mu}(f)
\]
As for standard $\hbar-PDO$ symbol we can define asymptotic expansions:%
\begin{framed}%
\begin{defn}
Let $a_{j}\in S_{\mu}(A_{\hbar})$ for $j=0,1,\dots$ then we call
$\sum\limits _{j}\hbar^{j}a_{j}$ an asymptotic expansion of $a\in S_{\mu}(A_{\hbar})$
(writing $a\sim\sum\limits _{j}\hbar^{j}a_{j}$) if and only if: 
\[
a-\sum\limits _{j<N}\hbar^{j}a_{j}\in\hbar^{N}S_{\mu}(A_{\hbar})
\]
\end{defn}
\end{framed}As in for the standard $\hbar$-PDOs we have some sort of Borel's
theorem also for symbols in $S_{\mu}(A_{\hbar})$ 

\begin{framed}%
\begin{prop}
Let $a_{i}\in S_{\mu}(A_{\hbar})$ then there is a symbol $a\in S_{\mu}(A_{\hbar})$
such that 
\begin{equation}
a-\sum\limits _{j<k}\hbar^{j}a_{j}\in\hbar^{k}S_{\mu}(A_{\hbar})\label{eq:AssExp1}
\end{equation}
\end{prop}
\end{framed} 
\begin{proof}
Once more we can use the inclusion (\ref{eq:IncSymbClass}) into the
standard $h-PDO$ classes and obtain the existence of a symbol $a\in S_{\mu}(f)$
such that 
\begin{equation}
a-\sum\limits _{j<k}\hbar^{j}a_{j}\in\hbar^{k}S_{\mu}(f)\label{eq:AssExp2}
\end{equation}
and we will show that this symbol belongs to $S_{\mu}(A_{\hbar})$
and that (\ref{eq:AssExp1}) holds: For the first statement we write
\[
a=\underbrace{a-\sum\limits _{j<c}\hbar^{j}a_{j}}_{\in\hbar^{c}S_{\mu}(f)}+\underbrace{\sum\limits _{j<c}\hbar^{j}a_{j}}_{\in S_{\mu}(A_{\hbar})}
\]
and use the inverse inclusion (\ref{eq:IncSymbClass2}).

In order to prove (\ref{eq:AssExp1}) write 
\[
a-\sum\limits _{j<k}\hbar^{j}a_{j}=\underbrace{a-\sum\limits _{j<k+c}\hbar^{j}a_{j}}_{\in\hbar^{c+k}S_{\mu}(f)}+\underbrace{\sum\limits _{j=k}^{k+c-1}\hbar^{j}a_{j}}_{\in\hbar^{k}S_{\mu}(A_{\hbar})}
\]
and use once more (\ref{eq:IncSymbClass2}).
\end{proof}
The advantage of this new symbol class is, that the order function
$A_{\hbar}(x,\xi)$ itself can depend on $\hbar$ and thus the control
in $\hbar$ can be localized. A simple example for such an order function
would be $A_{\hbar}=\hbar^{m\mu}\langle\frac{\xi}{\hbar^{\mu}}\rangle^{m}\in\mathcal{OF}^{c}(\langle\xi\rangle^{m})$.
For $\xi\neq0$ this function is of order $\hbar^{0}$ whereas for
$\xi=0$ it is of order $\hbar^{m\mu}$. Thus also all symbols in
$S_{\mu}(A_{\hbar})$ have to show this behavior.

\subsection{Composition of symbols}

By using the inclusion (\ref{eq:IncSymbClass}) we will show a result
for the composition of Symbols absolutely analogous to the one in
the standard case Theorem 4.18 in \cite{zworski-03}. We first note
that for $A_{\hbar}\in\mathcal{OF}^{c_{A}}(f{}_{A})$ and $B_{\hbar}\in\mathcal{OF}^{c_{B}}(f_{B})$
the product formula for derivative yields that $A_{\hbar}B_{\hbar}\in\mathcal{OF}^{c_{A}+c_{B}}(f_{A}f_{B})$
and can now formulate the following theorem:%
\begin{framed}%
\begin{thm}
\label{thm:Comp} Let $A_{\hbar}\in\mathcal{OF}^{c_{A}}(f_{A})$ and
$B_{\hbar}\in\mathcal{OF}^{c_{B}}(f_{B})$ be two $\hbar$-dependent
order functions and $a\in S_{\mu}(A_{\hbar})$ and $b\in S_{\mu}(B_{\hbar})$
two $\hbar$-local symbols. Then there is a symbol 
\[
a\#b\in S_{\mu}(A_{\hbar}B_{\hbar})
\]
such that 
\begin{equation}
Op_{\hbar}^{w}(a)Op_{\hbar}^{w}(b)=Op_{\hbar}^{w}(a\#b)\label{eq:CompOp}
\end{equation}
as operators on $\mathcal{S}$ and the at first order we have 
\begin{equation}
a\#b-ab\in\hbar^{1-2\mu}S_{\mu}(A_{\hbar}B_{\hbar})\label{eq:CompFirstOrder}
\end{equation}
\end{thm}
\end{framed}
\begin{proof}
The standard theorem of composition of $\hbar$-PDOs (see e.g. Th
4.18 in \cite{zworski-03}) together with the inclusion of symbol-classes
(\ref{eq:IncSymbClass}) provides us a symbol $a\#b\subset S_{\mu}(f_{A}\cdot f_{B})$
that fulfills equation (\ref{eq:CompOp}). Furthermore it provides
us with a complete asymptotic expansion for $a\#b$: 
\begin{equation}
a\#b-\sum\limits _{k=0}^{N-1}\left(\frac{1}{k!}\left[\frac{i\hbar(\langle D_{x},D_{\eta}\rangle-\langle D_{y},D_{\xi}\rangle)}{2}\right]^{k}a(x,\xi)b(y,\eta)\right)_{|y=x,\eta=\xi}\in\hbar^{N(1-2\mu)}S_{\mu}(f_{A}\cdot f_{B})\label{eq:CompAsymptExp}
\end{equation}
In order to prove our theorem it thus only rests to show, that $a\#b\in S_{\mu}(A_{\hbar}B_{\hbar})$
and that equation (\ref{eq:CompFirstOrder}) holds. We start with
the second one. First let $N\in\mathbb{N}$ be such that $(N-1)(1-2\mu)\geq c_{A}+c_{B}$,
then equation (\ref{eq:CompAsymptExp}) and inclusion (\ref{eq:IncSymbClass2})
assure that the remainder term in (\ref{eq:CompAsymptExp}) is in
$\hbar^{1-2\mu}S_{\mu}(A_{\hbar}B_{\hbar})$. For $0\leq k\leq N-1$
each term in (\ref{eq:CompAsymptExp}) can be written as a sum of
finitely many terms of the form 
\[
\frac{(i\hbar)^{k}}{2^{k}k!}\left(D_{x}^{\alpha}D_{\xi}^{\beta}a(x,\xi)\right)\cdot\left(D_{x}^{\gamma}D_{\xi}^{\delta}b(x,\xi)\right)
\]
where $\alpha,\beta,\gamma,\delta\in\mathbb{N}^{n}$ are multiindices
fulfilling $|\alpha|+|\beta|+|\gamma|+|\delta|=2k$. Via the product
formula one easily checks, that these terms are all in $\hbar^{k(1-2\mu)}S_{\mu}(A_{\hbar}B_{\hbar})$
which proves that $a\#b\in S_{\mu}(A_{\hbar}B_{\hbar})$.
\end{proof}

\subsection{Ellipticity and inverses}

In this section we will define ellipticity for our new symbol classes
and will prove a result on $L^{2}$-invertibility.%
\begin{framed}%
\begin{defn}
We call a symbol $a\in S_{\mu}(A_{\hbar})$ elliptic if there is a
constant $C$ such that: 
\begin{equation}
|a(x,\xi)|\geq CA_{\hbar}(x,\xi)\label{eq:ElipCond}
\end{equation}
\end{defn}
\end{framed}For an $\hbar$-dependent order function $A_{\hbar}\in\mathcal{OF}^{c}(f)$,
from (\ref{eq:dxxiEstimate_for_OF}) and (\ref{eq:AlowerBd}) it follows,
that $\hbar^{c}A_{\hbar}^{-1}\in\mathcal{OF}^{c}(f^{-1})$ is again
a $\hbar$- dependent order function and we can formulate the following
proposition:

\begin{framed}%
\begin{prop}
If $a\in S_{\mu}(A_{\hbar})$ is elliptic then $a^{-1}\in\hbar^{-c}S_{\mu}(\hbar^{c}A_{\hbar}^{-1})$\end{prop}
\end{framed} 
\begin{proof}
We have to show, that $|\partial_{x}^{\alpha}\partial_{\xi}^{\beta}a^{-1}(x,\xi)|\leq C\hbar^{-\mu(|\alpha|+|\beta|)}A_{\hbar}^{-1}(x,\xi)$
uniformly in $\hbar,x$ and $\xi$. For some first derivative (i.e.
for $\alpha\in\mathbb{N}^{2n},|\alpha|=1$) we have 
\[
|\partial_{x,\xi}^{\alpha}a^{-1}|=\frac{|\partial_{x,\xi}^{\alpha}a|}{|a^{2}|}\leq C\frac{\hbar^{-\mu}A_{\hbar}}{A_{\hbar}^{2}}=C\hbar^{-\mu}A_{\hbar}^{-1}
\]
where the inequality is obtained by (\ref{eq:dxxiEstimate_for_OF})
and (\ref{eq:ElipCond}). The estimates of higher order derivatives
can be obtained by induction.
\end{proof}
As for standard $\hbar$-PDOs this notion of ellipticity implies that
the corresponding operators are invertible for sufficiently small
$\hbar$.%
\begin{framed}%
\begin{prop}
\label{prop:L2inverse}Let $A_{\hbar}\in\mathcal{OF}^{c}(1)$ and
$a\in S_{\mu}(A_{\hbar})$ be an elliptic symbol, then $Op_{\hbar}^{w}(a):L^{2}(\mathbb{R}^{n})\to L^{2}(\mathbb{R}^{n})$
is a bounded operator. Furthermore there exists $\hbar_{0}>0$ such
that $Op_{\hbar}^{w}(a)$ is invertible for all $\hbar\in]0,\hbar_{0}]$.
Its inverse is again bounded and a pseudodifferential operator $Op_{\hbar}^{w}(b)$
with symbol $b\in S_{\mu}(A_{\hbar}^{-1})$. At leading order its
symbol is given by 
\[
b-a^{-1}\in\hbar^{1-2\mu}S_{\mu}(A_{\hbar}^{-1})
\]
\end{prop}
\end{framed} 
\begin{proof}
As $a\in S_{\mu}(A_{\hbar})\subset S_{\mu}(1)$ the boundedness of
$Op_{\hbar}^{w}(a)$ follows from theorem 4.23 in \cite{zworski-03}.
By theorem \ref{thm:Comp} we calculate 
\[
Op_{\hbar}^{w}(a)Op_{\hbar}^{w}(a^{-1})=Id+R
\]
where $R=Op_{\hbar}^{w}(r)$ is a PDO with symbol $r\in\hbar^{1-2\mu}S_{\mu}(1)$.
Again from theorem 4.23 in \cite{zworski-03} we obtain $\|R\|_{L^{2}}\leq C\hbar^{1-2\mu}$
thus there is $\hbar_{0}$ such that $\|R\|_{L^{2}}<1$ for $\hbar\in]0,\hbar_{0}]$.
According to theorem C.3 in \cite{zworski-03} we can conclude that
$Op_{\hbar}^{w}(a)$ is invertible and that the inverse is given by
$Op_{\hbar}^{w}(a^{-1})(Id+R)^{-1}$. The semiclassical version of
Beals theorem allows us to conclude that $(Id+R)^{-1}=\sum\limits _{k=0}^{\infty}(-R)^{k}$
is a PDO with symbol in $S_{\mu}(1)$ (cf. theorem 8.3 and the following
remarks in \cite{zworski-03}). The representation of $(Id-R)^{-1}$
as a series finally gives us the symbol of the inverse operator at
leading order. 
\end{proof}

\subsection{Egorov's theorem for diffeomorphisms}

In this section we will study the behavior of symbols $a\in S_{\mu}(A_{\hbar})$
under variable changes. Let $\gamma:\mathbb{R}^{n}\to\mathbb{R}^{n}$
be a diffeomorphism that equals identity outside some bounded set
then the pullback with this coordinate change acts as a continuous
operator on $\mathcal{S}(\mathbb{R}^{n})$ by: 
\[
(\gamma^{*}u)(x):=u(\gamma(x))
\]
Which can be extended by its adjoint to a continuous operator $\gamma^{*}:\mathcal{S}'(\mathbb{R}^{n})\to\mathcal{S}'(\mathbb{R}^{n})$.
By a variable change of an operator we understand its conjugation
by $\gamma$ and we are interested for which $a\in S_{\mu}(A_{\hbar})$
the conjugated operator $(\gamma^{*})^{-1}Op_{\hbar}(a)\gamma^{*}$
is again a $\hbar$-PDO with symbol $a_{\gamma}$. At leading order
this symbol will be the composition of the original symbol with the
so called canonical transformation
\[
T:\mathbb{R}^{2n}\to\mathbb{R}^{2n},(x,\xi)\mapsto(\gamma^{-1}(x),(\partial\gamma(\gamma^{-1}(x)))^{T}\xi)
\]
and the symbol class of $a_{\gamma}$ will be $S_{\mu}(A_{\hbar}\circ T)$.
For the $A_{\hbar}\in\mathcal{OF}^{c}(f)$ defined in Definition \ref{def:h-OrderFunction}
the composition $A_{\hbar}\circ T$ will in general however not be
a $\hbar$-dependent order function itself because the derivatives
in $x$ create a supplementary $\xi$ factor which has to be compensated
(cf. discussion in chapter 9.3 in \cite{zworski-03}). We therefore
demand in this section that our order function $A_{\hbar}$ satisfies:
\[
\left|\partial_{x}^{\alpha}\partial_{\xi}^{\beta}A_{\hbar}(x,\xi)\right|\leq C_{\alpha,\beta}\hbar^{\mu(|\alpha|+|\beta|)}\langle\xi\rangle^{-|\beta|}A_{\hbar}(x,\xi)
\]
A straightforward calculation shows then, that $A_{\hbar}\circ T\in\mathcal{OF}^{c}(f\circ T)$
is again a $\hbar$- dependent order function. The same condition
has to be fulfilled by the symbol of the conjugated operator: 
\begin{equation}
|\partial_{x}^{\alpha}\partial_{\xi}^{\beta}a(x,\xi)|\leq\hbar^{-\mu(|\alpha|+|\beta|)}\langle\xi\rangle^{-|\beta|}A_{\hbar}(x,\xi)\label{eq:EgoDecayXiCondition}
\end{equation}
\begin{framed}%
\begin{thm}
\label{thm:Ego}Let $a\in S_{\mu}(A_{\hbar})$ be an symbol which
fulfills (\ref{eq:EgoDecayXiCondition}) and has compact support in
$x$ (i.e. $\overline{\{x\in\mathbb{R}^{n}|\exists\xi\in\mathbb{R}^{n}:a(x,\xi)\neq0\}}$
is compact) and let $\gamma:\mathbb{R}^{n}\to\mathbb{R}^{n}$ be a
diffeomorphism. Then there is a symbol $a_{\gamma}\in S_{\mu}(A_{\hbar}\circ T)$
such that 
\begin{equation}
(Op_{\hbar}(a_{\gamma})u)(\gamma(x))=(Op_{\hbar}(a)(u\circ\gamma))(x)\label{eq:EgoOpIdentity}
\end{equation}
for all $u\in\mathcal{S}'(\mathbb{R}^{n})$. Furthermore $a_{\gamma}$
has the following asymptotic expansion. 
\begin{equation}
a_{\gamma}(\gamma(x),\eta)\sim\sum\limits _{n=0}^{k-n}\frac{1}{\nu!}\langle i\frac{\hbar}{\langle\eta\rangle}D_{y},D_{\xi}\rangle^{\nu}e^{\frac{i}{\hbar}\langle\rho_{x}(y),\eta\rangle}a(x,\xi)_{\big|y=0,\xi=(\partial\gamma(x))^{T}\eta}\label{eq:EgoAssExp}
\end{equation}
where $\rho_{x}(y)=\gamma(y+x)-\gamma(x)-\gamma'(x)y$. The terms
of the series are in $\hbar^{\frac{\nu(1-2\mu)}{2}}S_{\mu}(\langle\eta\rangle^{\frac{\nu}{2}}A_{\hbar}\circ T(\gamma(x),\eta))$.\end{thm}
\end{framed}We will prove this theorem similar to theorem 18.1.17 in \cite{hormander_3}
by using a parameter dependent stationary phase approximation (Thm7.7.7
in \cite{hormander_1}) as well as the following proposition which
forms the analog to Proposition 18.1.4 of \cite{hormander_3} for
our symbol classes and which we will prove first.

\begin{framed}%
\begin{prop}
\label{prop:AssExp}Let $a(x,\xi;\hbar)\in C^{\infty}(\mathbb{R}^{2n})$
a family of smooth functions that fulfills 
\begin{equation}
|\partial_{x}^{\alpha}\partial_{\xi}^{\beta}a(x,\xi)|\leq C\hbar^{-l}\langle\xi\rangle^{l}f(x,\xi)\label{eq:AssCond1}
\end{equation}
where $C$ and $l$ may depend on $\alpha$ and $\beta$. Let $a_{j}\in S_{\mu}(A_{\hbar})$,
$j=0,1,\dots$ be a sequence of symbols such that 
\begin{equation}
|a(x,\xi)-\sum\limits _{j<k}\hbar^{j}a_{j}(x,\xi)|\leq C\hbar^{\tau k}\langle\xi\rangle^{-\tau k}f(x,\xi)\label{eq:AssCond2}
\end{equation}
where $\tau>0$. Then $a\in S_{\mu}(A_{\hbar})$ and $a\sim\sum\hbar^{j}a_{j}$.\end{prop}
\end{framed}
\begin{proof}
We have to show that for all $k\geq0$ and $g_{k}(x,\xi):=a(x,\xi)-\sum\limits _{j<k}\hbar^{j}a_{j}(x,\xi)$
we have $|\partial_{x}^{\alpha}\partial_{\xi}^{\beta}g_{k}|\leq C\hbar^{k-\mu(|\alpha|+|\beta|)}A_{\hbar}$.
This result can be obtained by iterating the following argument for
the first derivative in $x_{1}$:

Let $e_{1}\in\mathbb{R}^{n}$ be the first eigenvector and $0<\varepsilon<1$.
For arbitrary $j\in\mathbb{N}$ we can write by Taylor's Formula 
\[
|g_{j}(x+\varepsilon e_{1},\xi)-g_{j}(x,\xi)-\partial_{x_{1}}g_{j}(x,\xi)\varepsilon)|\leq C\varepsilon^{2}\sup_{t\in[0,\varepsilon]}|\partial_{x_{1}}^{2}g_{j}(x+te_{1},\xi)|
\]
From (\ref{eq:AssCond1}) and the property, that all $a_{j}$ are
in $S_{\mu}(A_{\hbar})$ we get 
\[
\sup_{t\in[0,\varepsilon]}|\partial_{x_{1}}^{2}g_{j}(x+te_{1},\xi)|\leq C\hbar^{-l}\langle\xi\rangle^{l}f(x,\xi)
\]
for some $l\in\mathbb{R}$ and get 
\[
|\partial_{x_{1}}g_{j}(x,\xi)|\leq C\varepsilon\hbar^{-l}\langle\xi\rangle^{l}m(x,\xi)+\frac{|g_{j}(x+\varepsilon e_{1},\xi)-g_{j}(x,\xi)|}{\varepsilon}
\]
which turns for $j>\frac{2k+2c+l}{\tau}$ and $\varepsilon=\hbar^{k+l+c}\langle\xi\rangle^{-(k+l+c)}$
into: 
\[
|\partial_{x_{1}}g_{j}(x,\xi)|\leq C\hbar^{c+k}\langle\xi\rangle^{-(c+k)}f(x,\xi)\leq C\hbar^{k}A_{\hbar}(x,\xi)
\]
where we used (\ref{eq:IncSymbClass2}) in the second equation. Thus
\[
|\partial_{x_{1}}g_{k}(x,\xi)|\leq C\hbar^{k}A_{\hbar}(x,\xi)+|\sum\limits _{i=k}^{j}\hbar^{i}\partial_{x_{1}}a_{i}(x,\xi)|\leq C\hbar^{k-\mu}A_{\hbar}(x,\xi)
\]
which finishes the proof. 
\end{proof}
After having proven this proposition we can start with the proof of
theorem \ref{thm:Ego}: 
\begin{proof}
If we define 
\begin{equation}
a_{\gamma}(\gamma(x),\eta):=e^{-\frac{i}{\hbar}\gamma(x)\eta}Op_{\hbar}(a)e^{\frac{i}{\hbar}\gamma(\cdot)\eta}\label{eq:EgoDefaGamma}
\end{equation}
then equation (\ref{eq:EgoOpIdentity}) holds for all $e^{\frac{i}{\hbar}x\eta}$
which form a dense subset of $\mathcal{S}'(\mathbb{R}^{n})$. We thus
have to show that $a_{\gamma}$ defined in (\ref{eq:EgoDefaGamma})
is in $S_{\mu}(A_{\hbar})$ and that (\ref{eq:EgoAssExp}) holds.

We will first write $a_{\gamma}$ as an oscillating integral in order
to apply the stationary phase theorem. By definition of $Op_{\hbar}(a)$
one obtains 
\[
a_{\gamma}(\gamma(x),\eta)=\frac{1}{(2\pi\hbar)^{n}}\iint a(x,\tilde{\xi})e^{\frac{i}{\hbar}((x-\tilde{y})\tilde{\xi}+(\gamma(\tilde{y})-\gamma(x))\eta)}d\tilde{y}d\tilde{\xi}
\]
which we can transform by a variable transformation $\tilde{\xi}=\langle\eta\rangle\xi$
and $\tilde{y}=y+x$ into 
\[
a_{\gamma}(\gamma(x),\eta)=\frac{1}{(2\pi\tilde{\hbar})^{n}}\iint a(x,\langle\eta\rangle\xi)e^{\frac{i}{\tilde{\hbar}}(-y\xi+(\gamma(y+x)-\gamma(x))\frac{\eta}{\langle\eta\rangle})}dyd\xi
\]
where $\tilde{\hbar}=\frac{\hbar}{\langle\eta\rangle}$.

The critical points of the phase function are given by 
\[
y=0\textup{ and }\xi=(\partial\gamma(x))^{T}\frac{\eta}{\langle\eta\rangle}
\]
Let $\chi\in C_{c}^{\infty}([-2,2]^{n})$ such that $\chi=1$ on $[-1,1]^{n}$
then we can write 
\[
a_{\gamma}(\gamma(x),\eta)=I_{1}(\tilde{\hbar})+I_{2}(\tilde{\hbar})
\]
with 
\[
I_{1}(\tilde{\hbar})=\frac{1}{(2\pi\tilde{\hbar})^{n}}\iint\chi\left(y\right)\chi\left(\xi-(\partial\gamma(x))^{T}\frac{\eta}{\langle\eta\rangle}\right)a(x,\langle\eta\rangle\xi)e^{\frac{i}{\tilde{\hbar}}(-y\xi+(\gamma(y+x)-\gamma(x))\frac{\eta}{\langle\eta\rangle})}dyd\xi
\]
and 
\[
I_{2}(\tilde{\hbar})=\frac{1}{(2\pi\tilde{\hbar})^{n}}\iint\left(1-\chi\left(y\right)\chi\left(\xi-(\partial\gamma(x))^{T}\frac{\eta}{\langle\eta\rangle}\right)\right)a(x,\langle\eta\rangle\xi)e^{\frac{i}{\tilde{\hbar}}(-y\xi+(\gamma(y+x)-\gamma(x))\frac{\eta}{\langle\eta\rangle})}dyd\xi.
\]
While $I_{1}(\hbar)$ still contains critical points, for $I_{2}(\hbar)$
there are no critical points in the support of the integrand anymore.

$I_{1}$ is of the form studied in theorem 7.7.7 in \cite{hormander_1}.
Here the role of $x$ and $y$ is interchanged and there is an additional
parameter $\frac{\eta}{\langle\eta\rangle}$. We thus get from this
stationary phase theorem 
\begin{equation}
\begin{array}{lc}
\left|I_{1}(\tilde{\hbar})-\sum\limits _{\nu=0}^{k-n}\frac{1}{\nu!}\langle i\tilde{\hbar}D_{y},D_{\xi}\rangle^{\nu}e^{\frac{i}{\tilde{\hbar}}\langle\rho_{x}(y),\frac{\eta}{\langle\eta\rangle}\rangle}u(x,\xi,y,\eta)_{\big|y=0,\xi=(\partial\gamma(x))^{T}\frac{\eta}{\langle\eta\rangle}}\right|\\
\leq C\tilde{\hbar}^{\frac{k+n}{2}}\sum\limits _{|\alpha|\leq2k}\sup_{y,\xi}|D_{y,\xi}^{\alpha}u(x,\xi,y,\eta)|
\end{array}\label{eq:EgoExp1}
\end{equation}
where $u(x,\xi,y,\eta)=\chi\left(y\right)\chi\left(\xi-(\partial\gamma(x))^{T}\frac{\eta}{\langle\eta\rangle}\right)a(x,\langle\eta\rangle\xi)$.
Because of (\ref{eq:EgoDecayXiCondition}) and (\ref{eq:OrderFunctionClassical})
we can estimate 
\[
\sup_{y,\xi}|D_{y,\xi}^{\alpha}u(x,\xi,y,\eta)|\leq C\hbar^{-\mu|\alpha|}f(x,(\partial\gamma(x))^{T}\eta)=C\hbar^{-\mu|\alpha|}f\circ T(\gamma(x),\eta)
\]
Thus transforming the expansion (\ref{eq:EgoExp1}) back to an expansion
in $\hbar$ we get 
\[
\begin{array}{lc}
\left|I_{1}(\hbar)-\sum\limits _{\nu=0}^{k-n}\frac{1}{\nu!}\langle i\frac{\hbar}{\langle\eta\rangle}D_{y},D_{\xi}\rangle^{\nu}e^{\frac{i}{\hbar}\langle\rho_{x}(y),\eta\rangle}u(x,\xi,y,\eta)_{\big|y=0,\xi=(\partial\gamma(x))^{T}\frac{\eta}{\langle\eta\rangle}}\right|\\
\leq C\hbar^{\frac{k(1-2\mu)+n}{2}}\langle\eta\rangle^{-\frac{k+n}{2}}f\circ T(\gamma(x),\eta)
\end{array}
\]
As the stationary points for $I_{2}$ are not contained in the support
of the integrand we get by the non stationary phase theorem: 
\[
|I_{2}(\hbar)|\leq C\left(\frac{\hbar}{\langle\eta\rangle}\right)^{N}f\circ T(\gamma(x),\eta)
\]
for all $N\in\mathbb{N}$. Thus we finally get 
\begin{equation}
\begin{array}{lc}
\left|a_{\gamma}(\gamma(x),\eta)-\sum\limits _{\nu=0}^{k-n}\frac{1}{\nu!}\langle i\frac{\hbar}{\langle\eta\rangle}D_{y},D_{\xi}\rangle^{\nu}e^{\frac{i}{\hbar}\langle\rho_{x}(y),\eta\rangle}u(x,\xi,y,\eta)_{\big|y=0,\xi=(\partial\gamma(x))^{T}\frac{\eta}{\langle\eta\rangle}}\right|\\
\leq C\hbar^{\frac{k(1-2\mu)+n}{2}}\langle\eta\rangle^{-\frac{k+n}{2}}f\circ T(\gamma(x),\eta)
\end{array}\label{eq:EgoExp2}
\end{equation}
If we show that the elements of the series are in $\hbar^{\frac{\nu(1-2\mu)}{2}}S_{\mu}(\langle\eta\rangle^{\frac{\nu}{2}}A_{\hbar}\circ T(\gamma(x),\eta))$
then this equation is of the form (\ref{eq:AssCond2}). The terms
of order $\nu$ in the series are of the form 
\[
\left.\left(\frac{i\hbar}{\langle\eta\rangle}\right)^{\nu}\partial_{y}^{\alpha}e^{\frac{i}{\hbar}\langle\rho_{x}(y),\eta\rangle}(\partial_{\xi}^{\alpha}a)(x,(\partial\gamma(x))^{T}\eta)\langle\eta\rangle^{\nu}\right._{\big|y=0}
\]
Where $\alpha\in\mathbb{N}^{n}$ with $|\alpha|=\nu$. The second
factor $(\partial_{\xi}^{\alpha}a)(x,(\partial\gamma(x))^{T}\eta)\langle\eta\rangle^{\nu}$
is in $\hbar^{-\mu\nu}S_{\mu}(A_{\hbar}\circ T(\gamma(x),\eta))$
as we demanded the condition (\ref{eq:EgoDecayXiCondition}) on our
symbol $a$. Thus it remains to show that the other factor is of order
$\left(\frac{\hbar}{\langle\eta\rangle}\right)^{\frac{\nu}{2}}$ on
the support of $a$. This is the case because $\rho_{x}(y)$ vanishes
at second order in $y=0$. Each derivative of $e^{\frac{i}{\hbar}\langle\rho_{x}(y),\eta\rangle}$
produces a factor $\frac{i}{\hbar}\langle\partial_{y_{i}}\rho_{x}(0),\eta\rangle$.
But as $\partial_{y_{i}}\rho_{x}(0)$ vanishes we need a second derivative,
now acting on $\partial_{y_{i}}\rho_{x}(y)$, in order to get a contribution.
Thus in the worst case $\partial_{y}^{\alpha}e^{\frac{i}{\hbar}\langle\rho_{x}(y),\eta\rangle}$
is of order $\left(\frac{\hbar}{\langle\eta\rangle}\right)^{-\frac{\nu}{2}}$.
Thus we have shown that (\ref{eq:EgoExp2}) is of the form (\ref{eq:AssCond2}).

The last thing that we have to show is thus, that $a_{\gamma}$ fulfills
(\ref{eq:AssCond1}). If we consider the definition (\ref{eq:EgoDefaGamma})
of $a_{\gamma}$ we see that $\partial_{x}^{\alpha}\partial_{\xi}^{\beta}a_{\gamma}(\gamma(x),\eta)$
can be written as a sum of terms of the form $\frac{P(\eta)}{\hbar^{k}}e^{-\frac{i}{\hbar}\gamma(x)\eta}Op_{\hbar}(b)e^{\frac{i}{\hbar}\gamma(\cdot)\eta}$
where $b\in S_{\mu}(A_{\hbar}\langle\xi\rangle^{j})$ and $P(\eta)$
is a polynomial in $\eta$. The constants $j,k$ and the degree of
$P(\eta)$ depend on $\alpha$ and $\beta$. Thus writing these terms
as oscillating integrals and applying the same arguments as above
one gets (\ref{eq:AssCond1}).

We have thus shown that all the conditions for proposition \ref{prop:AssExp}
are fulfilled and can conclude that $a_{\gamma}$ belongs to $S_{\mu}(A_{\hbar})$
and that (\ref{eq:EgoExp2}) is also an asymptotic expansion w.r.t.
the order function $A_{\hbar}$.\end{proof}

\bibliographystyle{plain}
\bibliography{/home/faure/articles/articles}

\end{document}